\documentclass[aps,jmp,amsmath,amssymb,reprint]{revtex4-1}

\usepackage{graphicx}
\usepackage{dcolumn}
\usepackage{bm}
\usepackage{lineno}
\usepackage{lipsum}
\usepackage[colorlinks=true,citecolor=blue]{hyperref}
\usepackage{resizegather}

\newcommand{\leftquote}{\textquotedblleft}

\begin{document}

\date{\today}

\title{Structure and dynamics of a polymer-nanoparticle composite: Effect of nanoparticle size and volume fraction}

\author{Valerio Sorichetti}
\affiliation{Laboratoire Charles Coulomb (L2C), Univ Montpellier, CNRS, Montpellier, France and SPO, Univ Montpellier, INRA, Montpellier SupAgro, Montpellier, France}
\author{Virginie Hugouvieux}
\affiliation{SPO, Univ Montpellier, INRA, Montpellier SupAgro, Montpellier, France}
\author{Walter Kob}
\affiliation{Laboratoire Charles Coulomb (L2C), Univ Montpellier, CNRS, Montpellier, France}

\begin{abstract}

We use molecular dynamics simulations to study a semidilute, unentangled polymer solution containing well dispersed, weakly attractive nanoparticles (NP) of size ($\sigma_N$) smaller than the polymer radius of gyration $R_g$. We find that if $\sigma_N$ is larger than the monomer size the polymers swell, while smaller NPs cause chain contraction. The diffusion coefficient of polymer chains ($D_p$) and NPs ($D_N$) decreases if the volume fraction $\phi_N$ is increased. The decrease of $D_p$ can be well described in terms of a dynamic confinement parameter, while $D_N$ shows a more complex dependence on $\sigma_N$, which results from an interplay between energetic and entropic effects. When $\phi_N$ exceeds a $\sigma_N$-dependent value, the NPs are no longer well dispersed and $D_N$ and $D_p$ increase if $\phi_N$ is increased.

\end{abstract}


\maketitle

\section{Introduction}
Understanding the motion of nanoparticles (NP) and macromolecules in complex fluids, such as polymer solutions and melts, is a problem of broad importance, with applications to many different fields. In material science, understanding how NPs move in a polymer matrix is fundamental for the production of nanocomposites with mechanical, thermal, optical or electrical properties superior to those of pure polymeric materials \cite{balazs2006nanoparticle,winey2007polymer,jancar2010current}. In biophysics, the dynamics of macromolecules in the cytoplasmic environment can have a strong influence on cellular functions, such as enzymatic reactions and self-assembly of cellular structures \cite{zhou2008macromolecular,woodrow2009intravaginal,zhou2013influence}. Also in medicine there is a growing interest in the topic, with the objective to develop new and more efficient forms of NP-mediated drug delivery \cite{soppimath2001biodegradable,lai2007rapid}, a practice which is already in use for cancer treatment \cite{brigger2002nanoparticles,cho2008therapeutic}. 

In the past years, a lot of attention has been dedicated to the study of the motion of polymers and NPs in polymer solutions and melts, using theoretical \cite{cai2011mobility,egorov2011anomalous,yamamoto2011theory,yamamoto2014microscopic,dong2015diffusion} and experimental  \cite{tuteja2007breakdown, grabowski2009dynamics, kohli2012diffusion,gam2011macromolecular,gam2012polymer,lin2013attractive,choi2013universal,kim2012polymer,poling2015size,babaye2014mobility} approaches, as well as computer simulations \cite{bedrov2003matrix,liu2008molecular,kalathi2014nanoparticle,patti2014molecular,li2014dynamic,volgin2017molecular,karatrantos2017polymer}. Polymer-NP mixtures represent a tough challenge for theoretical physics mainly because of the large number of different length scales present: the NP diameter $\sigma_N$, the monomer diameter $\sigma$, the Kuhn length $\ell_K$, the radius of gyration $R_{g}$, and, in the case of concentrated solutions and melts, the mesh size $\xi$ and the diameter of the Edwards tube $a$ \cite{rubinstein2003polymer,teraoka2002polymer}. The behavior of these systems strongly depends on the length and time scale at which they are probed, and in certain conditions it is possible to observe interesting dynamical phenomena, like anomalous diffusion \cite{banks2005anomalous,metzler2016non,cai2011mobility,babaye2014mobility,poling2015size} or the breakdown of the Stokes-Einstein relation \cite{wyart2000viscosity, ould2000molecular, tuteja2007breakdown, liu2008molecular, grabowski2009dynamics, cai2011mobility, kohli2012diffusion, kalathi2014nanoparticle}. Also the interaction between the different components, which depend on the microscopic details, can have a great impact on the system's structure and dynamics \cite{desai2005molecular,yamamoto2011theory,patti2014molecular,liu2011nanoparticle,hooper2006theory,meng2013simulating}.  Understanding how all these factors affect the static and dynamic properties of the NPs in polymer solutions and melts is thus crucial for practical applications. 

When studying polymer-NP mixtures, two main regimes can be identified depending on the NP diameter $\sigma_N$: the \leftquote colloid limit", where the polymers are much smaller than the NPs ($2 R_g/\sigma_N \ll 1$) and the \leftquote protein limit" or \leftquote nanoparticle limit" \cite{bolhuis2003colloid}, where the size of the polymers is larger or comparable to that of the NPs ($2 R_g/\sigma_N \gtrsim 1$). The colloid limit has been studied extensively and it is nowadays well understood in terms of effective depletion pair potentials \cite{asakura1954interaction,poon2002physics}. The protein limit, on the other hand, is much more problematic, since  an accurate description in terms of effective pair potentials is not possible \cite{meijer1994colloids,dijkstra1999phase}. In the present work, we will focus on the protein limit, using molecular dynamics simulations of a coarse-grained model.

With few exceptions \cite{liu2008molecular,li2014dynamic,karatrantos2017polymer}, most of the previous simulation studies of polymer-NP mixtures have focused on the dilute NP regime, in which the NPs can be assumed not to interact with each other and the properties of the polymer solution are expected to be unchanged by the presence of the NPs. Thus, the purpose of the present work is to study the diffusion of polymers and NPs in an unentangled, semidilute polymer solution in a wide range of NP volume fractions and NP diameters, up to values where the interaction between NPs cannot be neglected. 

The paper is organized as follows: In Section~\ref{model}, the model and the simulation method and details are presented. In Section~\ref{structure} we discuss the structural properties of the polymer-NP solution for different NP sizes and volume fractions, with a special focus on the structure of single polymer chains. In Section \ref{dynamics}, we study the dynamical properties of the system in the presence of good NP dispersion, and in particular the diffusion coefficient of the centers of mass of the chains and of the NPs. Finally, in Section~\ref{high_nppf} we investigate the behavior of the system at high NP volume fraction, where the NP dispersion becomes progressively poorer until large polymer-free regions are formed. We conclude with a summary in Section~\ref{summary}.

\section{Model and simulation method} \label{model}

We performed $NVT$ molecular dynamics simulations of a system of $N_p=500$ polymer chains of length (degree of polymerization) $N=100$ and a variable number $N_N$ of nanoparticles of different diameters $\sigma_N$. To simulate the polymer chains, we used the bead-spring model of Kremer and Grest \cite{kremer1990dynamics}. All monomers interact via a Weeks-Chandler-Andersen (WCA) potential \cite{weeks1971role}, 

\begin{equation}
U_{mm}(r) = 
\begin{cases}
4 \epsilon \left[ \left(\frac \sigma r\right)^{12} -\left(\frac \sigma r\right)^6+ \frac 1 4 \right] & r \leq 2^{1/6} \sigma\\
0 & \text{otherwise}.\\
\end{cases}
\label{lj}
\end{equation}

\noindent In addition, bonded monomers interact via a finite extensible nonlinear elastic (FENE) potential,

\begin{equation}
U_{\text{bond}}(r)= -\frac 1 2 k r_0^2 \ln [1-(r/r_0)^2],
\end{equation}

\noindent where $k=30 \epsilon/\sigma^2$ and $r_0 = 1.5 \sigma$. With this choice of parameters the bond length at the minimum of the potential  is $r_b = 0.961$. The combined effect of the FENE and the WCA potentials prevents the chains from crossing each other at the thermodynamic conditions considered here \cite{kremer1990dynamics}.

In the following, all quantities are given in Lennard-Jones (LJ) reduced units. The units of energy, length and mass are respectively $\epsilon$, $\sigma$ and $m$, where $\epsilon$, and $\sigma$ are defined by Eq.~\eqref{lj} and $m$ is the mass of a monomer. The units of temperature, pressure, volume fraction and time are respectively $[T]=\epsilon/k_B, [P]=\epsilon\sigma^{-3},[\phi]= \sigma^{-3}$ and $[t]=\sqrt{m \sigma^2/\epsilon}$.

\begin{figure}
\centering
\includegraphics[width=0.45 \textwidth]{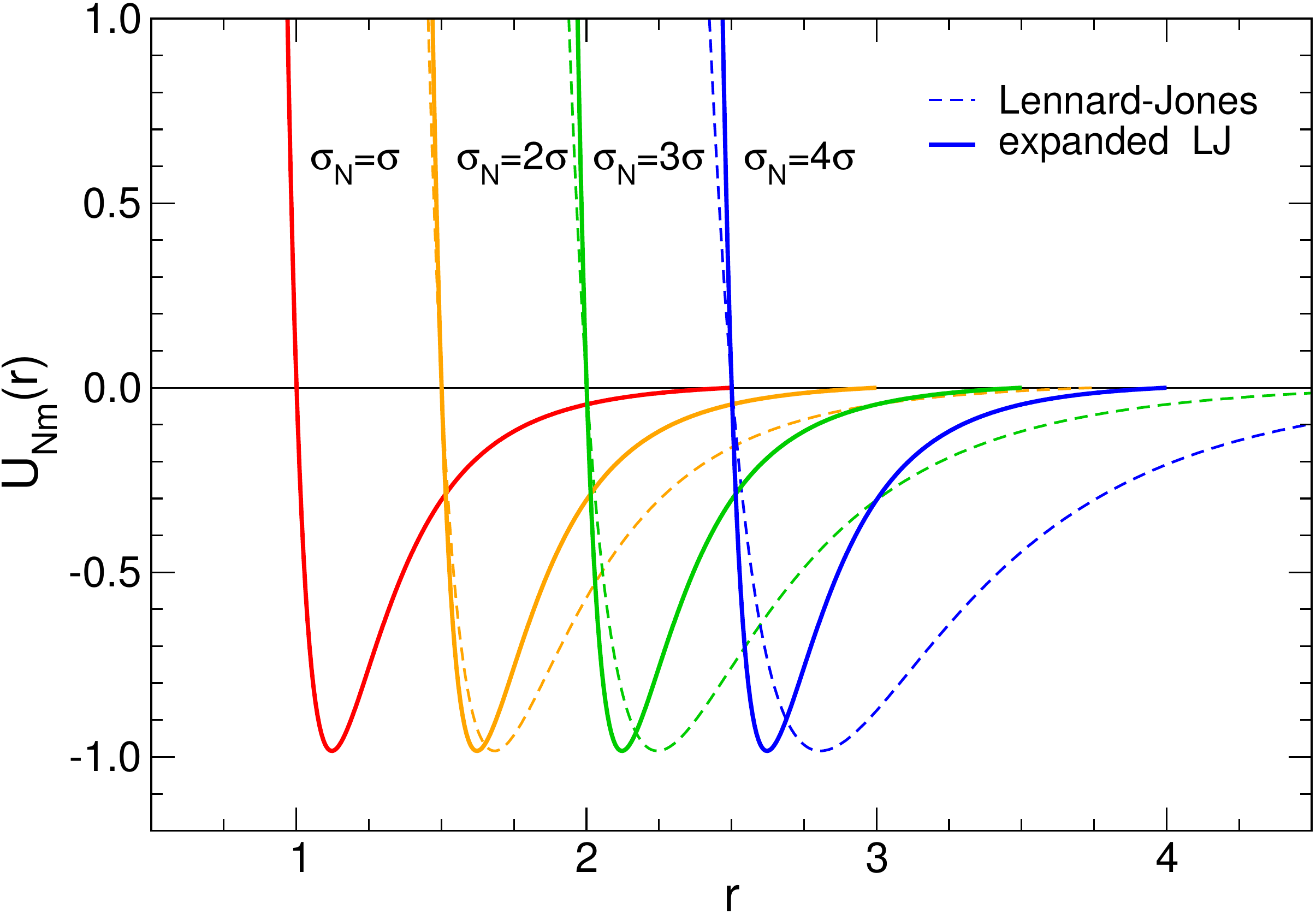}
\caption{Monomer-NP potential $U_{Nm}(r)$ (expanded LJ) compared to a standard LJ potential $U_{LJ}(r)=4 [(\sigma_{Nm}/r)^{12}-(\sigma_{Nm}/r)^6]$, where $\sigma_{Nm}=(\sigma_N+\sigma)/2$. The expanded LJ potential is cut and shifted at $2.5\sigma+(\sigma_N-\sigma)/2$ and the LJ potential is cut and shifted at $2.5 \sigma_{Nm}$. 
For $\sigma_N=\sigma$, the two potentials coincide.}
\label{potentials_cfr}
\end{figure}

For the interaction potentials involving the NPs, we use an \leftquote expanded Lennard-Jones" (expanded LJ) potential, which is a LJ potential shifted to the right by a quantity $\Delta_{ij}$: Thus, as opposed to the standard LJ potential, in the expanded LJ potential the \leftquote softness" (slope) of the potential does not change when the NP size varies, as one can see in Fig.~\ref{potentials_cfr}. Since experiments have shown that the thickness of the interfacial region surrounding a NP in a polymer matrix changes only weakly with the size of the NP \cite{gong2014segmental},  the expanded LJ is a better choice than the standard LJ potential when simulating polymer-NP mixtures \cite{bedrov2003matrix, liu2008molecular, frischknecht2010expanded}.

The interaction between monomers and NPs and between two NPs has thus the following general form:

\begin{gather}
U_{ij}(r) = 
\begin{cases}
4 \epsilon\left[ \left(\frac \sigma {r-\Delta_{ij}}\right)^{12} -\left(\frac \sigma  {r-\Delta_{ij}}\right)^6 \right] +\delta_{ij} &  r \leq r^c_{ij} + \Delta_{ij}\\
0 & \text{otherwise},\\
\end{cases}
\label{uij}
\end{gather}

\noindent where $\Delta_{Nm} =(\sigma_N+\sigma)/2-\sigma=(\sigma_N - \sigma)/2$ and $\Delta_{NN} = \sigma_N - \sigma$. The quantity $\delta_{ij}$ is such that $U_{ij}(r^c_{ij}+\Delta_{ij}) = 0$. The cutoff distances are $r^c_{Nm}=2.5$ for the NP-monomer interaction and  $r^c_{NN}=2^{1/6}$ for the NP-NP interaction. The interaction between monomers and NPs is therefore attractive, while the interaction between NPs is purely repulsive. A moderate attractive interaction between polymers and NPs is required in order to prevent aggregation (and eventually phase separation) of the NPs \cite{hooper2006theory,liu2011nanoparticle,meng2013simulating,karatrantos2015polymer}. 

In this study, we consider NP diameters $\sigma_N=1, 2 ,3 ,4,5$, and $7\sigma$. We assume that the NPs have the same mass density as the monomers, $\rho_\text{mass} =6 m/\pi \sigma^3$, and therefore the mass of the NPs is $m_N = m (\sigma_N/\sigma)^3$. 

We define the NP volume fraction as ${\phi_N=\pi \sigma_N^3 N_N / 6 V}$, where $V$ is the total volume of the simulation box; the monomer volume fraction $\phi_m$ is defined in an analogous way. In our simulations, $\phi_m$ is larger than the overlap volume fraction \cite{rubinstein2003polymer}, which can be estimated from the polymer's radius of gyration at infinite dilution (see below) and for the pure polymer system has the value $\phi_m^*=2.98 \cdot 10^{-2}$. Moreover, since the entanglement length $N_e$ \cite{rubinstein2003polymer} for this model is $N_e\approx 85$ at $\rho_m=0.85$ \cite{hoy2009topological} and since $N_e$ scales approximately as $\rho_m^{-2}$ \cite{fetters1994connection,fetters1999packing,kroger2000rheological}, we are always in the unentangled regime \footnote{In our simulations ${N=100}$ and ${\rho_m<0.32}$, therefore a crude estimate gives ${N_e \gtrsim 85 \cdot (0.85/0.32)^2 = 600}$.}.

All the simulations were carried out using the LAMMPS software \cite{plimpton1995fast}. The simulation box is cubic and periodic boundary conditions are applied in all directions. The initial configurations are prepared by randomly placing the polymers and the NPs in the box; initially, the NPs have diameter equal to that of the monomers ($\sigma_N = \sigma$) and overlaps between particles are allowed. The overlaps are then removed by using a soft potential whose strength is increased over a short amount of time (\leftquote fast push-off" method \cite{auhl2003equilibration}). After the overlaps are removed, the diameter of the NPs is gradually increased until the desired value is reached, and finally the system is allowed to adjust its density until we reach pressure $P=0.1$ at temperature $T=1.0$. In the pure polymer systems, these parameters correspond to a monomer volume fraction $\phi_m=0.147$ (monomer density $\rho_m=0.280$). Finally, we switch to the $NVT$ ensemble and perform an equilibration run before starting the production run. During the $NVT$ simulations, the pressure fluctuations are always less than $14\%$. 

The length of both the equilibration and the production runs is $10^8 \delta t = 3 \cdot 10^5$, where $\delta t = 3 \cdot 10^{-3}$ is the integration time step. In all cases, we verified that during the equilibration runs the NPs (polymers) diffused on average over a distance equal to several times their diameter (radius of gyration), and that their motion became diffusive (see below).

Both during the $NPT$ and the $NVT$ runs, the temperature is kept fixed by means of a Langevin thermostat, so that the force experienced by a particle $i$ (monomer or NP) is 
\begin{equation}
m_i \ddot {\mathbf r}_i = - \mathbf \nabla_i U (\{ \mathbf r_k \}) - m_i \Gamma_i \dot {\mathbf r}_i + \sqrt{2m_i\Gamma_i k_B T} \ \boldsymbol{\zeta}(t),
\label{langevin}
\end{equation}

\noindent where $\mathbf r_i$ is the position vector, $m_i$ the mass and $U (\{ \mathbf r_k \})$ is the total interaction potential acting on the particle, with $\{ \mathbf r_k \}$ representing the set of coordinates of all the particles in the system. The second term of the right side of Eq.~\eqref{langevin} represents viscous damping, with $\Gamma_i$ the friction coefficient, and the last term is a random, uncorrelated force representing the collisions with solvent particles. The Langevin thermostat acts therefore as an \emph{implicit solvent}, in which every particle interacts independently with the solvent \leftquote molecules", but hydrodynamic interactions between solute particles are not accounted for.  We note that it has been recently pointed out that hydrodynamic interactions can affect the long-time dynamics of NPs in a polymer solution even at high monomer volume fractions \cite{chen2017effect}, an observation which warrants further investigation.

The damping constant for the monomers is $\Gamma_m= 0.1$, while that of the NPs is chosen by imposing that the viscosity of the pure solvent  calculated via the Stokes formula, $\eta_s = \Gamma_i m_i/ 3 \pi \sigma_i$, is the same for both free monomers and NPs. Therefore we have $\Gamma_N=\Gamma_m \sigma_N/m_N$. For a more detailed discussion, see Sec.~\ref{sec:thermostat} in the S.I.

Additional details about the simulations can be found in Tab.~\ref{tab:data1} in the S.I.

\section{Structure}\label{structure}

\subsection{Nanoparticles} \label{np_structure}

\begin{figure}
\centering
\includegraphics[width=0.5 \textwidth]{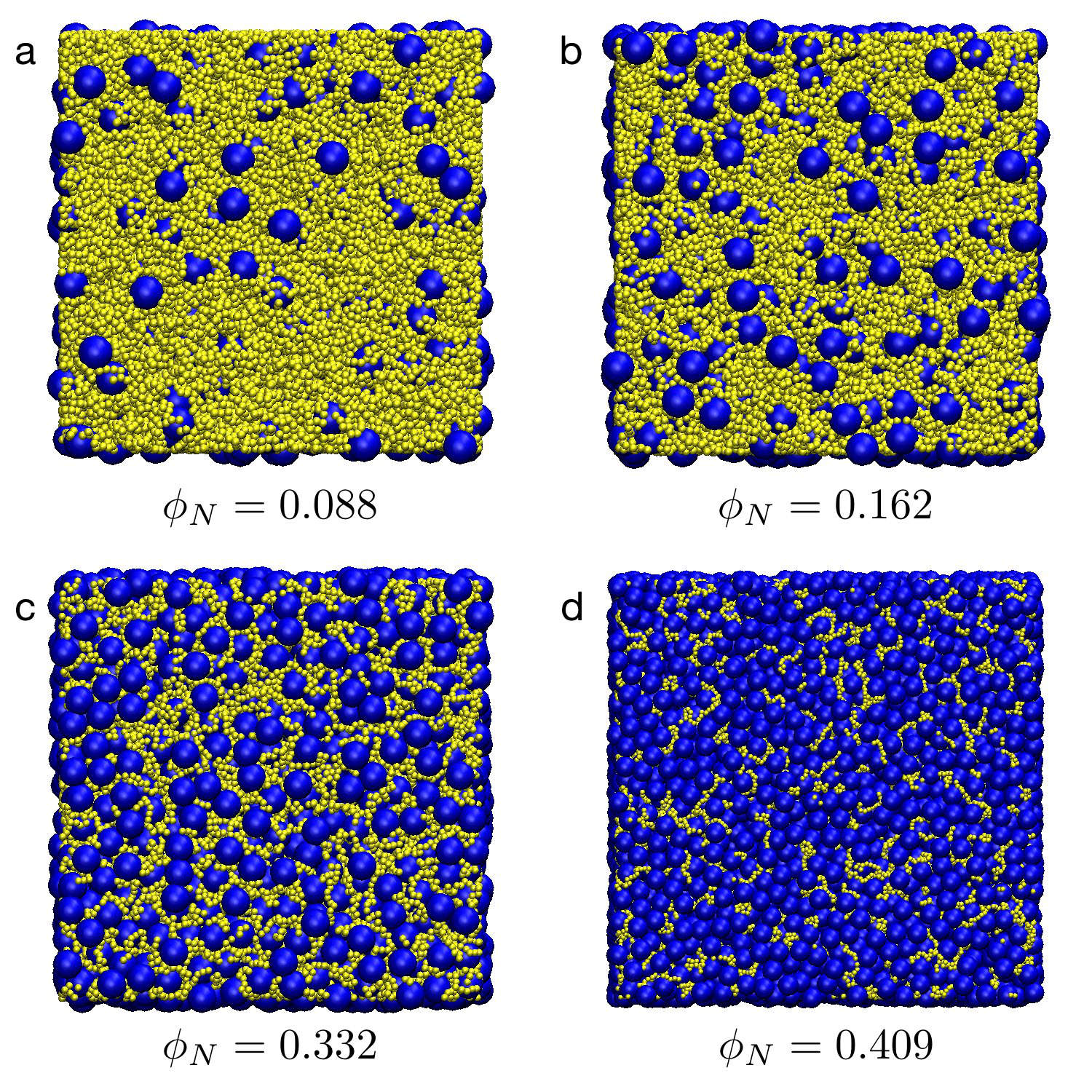}
\caption{Snapshots of systems containing NPs of diameter $\sigma_N=4$ at different volume fractions (yellow spheres: monomers, blue spheres: NPs). In systems (a) and (b), the dispersion of NPs is good, while in (c) and (d) it is poor. In system (d) the formation of large polymer-free regions is evident. The length of the simulation box edges are respectively $57.45$ (a), $59.15$ (b), $67.13$ (c), and $93.57$ (d). }
\label{snapshots}
\end{figure}

To give a feeling of what the simulated system looks like, we show in Fig.~\ref{snapshots} some snapshots for $\sigma_N=4$ and different values of the NP volume fraction $\phi_N$. We can see how the NP dispersion, which is initially good (Figs. \ref{snapshots}a-b), becomes progressively poorer as $\phi_N$ is increased (Fig.~\ref{snapshots}c), until eventually large polymer-free regions are formed (Fig.~\ref{snapshots}d). In order to characterize the structure of the systems when $\sigma_N$ and $\phi_N$ are varied, we start by analyzing some basic quantities, such as the radial distribution function $g(r)$ and the structure factor $S(q)$.

The radial distribution function $g(r)$ can be obtained from the pair correlation function $g(\mathbf r)$ by performing a spherical average \cite{tuckerman2010statistical}. We recall that the pair correlation function of a system of $M$ particles with number density $\rho$ is defined as \cite{hansen1990theory}

\begin{equation} 
\rho g(\mathbf r) = \frac{1}{M} \sum_{\substack{k=1\\ j\neq k}}^{M}  \langle \delta(\mathbf r + \mathbf r_k- \mathbf r_j) \rangle,
\label{rdf}
\end{equation}

\noindent where $\langle \cdot \rangle$ denotes the thermodynamic average.

\begin{figure}
\centering
\includegraphics[width=0.45 \textwidth]{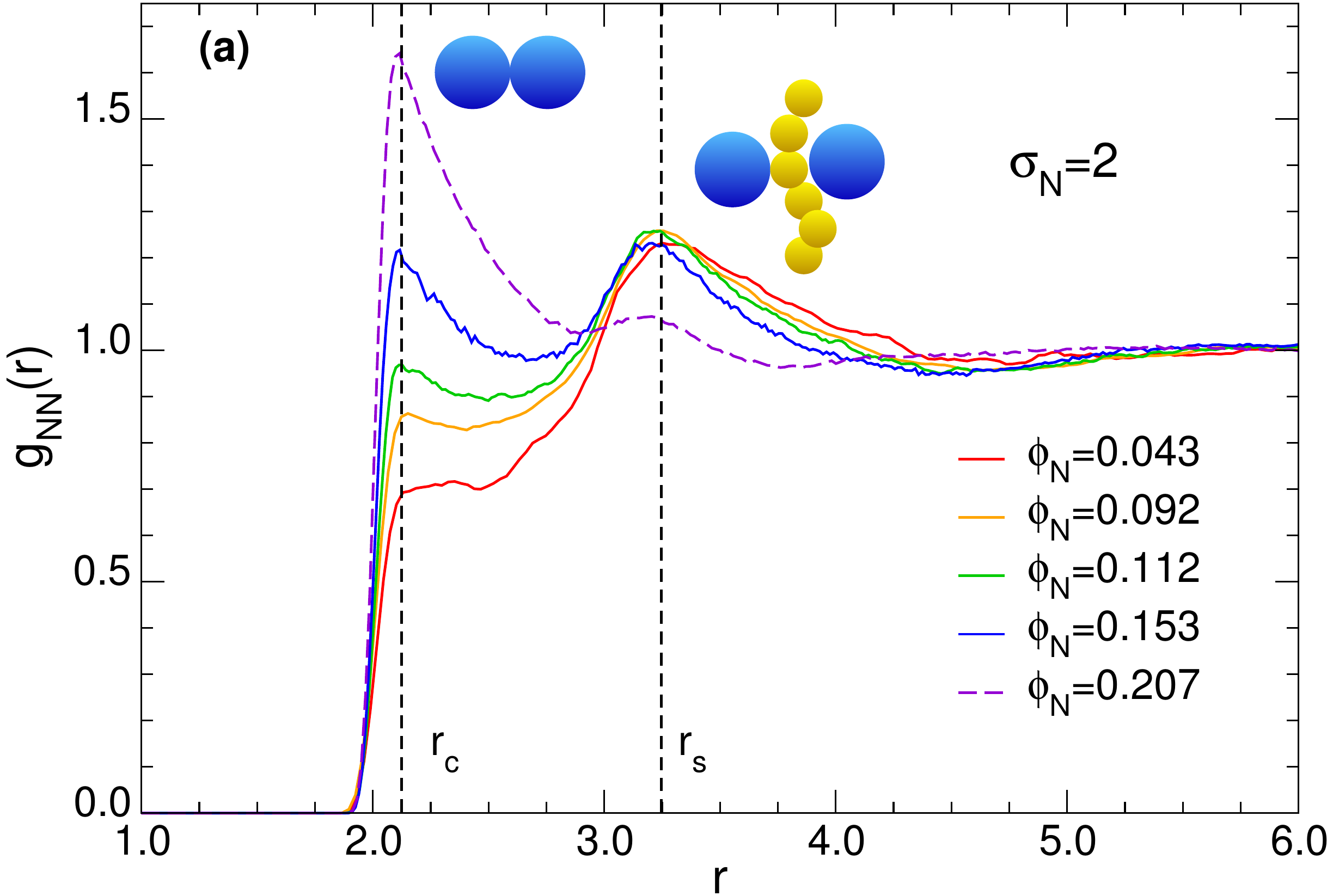}
\includegraphics[width=0.45 \textwidth]{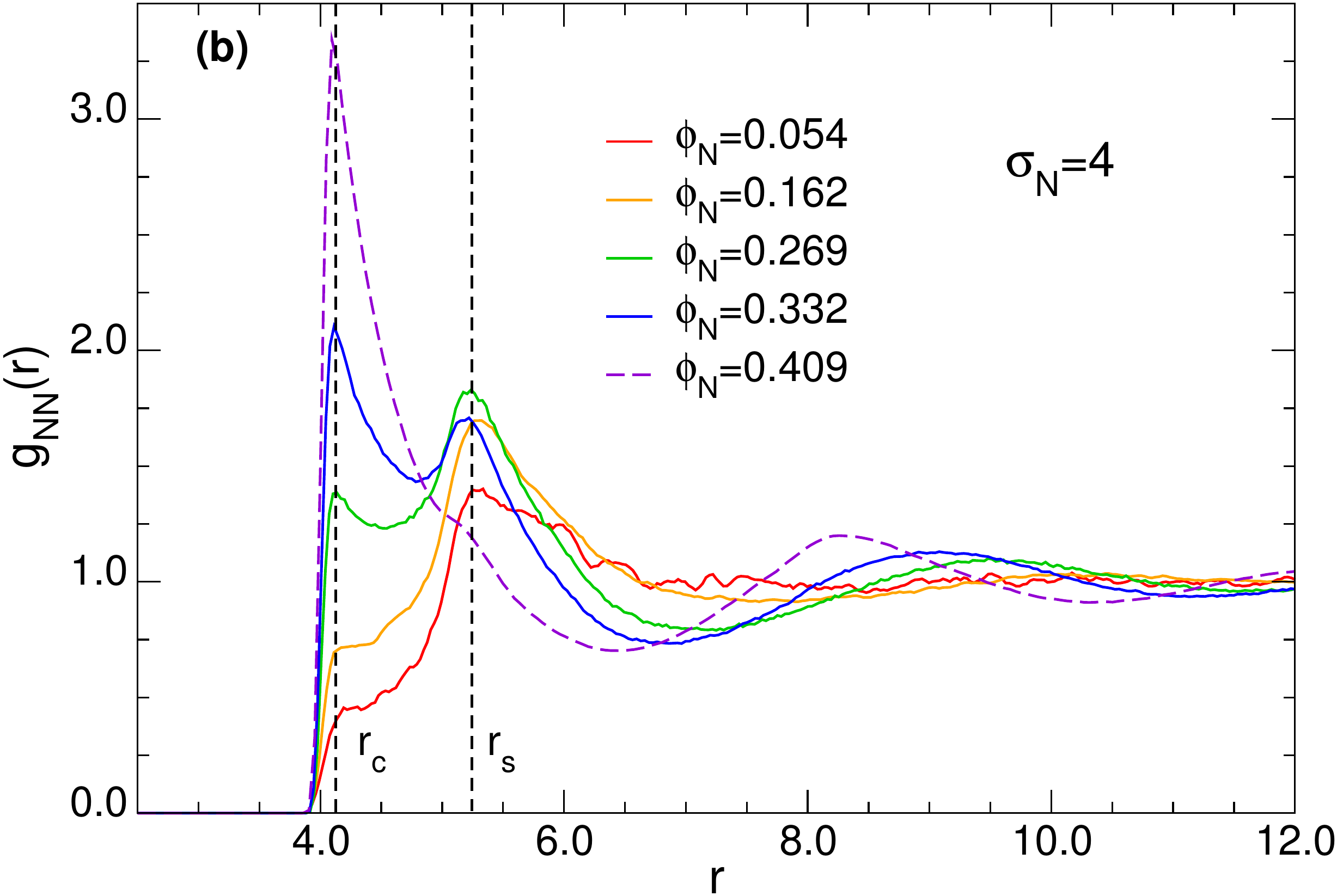}
\caption{(color online) NP-NP radial distribution function at different NP volume fractions $\phi_N$ for $\sigma_N=2$ (a) and $\sigma_N=4$ (b). The \emph{contact peak} at $r_c=\sigma_N+0.122$ corresponds to a configuration in which two NPs are touching, while the \emph{secondary peak} at $r_s=\sigma_N+1.245$ represents a configuration in which two NPs are separated by a polymer strand (a schematic representation of the two cases is shown in (a)): At low and intermediate $\phi_N$, this last configuration is favored.}
\label{rdf_pp}
\end{figure}

\noindent Figure~\ref{rdf_pp} shows the NP-NP radial distribution function $g_{NN}(r)$ for $\sigma_N=2$ and $4$ and different values of the NP volume fraction $\phi_N$. For low values of $\phi_N$, $g_{NN}$ shows a peak at $r_s=\sigma_N+(2^{7/6}-1)=\sigma_N+1.245$, which corresponds to twice the distance at the minimum of the monomer-NP potential. This indicates that the NPs are well dispersed in the polymer solution and configurations in which two neighboring NPs are separated by a polymer strand are favored (this kind of configuration is schematically represented in Fig.~\ref{rdf_pp}a). We call this peak \emph{secondary peak}.

When $\phi_N$ increases, another peak appears at $r_c=\sigma_N+(2^{1/6}-1)=\sigma_N+0.122$, which corresponds to the cutoff of the NP-NP potential and represents a configuration in which two NPs are touching; we therefore call it \emph{contact peak}. Eventually, the contact peak becomes higher than the secondary peak, an evidence of the formation of large polymer-free regions (Fig.~\ref{snapshots}d). 

The structure factor is defined as \cite{hansen1990theory}

\begin{equation} 
S(\mathbf q) = \frac 1 M\sum_{k,j=1}^{M} \langle \exp [-i \mathbf q \cdot (\mathbf r_k - \mathbf r_j)] \rangle,
\label{sq}
\end{equation}

\noindent where $\mathbf q$ is the wavevector. Since our configurations are isotropic, also in this case we will consider the spherically averaged structure factor $S(q)$ \cite{tuckerman2010statistical}.

\begin{figure}
\centering
\includegraphics[width=0.44 \textwidth]{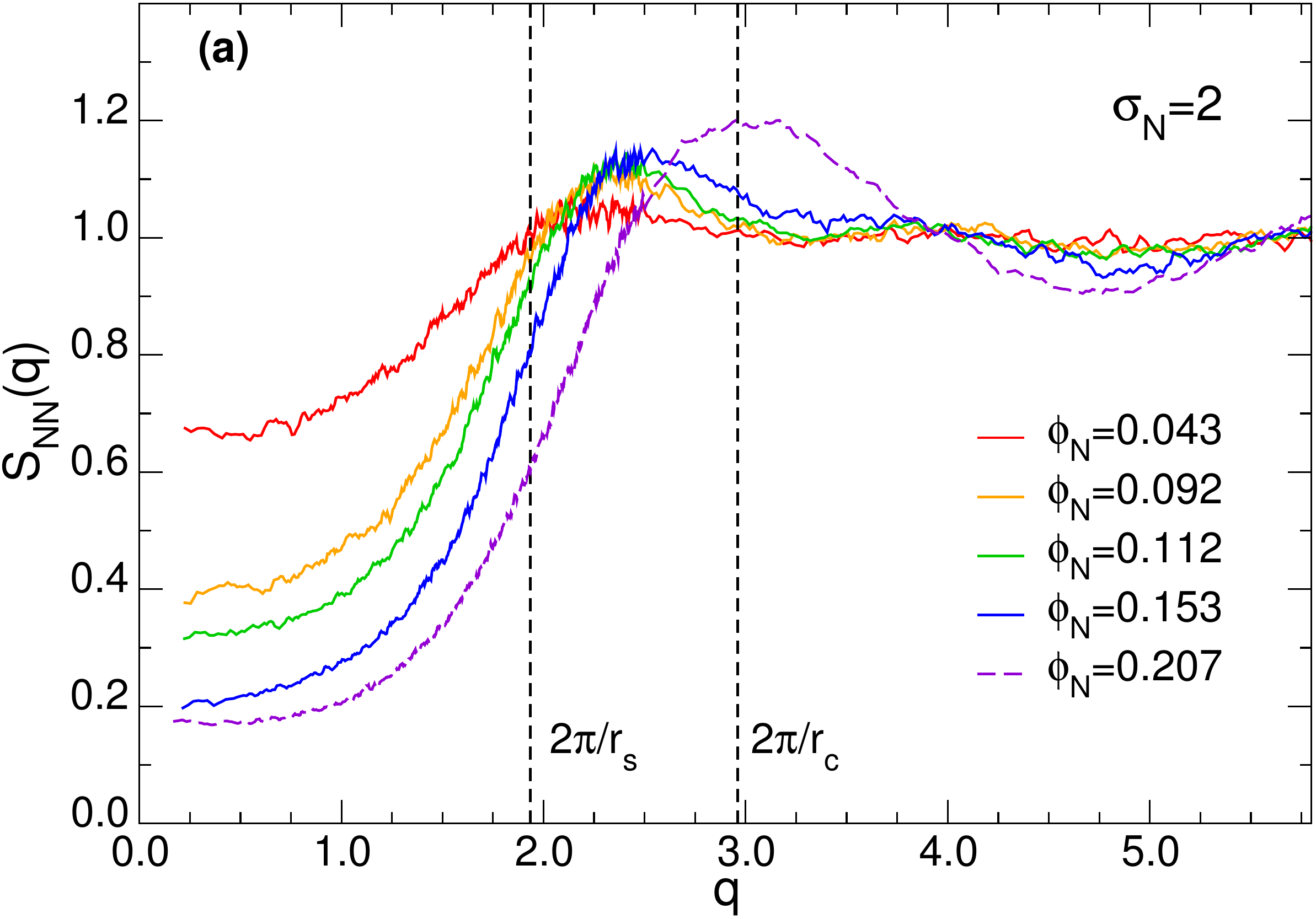}
\includegraphics[width=0.45 \textwidth]{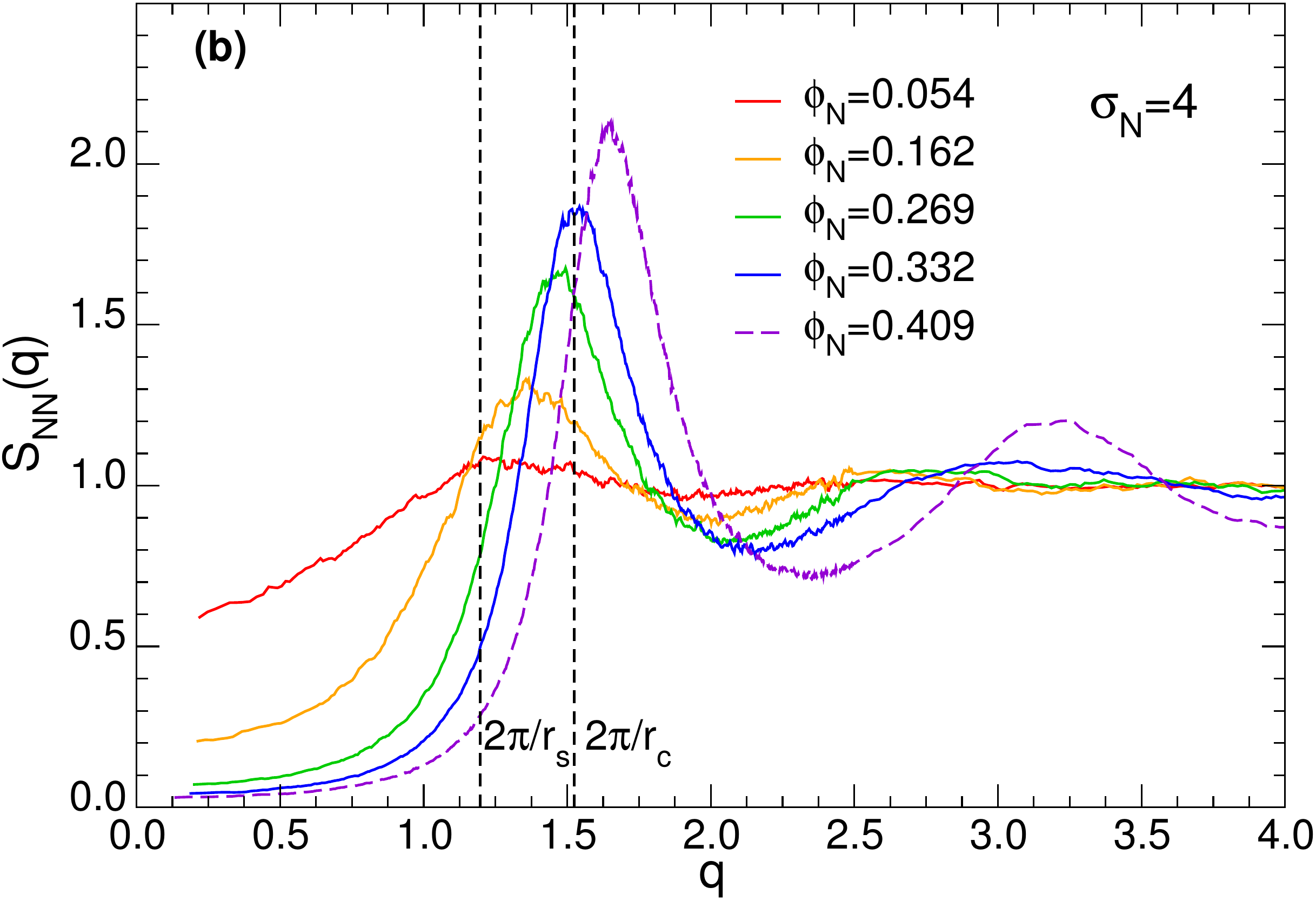}
\caption{(color online) NP-NP structure factor at different NP volume fractions $\phi_N$ for $\sigma_N=2$ (a) and $\sigma_N=4$ (b).}
\label{sq_pp}
\end{figure}

Because $S(q)-1$ is the Fourier transform of $\rho g(r)$ \cite{hansen1990theory},  we can in principle find in the NP-NP structure factor $S_{NN}(q)$ the same information that we find in $g_{NN}(r)$. If the position of the main peak of $g(r)$ is $r=r_0$, the main peak of $S(q)$ will be at $q_0 \approx 2 \pi / r_0$, although the precise value of $q_0$ depends on temperature and density \cite{verlet1968computer}. Hence, we expect to find the main peak of $S_{NN}(q)$ at $q \approx 2 \pi / r_s$ at low NP volume fraction and at $q \approx 2 \pi / r_c$ at high NP volume fraction, as we indeed observe in Figs. \ref{sq_pp}a-b. We also notice that, while in the $g(r)$ we can clearly distinguish two peaks at intermediate values of $\phi_N$ (Fig.~\ref{rdf_pp}), in the $S(q)$ their contributions interfere with each other and result in a single peak that is shifted towards higher wavevectors as $\phi_N$ is increased. Therefore, interpretation of $S_{NN}(q)$ might not always be straightforward.

In what follows, we will mainly consider those systems in which the NPs are well dispersed in the polymer solution (Fig.~\ref{snapshots}a-b). As a qualitative criterion, we define a system with good NP dispersion as one where the secondary peak of $g_{NN}(r)$ is higher than or comparable to the contact peak. It should be noticed that the maximum volume fraction that we can reach while keeping a good NP dispersion depends on the NP diameter $\sigma_N$. To see this, we consider the \emph{interparticle distance} $h$ \cite{gam2011macromolecular,gam2012polymer,lin2013attractive,choi2013universal}, which represents the average spacing between the surfaces of neighboring nanoparticles. In the literature, the following expression for $h$ has often been used \cite{gam2011macromolecular,gam2012polymer,lin2013attractive,choi2013universal}:

\begin{equation}
h^\text{th.}=\sigma_N\left[\left(\frac{\phi_{N}^\text{\tiny M}}{\phi_N}\right)^{1/3}-1\right],
\label{id}
\end{equation}

\noindent where $\phi_{N}^\text{\tiny M}$ represents the maximum NP volume fraction, at which $h=0$. The value ${\phi_{N}^\text{\tiny M} = 2 / \pi \approx 0.637}$, corresponding to the (ill-defined \cite{torquato2000random}) random close packing, is often employed \cite{gam2011macromolecular,gam2012polymer,lin2013attractive,choi2013universal}. However, this definition presents some issues (Sec.~\ref{sec:pore_size} in the Appendix), and therefore we have chosen to measure $h$ directly from the data using the pore-size distribution \cite{torquato2013random}. This approach is similar to the one used by Li \emph{et al.} in Ref.~\citenum{li2014dynamic}, with the difference that they used an Euclidean distance map. For the details on how $h$ can be extracted from the pore-size distribution, see Sec.~\ref{sec:pore_size} in the Appendix.
 
In Fig.~\ref{id_vs_nppf}, we show the interparticle distance calculated from the pore-size distribution, $h$, versus the NP volume fraction: filled (open) symbols represent systems with a good (poor) NP dispersion (according to the above defined criterion). We also report for comparison the \leftquote theoretical" interparticle distance $h^\text{th.}$, Eq.~\eqref{id}, with $\phi_{N}^\text{\tiny M}=0.637$ (continuous lines); we note that the two quantities are very similar, with $h^\text{th.}$ being on average slightly larger than $h$. As we can see, NP dispersion starts to become poor when $h \approx 1$, i.e., when the average distance between the surface of neighboring NPs becomes comparable with the monomer size, in qualitative agreement with the snapshots shown in Fig.~\ref{snapshots}c-d. 

\begin{figure}
\centering
\includegraphics[width=0.45 \textwidth]{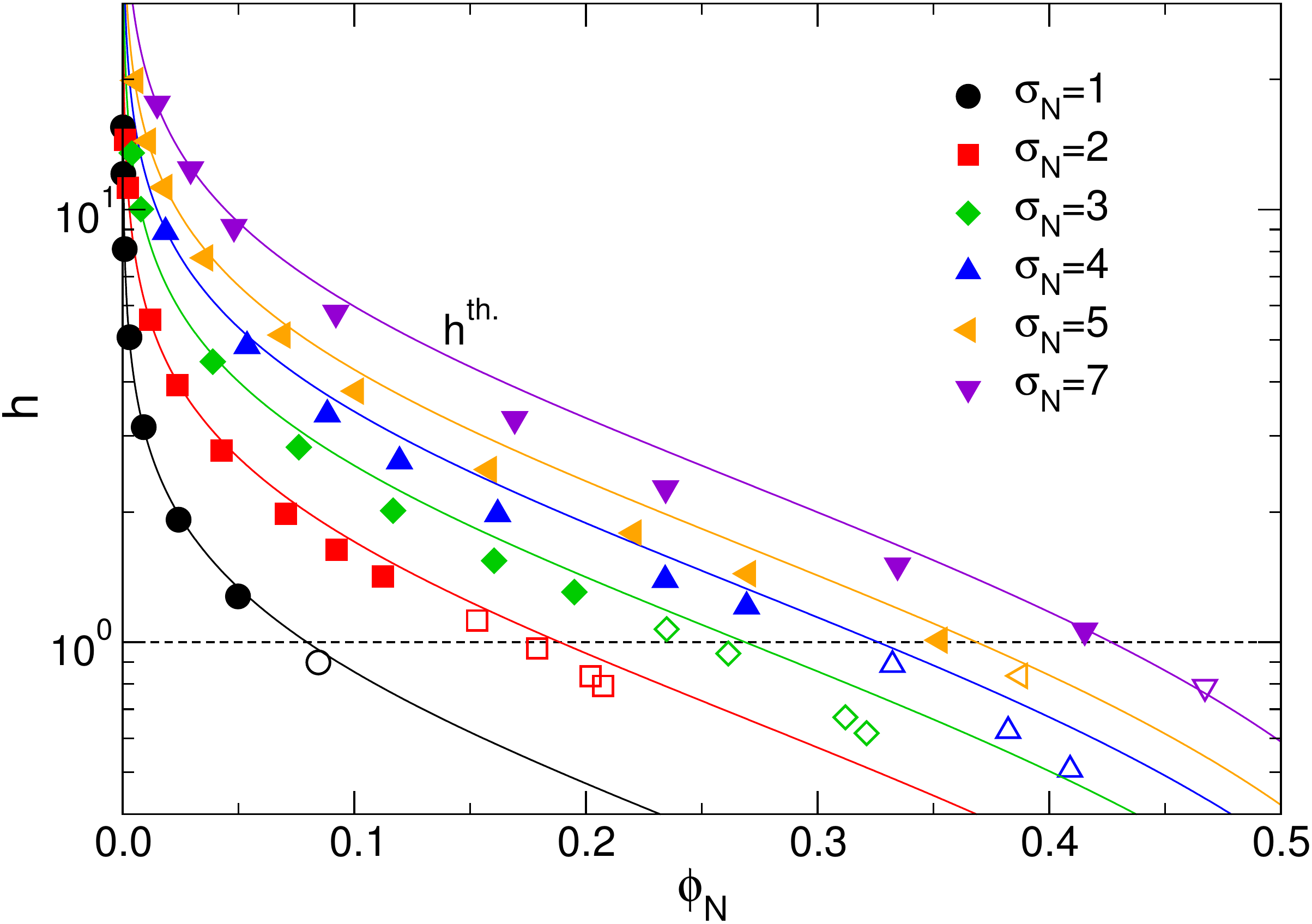}
\caption{Interparticle distance as a function of the NP volume fraction $\phi_N$. Filled symbols: systems with good NP dispersion. Open symbols: systems with poor NP dispersion. Continuous lines: \leftquote theoretical" interparticle distance $h^\text{th.}$, calculated from Eq.~\eqref{id}, with $\phi_{N}^\text{\tiny M}=0.637$.}
\label{id_vs_nppf}
\end{figure}

\subsection{Polymers}\label{polymer_structure}

\begin{figure}
\centering
\includegraphics[width=0.45 \textwidth]{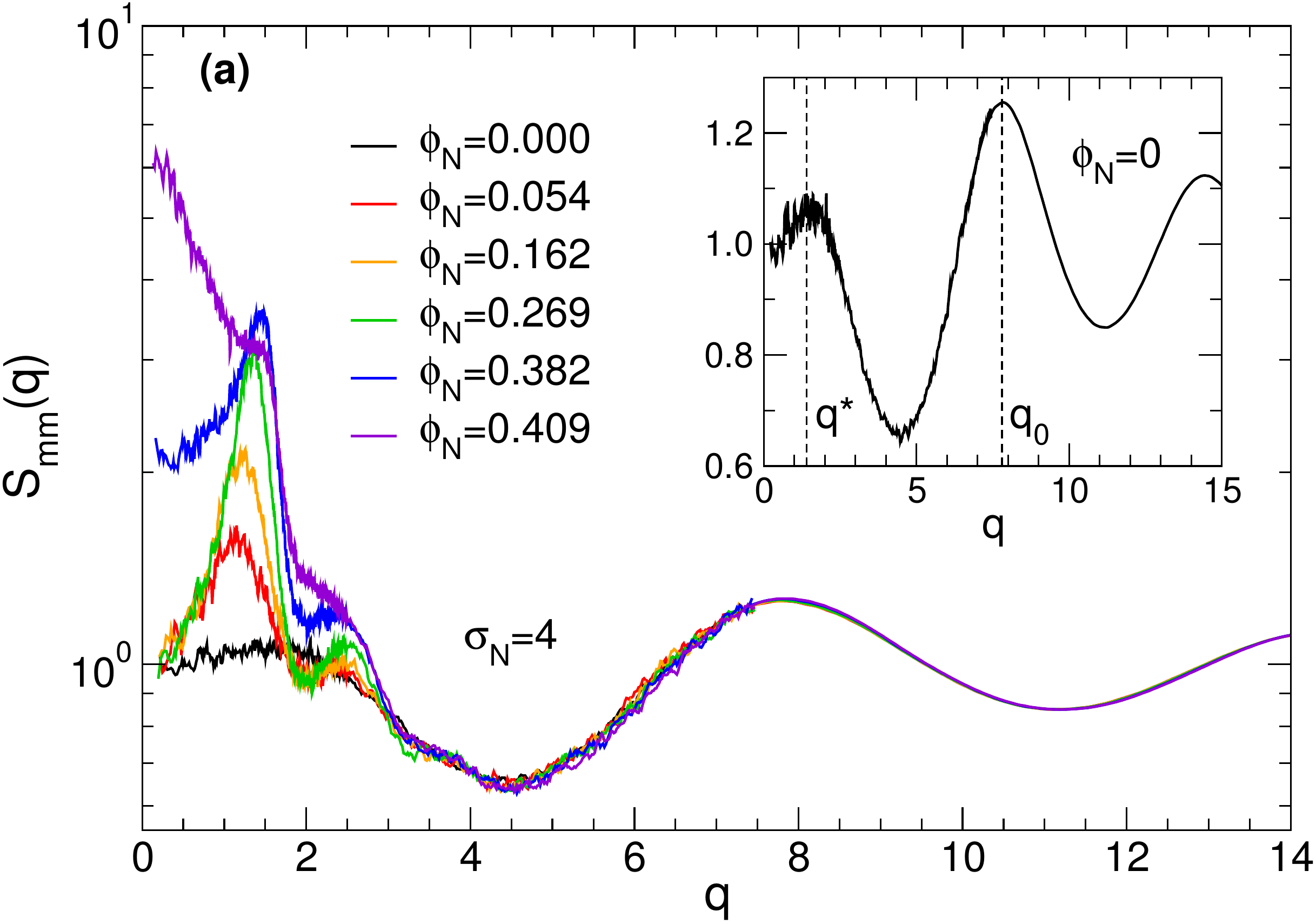}
\includegraphics[width=0.45 \textwidth]{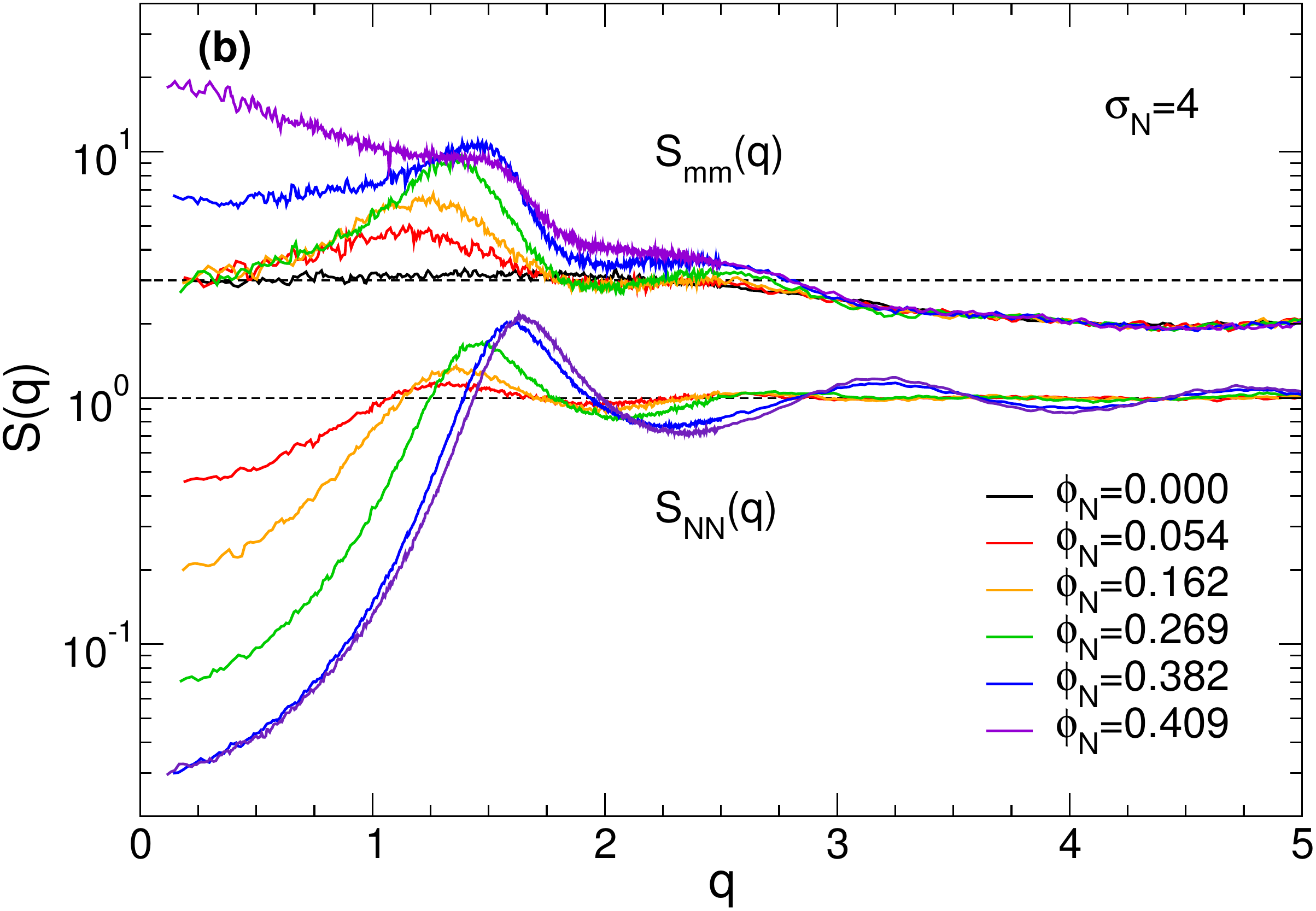}
\caption{(color online)  (a) Monomer-monomer structure factor $S_{mm}(q)$ (log-lin scale) at different NP volume fractions for $\sigma_N=4$. \emph{Inset}: $S_{mm}(q)$ in the pure polymer solution (linear scale) (b) Comparison between the structure factors of monomers and NPs for $\sigma_N=4$. For clarity, the curves representing $S_{mm}(q)$ have been shifted up by a factor of $3$. }
\label{sq_mm}
\end{figure}

The radial distribution function of the monomers, $g_{mm}(r)$, is strongly dominated by the short-distance signals coming from the chain bonds (see Fig.~\ref{rdf_mm} in S.I.), and therefore it is not easy to extract from it information about the medium and long range distribution of the monomers. Hence we focus our attention on the monomer-monomer structure factor, $S_{mm}(q)$, shown in Fig.~\ref{sq_mm}. 

Figure~\ref{sq_mm}a, shows $S_{mm}(q)$ for $\sigma_N=4$. At $\phi_N=0$ (pure polymer solution), there is a small peak at ${q^*\approx 1.4}$ (inset of Fig.~\ref{sq_mm}a), which in real space corresponds to a distance ${r^*=2 \pi / q^*\approx4.5}$. This peak reveals the presence of a typical length scale in the NP-free system, which can be interpreted as the average size of the holes in the polymer matrix \cite{testard2014intermittent}. The main peak of $S_{mm}(q)$ is at ${q_0=7.8 \approx 2 \pi / r_b}$, where $r_b = 0.96$ is the average monomer-monomer bond length.

For ${\phi_N>0}$, the spatial arrangement of the NPs starts to be visible as a modulation in $S_{mm}(q)$, with a main peak appearing approximately at the same wavevector as the main peak of $S_{NN}(q)$, as we can see from Fig.~\ref{sq_mm}b, where $S_{mm}(q)$ is compared to $S_{NN}(q)$. At even higher NP volume fraction, a signal starts to appear at $q=0$, due to the fact that the polymers are getting far from each other (see Fig.~\ref{snapshots}d). If $\phi_N$ was increased even more, eventually the monomer volume fraction would become smaller than the overlap volume fraction (dilute regime) and $S_{mm}(0)$ would saturate to $N$ \cite{rubinstein2003polymer}.

Another important quantity to characterize the structure of the polymer mesh is the correlation length or mesh size $\xi$, which for the pure polymer solution ($\phi_N=0$) can be estimated via scaling considerations \cite{teraoka2002polymer,rubinstein2003polymer}:

\begin{equation}
\xi = 
\begin{cases}
\left({ R_g^*}/\sqrt{3}\right)& \rho_m < \rho_m^*\\
\left({ R_g^*}/\sqrt{3}\right)(\rho_m/ \rho_m^*)^{-\nu/(3\nu -1)} & \rho_m > \rho_m^*,\\
\end{cases}
\label{xi}
\end{equation}

\noindent where ${R_g^*}$ is the radius of gyration of an isolated chain, $\rho_m^*$ is the overlap monomer concentration and  $\nu=0.588$ is the Flory exponent \cite{rubinstein2003polymer} \footnote{The factor $1/\sqrt{3}$ is justified by considering that when $\rho_m$ is not too high the mesh size can be extracted from the low-$q$ limit of the monomer-monomer structure factor, which is given by the Ornstein-Zernike relation ${S_{mm}(q) = S_{mm}(0)/[1+(\xi q)^2]}$, and that at low monomer densities and low wavevectors ${S_{mm}(q) \approx S_1(q)  = N /[1+(R_g^* q/\sqrt{3})^2]}$, where $S_1(q)$ is the single chain monomer-monomer structure factor \cite{rubinstein2003polymer}. From the previous relations, it follows that for ${\rho_m < \rho_m^*}$ we have ${\xi = R_g^*/\sqrt{3}}$.}.

In this work, ${R_g^*}=7.48$ and consequently, defining ${\rho_m^*=N (4 \pi R_g^{*3}/3)^{-1}}$ (other definitions are possible  \cite{teraoka2002polymer}), we get ${\rho_m^*= 5.70 \cdot 10^{-2}}$. Using these values we obtain from Eq.~\eqref{xi} ${\xi = 1.27}$ for the pure polymer solution ($\rho_m=0.280$). 

The structure of the individual polymer chains can be characterized by the function ${p(r)=4\pi r^2 \rho g_1(r) / (N-1)}$, where $\rho g_1(r)$ is obtained by applying Eq.~\eqref{rdf} to a single polymer chain (and, as usual, taking the spherical average). The quantity $p(r)dr$ represents the probability to find a monomer belonging to the same chain at distance between $r$ and $r+dr$ from a given monomer.

\begin{figure}
\centering
\includegraphics[width=0.45 \textwidth]{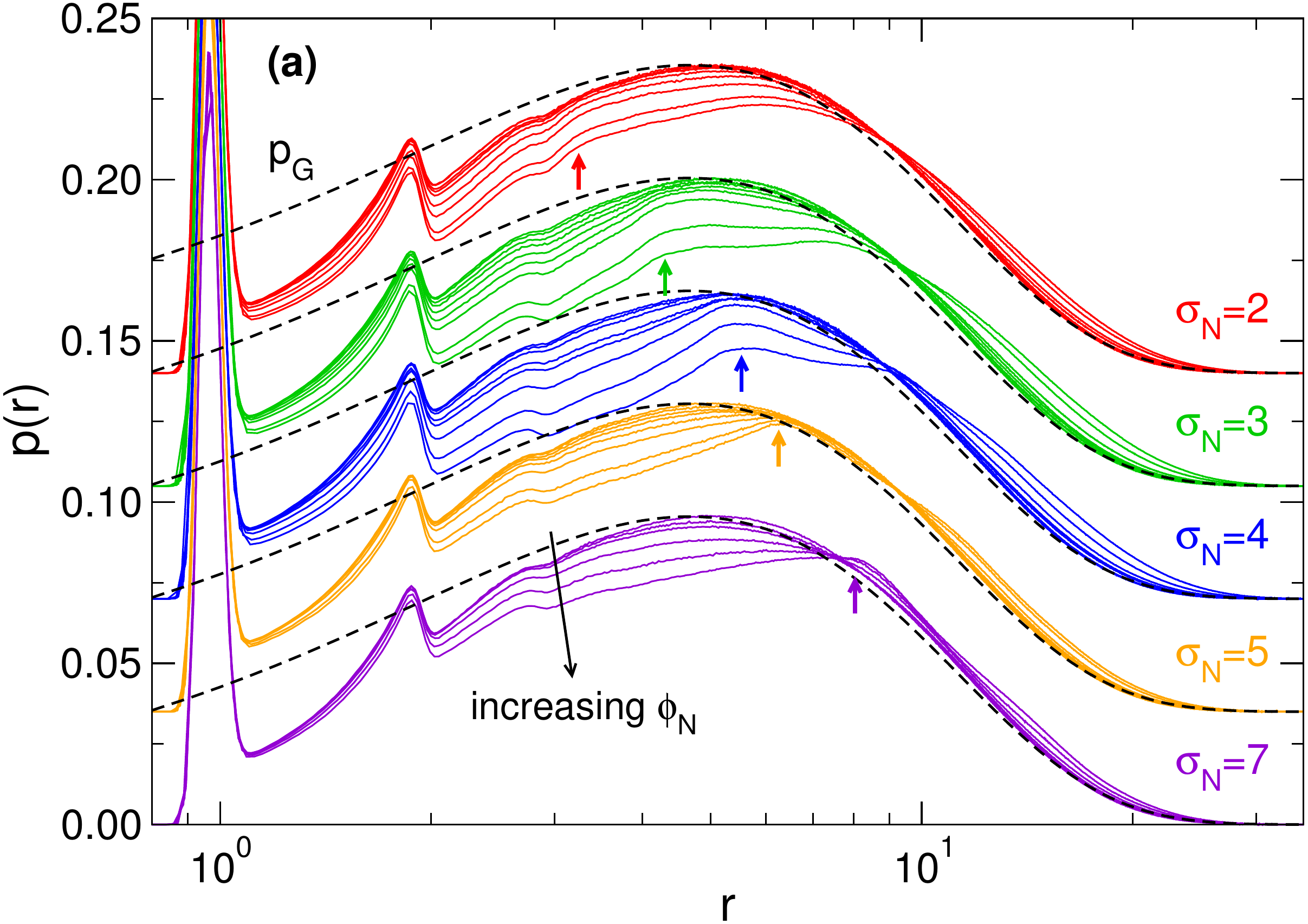}
\includegraphics[width=0.45 \textwidth]{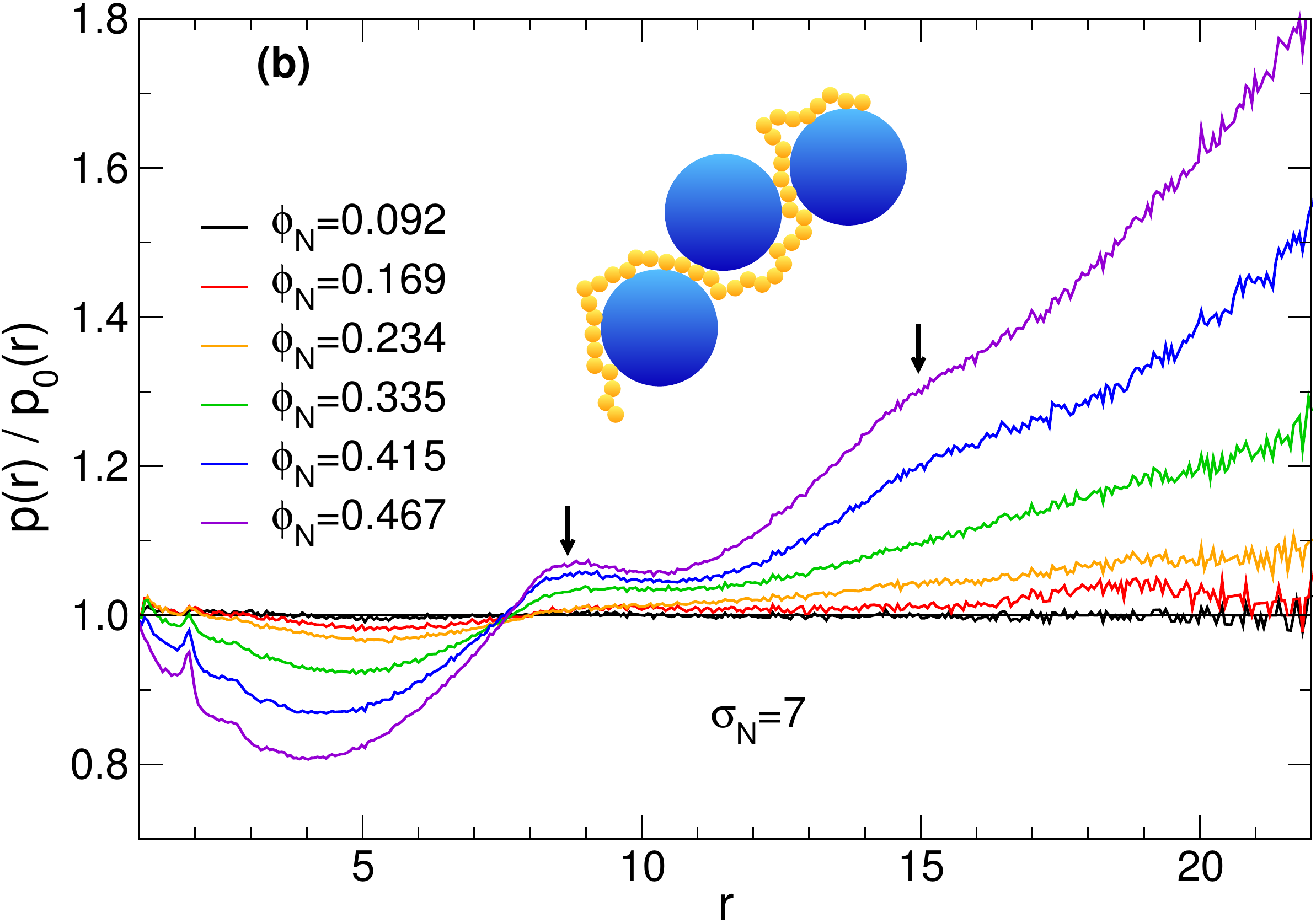}
\caption{(color online) (a) The function $p(r)$ for different values of the NP diameter $\sigma_N$ and  volume fraction $\phi_N$. The long black arrow indicates the direction of increasing $\phi_N$. The small arrows indicate the \leftquote bump" at $r \approx \sigma_N+1$. Dashed lines: Gaussian chain approximation $p(x)=p_G(x)$ with ${R_g=R_g(\phi_N=0)}$. For the sake of clarity, for $\sigma_N\leq5$ every set of curves has been shifted upwards with respect to the previous set by $0.035$. (b) The ratio $p(r)/p_0(r)$ for $\sigma_N=7$ and different values of $\phi_N$. The small arrows help locating the two \leftquote bumps" at $r \approx \sigma_N+1$ and $2(\sigma_N+1)$. Cartoon: schematic representation of the polymer structure in the presence of the NPs.}
\label{rdf_poly}
\end{figure}

For a Gaussian chain, $p$ has the following expression \cite{teraoka2002polymer}:

\begin{equation}
p_G(x)= \frac{8Nx}{\sqrt \pi (N-1) R_g} \left[ \frac{\sqrt \pi} 2 (1+2x^2) \text{erfc}(x) - x e^{-x^2} \right],
\label{chain_prob}
\end{equation}

\noindent where $x=r/2R_g$ and $\text{erfc}(x)$ is the complementary error function. This probability density peaks at $r \approx 0.74 R_g$ \cite{teraoka2002polymer}.

In Fig.~\ref{rdf_poly}a, we show $p(r)$ for different values of $\sigma_N$ and $\phi_N$, along with $p_G$ for a Gaussian chain, Eq.~\eqref{chain_prob}, with ${R_{g0}=R_g(\phi_N=0)}$ (dashed line). We observe that, for small $\phi_N$, $p_G$ provides a good approximation of $p$ at intermediate and large $r$ (at small $r$, $p(r)$ is dominated by excluded volume interaction between nearest neighbors). For all values of $\phi_N$, $p(r)$ shows a very high peak at $r\approx r_b=0.96$, corresponding to the first nearest neighbor, and a smaller peak at $r\approx 2 r_b=1.92$, corresponding to the second nearest neighbor. For values of $r$ larger than $2 r_b$, this signal gets washed out, and ultimately $p(r)$ decays to zero. When $\phi_N$ increases, we observe two effects: The third nearest neighbor peak becomes more pronounced and the curve becomes broader. This indicates that the presence of the NPs stretches the chains, causing them to become locally more ordered. We also note that $p(r)$ shows a modulation of wavelength $\approx \sigma_N+1$, the first peak of which is clearly visible in Fig.~\ref{rdf_poly}a as a \leftquote bump" at $r \approx \sigma_N+1$ (colored arrows). The presence of this modulation can be better appreciated by plotting the ratio $p(r)/p_0(r)$, where ${p_0(r)=p(\phi_N=0,r)}$. In Fig.~\ref{rdf_poly}b, we report $p(r)/p_0(r)$ for $\sigma_N=7$: As we can see, the effect of the NPs is to produce a \leftquote hole" in the range $0\lesssim r \lesssim \sigma_N+1$, but also to stretch the chain, increasing $p(r)$ significantly at larger distances. The modulation is clearly visible, with two bumps appearing at $r \approx \sigma_N+1$ and $2(\sigma_N+1)$ (small arrows).

Chain swelling in the presence of NPs has already been predicted theoretically \cite{frischknecht2010expanded} and observed in both simulations \cite{karatrantos2015polymer} and experiments \cite{nakatani2001chain,mackay2006general,tuteja2008polymer}. In particular, Karatrantos \emph{et al.} \cite{karatrantos2015polymer} have shown that polymer chains are unperturbed by the presence of repulsive NPs, while attractive NPs cause the polymer chains to be stretched and flattened when $2R_g > \sigma_N$ (which is always the case for the systems that we considered). Using the SC/PRISM theory, Frischknecht \emph{et al.} \cite{frischknecht2010expanded} reached the same conclusions. For a recent review discussing the influence of NPs on polymer size and local structure in simulations, see Ref.~\citenum{karatrantos2016modeling}.

In order to quantify the expansion of the chains, we measure the radius of gyration $R_g$, defined as \cite{rubinstein2003polymer}

\begin{equation}
R_g^2=  \frac 1 { N}  \sum_{i=1}^N  \langle (\mathbf r_{i} - \mathbf R_{\text{CM}})^2 \rangle,
\label{rg}
\end{equation}

\begin{figure}
\centering
\includegraphics[width=0.45 \textwidth]{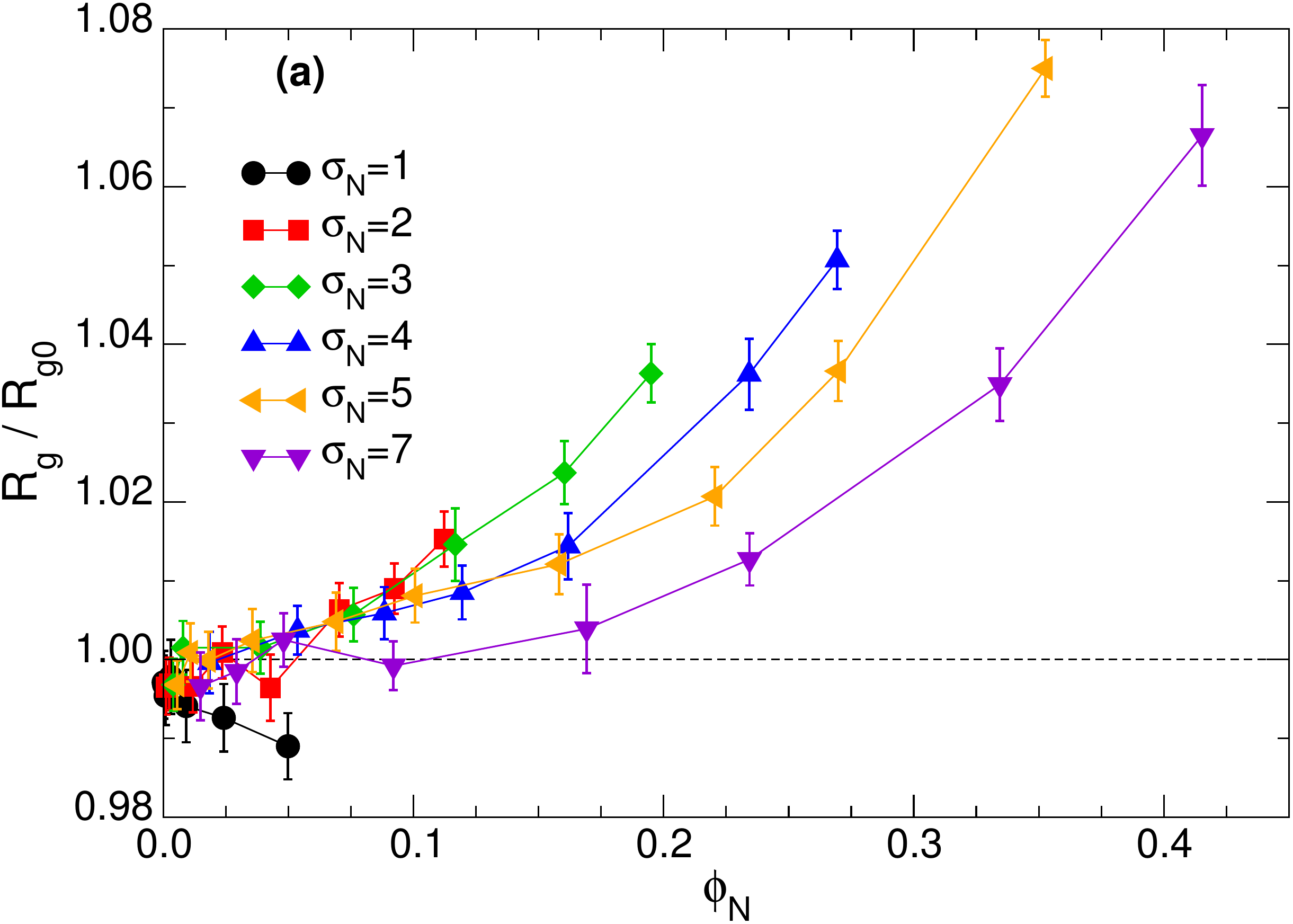}
\includegraphics[width=0.45 \textwidth]{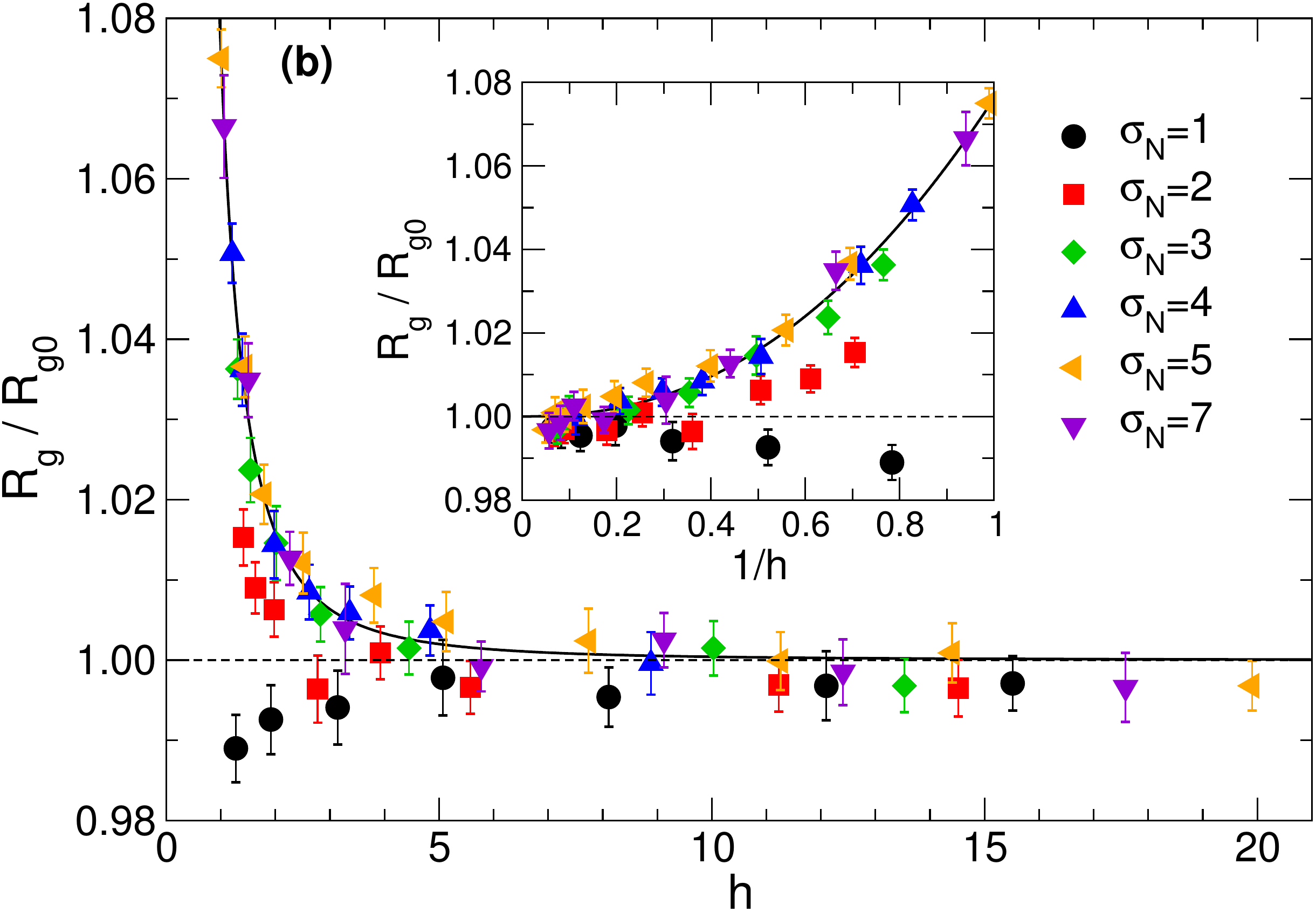}
\caption{Reduced radius of gyration of the polymers as a function of the NP volume fraction $\phi_N$ (a) and of interparticle distance $h$ (b).  \emph{Inset}:  $R_g/R_{g0}$ as a function of the inverse interparticle distance $1/h$. Continuous lines: ${R_g=R_{g0}(1+0.0762 \cdot h^{-2.26})}$.}
\label{gyr_norm}
\end{figure}

\noindent where $\mathbf R_{\text{CM}}$ is the position of the center of mass of the polymer. In the pure polymer solution, we have $R_{g0}={R_g(\phi_N=0)=6.28 \pm 0.02}$. In Fig.~\ref{gyr_norm}a we present the reduced radius of gyration $R_g/R_{g0}$ as a function of NP volume fraction for different values of $\sigma_N$.  With the exception of $\sigma_N=1$, there is a modest but clear increase of $R_g/R_{g0}$ with increasing NP volume fraction. We also notice that at fixed NP volume fraction the increase is stronger for smaller NPs, which suggests that chain expansion is mainly controlled by the NP excluded volume, which, at fixed $\phi_N$, is larger for smaller NPs \footnote{As a first approximation, if $\phi_N$ is low, the excluded volume can be estimated as $V_{\text{ex}}=\pi N_N(\sigma_N+1)^3/6V = \phi_N (1 + \sigma_N^{-3} + 3 \sigma_N^{-2}+3 \sigma_N^{-1})$.}.

Fig.~\ref{gyr_norm}a also shows that for $\sigma_N=1$, $R_g$ \emph{decreases} with increasing NP volume fraction. The reason is that NPs of this size have the largest surface-to-volume ratio, making the monomer-NP interaction (which scales approximately with the NP surface) very relevant. The consequence is that while in this range of $\phi_N$ the effect of the excluded volume is small, the effect of the interaction is large: Small NPs produce an effective attractive interaction between the monomers, which results in a decrease of $R_g$ and of the overall monomer volume fraction $\phi_m$ (we recall that all the simulations were performed at the same average pressure $P=0.1$; see also Sec. \ref{high_nppf}). We can therefore say that in this range of $\phi_N$, the NPs of size $\sigma_N=1$ act like a poor solvent, promoting chain contraction.

In Fig.~\ref{gyr_norm}b we plot the reduced radius of gyration as a function of the interparticle distance $h$. For $\sigma_N \geq 3$, the data fall on a master curve, which can be approximated by the empirical expression  ${R_g=R_{g0}(1+0.0762 \cdot h^{-2.26})}$ (continuous line in Fig.~\ref{gyr_norm}b). This confirms that in this range of NP size chain expansion is a geometrical effect, dominated by excluded volume: the NPs force the chains to take less tortuous paths, therefore increasing their effective size. The fact that larger particles have a locally \leftquote flatter" surface that could enhance chain expansion does not seem to play a role in this size range, as we can conclude from the fact that data for different $\sigma_N$ fall on the same master curve. For $\sigma_N =1$ and $2$ the data do not fall on the master curve, for the reasons explained above (high surface-to-volume ratio promotes chain contraction). To provide a better resolution for small values of $h$, in the inset of Fig.~\ref{gyr_norm}b we plot $R_g/R_{g0}$ as a function of $1/h$ \cite{frischknecht2010expanded}. 

We can summarize our results by saying that NPs of size $\sigma_N \geq 2$ act like a good solvent, swelling the polymer chains, while NPs of size $\sigma_N =1$ act like a poor solvent, causing them to contract. We note that this effect is expected to depend on the strength of the monomer-NP interaction: With stronger interactions, chain contraction could be observed also for $\sigma_N>1$. Further study is needed in order to clarify this point.

\section{Dynamics}\label{dynamics}

\subsection{Mean squared displacement} \label{sec_msd}

\begin{figure}
\centering
\includegraphics[width=0.45 \textwidth]{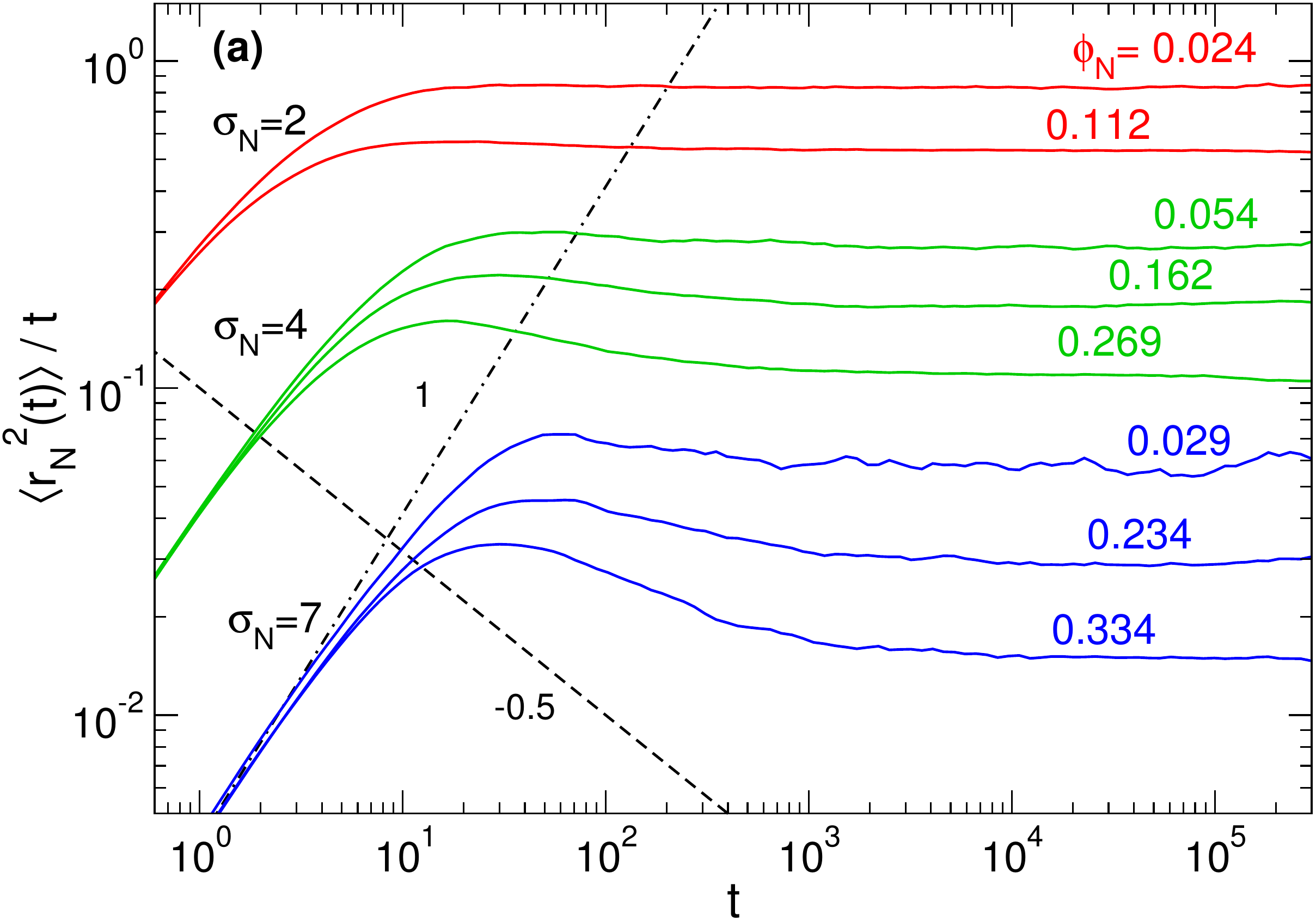}
\includegraphics[width=0.45 \textwidth]{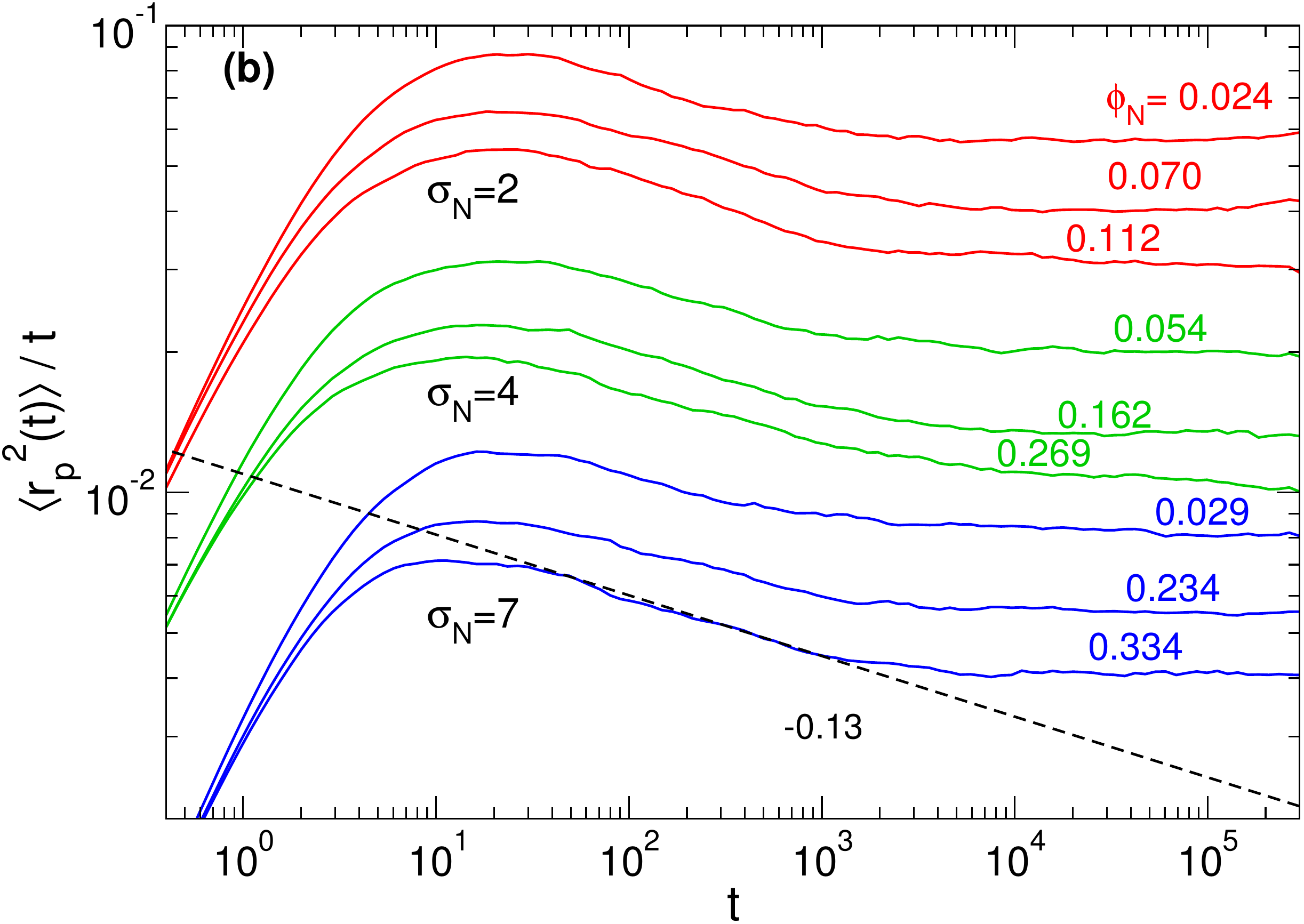}
\caption{(color online) (a) MSD of the NPs divided by time. Dashed line: $\langle r^2 (t)\rangle /t \propto t^{-1/2}$, predicted in Ref.~\citenum{cai2011mobility} for the subdiffusive regime of intermediate size NPs. Dash-dotted line:  $\langle r^2 (t)\rangle /t \propto t$ (ballistic regime). The curves for $\sigma_N=7$ have been shifted downwards by a factor $2$ to facilitate the visualization. (b) MSD of the centers of mass of the chains divided by time. Dashed line: $\langle r^2 (t)\rangle /t \propto t^{-0.13}$ (subdiffusive transient). The curves for $\sigma_N=4$ have been shifted down by a factor $2$ and those for $\sigma_N=7$ by a factor $8$ for clarity. }
\label{msd_vs_t}
\end{figure}

To characterize the dynamics of the system, we study the mean squared displacement (MSD) of the NPs and of the centers of mass (CM) of the chains. We recall that the MSD of a system of $M$ particles is defined as \cite{hansen1990theory}:

\begin{equation}
\langle r^2 (t)\rangle = \frac 1 M  \sum_{i=1}^M \langle [\mathbf r_i(t) - \mathbf r_i(0)]^2\rangle .
\label{msd}
\end{equation} 

\noindent In order to visualize more clearly the transition between the short-time ballistic regime, ${\langle r^2 (t)\rangle \propto t^2}$, and the long time diffusive regime, ${\langle r^2 (t)\rangle \propto t}$, we show in Fig.~\ref{msd_vs_t} the MSD divided by time $t$ for the NPs (Fig.~\ref{msd_vs_t}a) and for the polymers (Fig.~\ref{msd_vs_t}b).  At low $\phi_N$, the motion of the NPs shows the same qualitative behavior for all the values of the NP diameter $\sigma_N$ (Fig.~\ref{msd_vs_t}a): After the initial ballistic regime, the motion becomes almost immediately diffusive, with the exception of the system with $\sigma_N=7$, which shows a weak subdiffusive transient, $\langle r^2 (t)\rangle \propto t^\beta \ (\beta<1)$, between these two regimes. A clear transient subdiffusive regime appears between the ballistic and diffusive regimes at intermediate and high values of $\phi_N$. The MSD of the chains, on the other hand, shows a weak subdiffusive transient for all values of $\phi_N$ and $\sigma_N$, with an exponent $\beta \approx 0.87$ that is not much influenced by the value of $\phi_N$ (Fig.~\ref{msd_vs_t}b). This transient, which is most likely due to non-Gaussian dynamics caused by intermolecular correlations \cite{smith2001non}, has been previously observed in experiments \cite{paul1998chain,padding2001zero,smith2001non} and simulations \cite{kremer1990dynamics,paul1991dynamics,kopf1997dynamics,paul1998chain,smith2001non} of polymer melts, where the measured subdiffusive exponent was $\beta \approx 0.8$. The fact that  in our case the exponent is slightly larger than $0.8$ is likely due to the fact that the density considered here is significantly smaller than that of a melt, $\rho \approx 0.85$. The different regimes (ballistic, subdiffusive, diffusive) and the transitions between them can also be studied systematically through the function $d \log \langle r^2(t)\rangle / d \log t$, which is a generalization of the subdiffusive exponent $\beta$. For a detailed analysis of this quantity, see Sec.~\ref{sec:beta} in the S.I.

Using scaling arguments, Cai \emph{et al.} have formulated a theory for the diffusion of single nonsticky NPs in polymer liquids \cite{cai2011mobility}. For an unentangled polymer mixture, they predicted that NPs of diameter $\sigma_N<\xi$, where $\xi$ is the mesh size, should always move diffusively, whereas the MSD of larger NPs, $\xi<\sigma_N<2R_g$, should behave as follows:

\begin{equation}
\langle r_N^2(t) \rangle \propto
\begin{cases}
t & t<\tau_\xi\\
t^{1/2} & \tau_\xi < t < \tau_{\sigma_N}\\
t & \tau_{\sigma_N}<t,\\
\end{cases}
\label{msd_np_rubinstein}
\end{equation}

\noindent where $\tau_\xi \propto \eta_s \xi^3 / k_B T$ and $\tau_{\sigma_N} \propto \tau_\xi (\sigma_N/\xi)^4$ are, respectively, the relaxation times of polymer segments of size $\xi$ and $\sigma_N$ (here $\eta_s$ is the viscosity of the pure solvent). For NPs larger than the polymers, $\sigma_N>2R_g$, $\tau_{\sigma_N}$ must be replaced by $\tau_{R_g}\propto \tau_\xi (R_g/\xi)^4$. The crossover from subdiffusive to diffusive motion for NPs in polymer solutions has also been observed in experiments \cite{babaye2014mobility,poling2015size}.

Since at low NP volume fraction in our system $\xi \approx 1.3$, according to the scaling theory of Cai \emph{et al.} one expects the MSD of the NPs of diameter $\sigma_N > 2$ to show subdiffusive behavior with exponent $\beta=1/2$ at small $\phi_N$. However, no such behavior is observed for any value of $\sigma_N$. For small NPs, this may be due to the fact that the time window in which the subdiffusive behavior is expected to be present, i.e., $\tau_\xi<t<\tau_{\sigma_N}$, is too small, since $\tau_{\sigma_N} \propto \tau_\xi (\sigma_N/\xi)^4$. For larger NPs, this time window regime should be large enough to observe subdiffusion, and indeed for $\sigma_N = 7$ we observe a very weak subdiffusive transient, but the exponent $\beta$ is close to $1$.  This is in agreement with previous simulations, which have also found that $\beta$ is not always equal to $1/2$ in the subdiffusive regime, but rather gradually approaches this value as $\sigma_N$ is increased \cite{ge2017nanoparticle,chen2017coupling}.

In addition to the mean squared displacement, we have also studied the van Hove function \cite{hansen1990theory} and the non-Gaussian parameter $\alpha_2(t)$ \cite{kob1995testing} of the NPs, finding that their dynamics is with a good approximation Gaussian, in agreement with experiments \cite{babaye2014mobility,poling2015size} (see Sec.~\ref{sec:vanhove} in the S.I.). These results indicate that the dynamics of the NPs is not heterogeneous. 

\subsection{Polymer diffusion} \label{sec:poly_diff}

In order to make a more quantitative characterization of the dynamical properties of the polymers and the NPs, we now focus on the self diffusion coefficient $D$ (which for simplicity we will refer to as \leftquote diffusion coefficient"), which can be obtained from the MSD, Eq.~\eqref{msd}, through Einstein's relation \cite{hansen1990theory}:

\begin{equation}
D = \lim_{t \to \infty} \frac {\langle r^2 (t)\rangle }{6t}.
\label{d}
\end{equation} 

It is known that measurements of $D$ in systems with periodic boundary conditions suffer from finite-size effects because of long-ranged hydrodynamic interactions \cite{dunweg1993molecular,yeh2004system}. Although an analytical expression for the correction to $D$ is available \cite{dunweg1993molecular,yeh2004system}, it is not evident whether it can be applied to the motion of polymer chains and NPs in a concentrated polymer solution. For the NPs, such an expression is most likely not adequate when, as in our case, the NP size is smaller than the polymer size \cite{kalathi2014nanoparticle}. Therefore, for consistency we choose not to apply any finite size correction to the measured diffusion coefficients. 

\begin{figure}
\centering
\includegraphics[width=0.45 \textwidth]{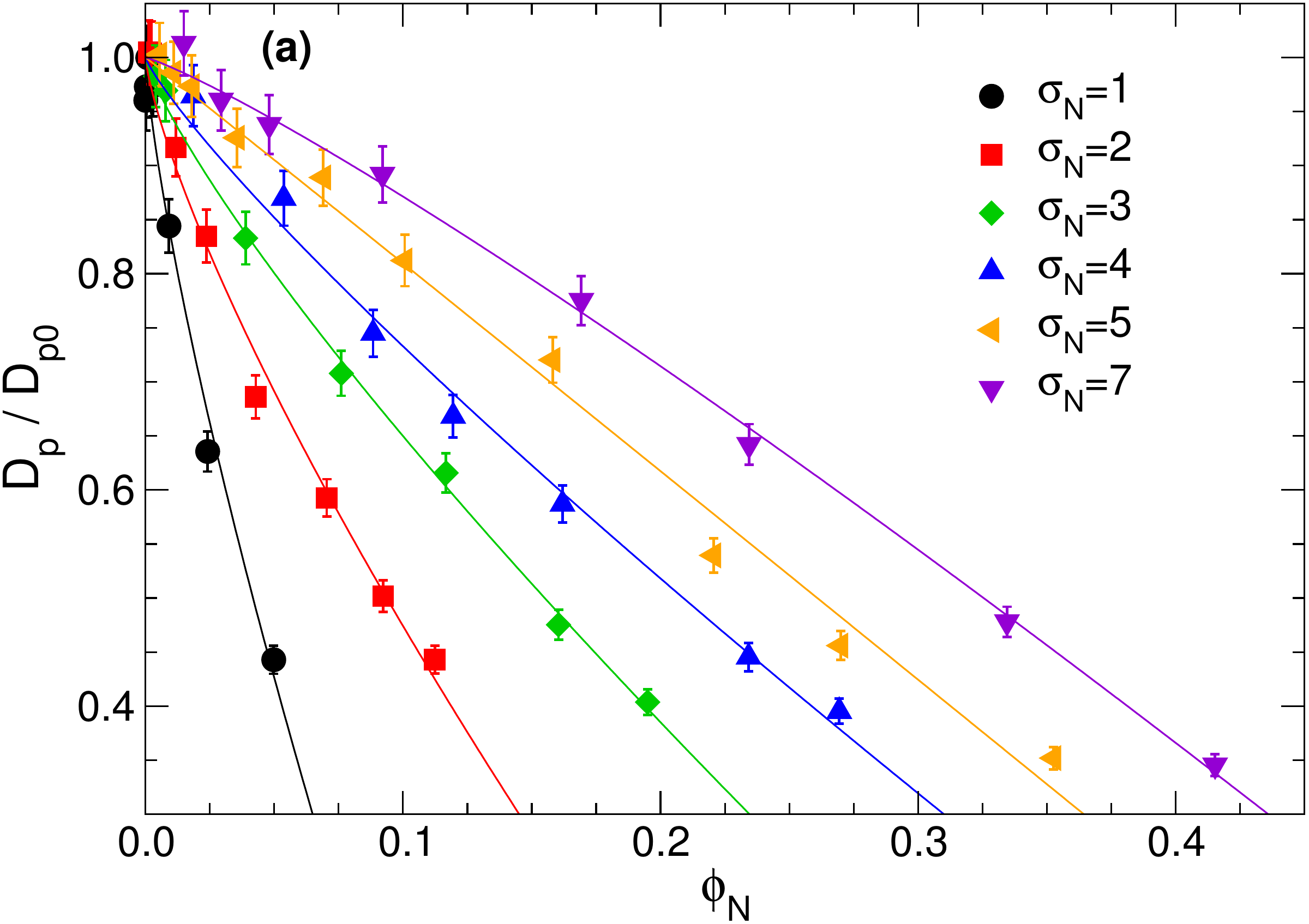}
\includegraphics[width=0.45 \textwidth]{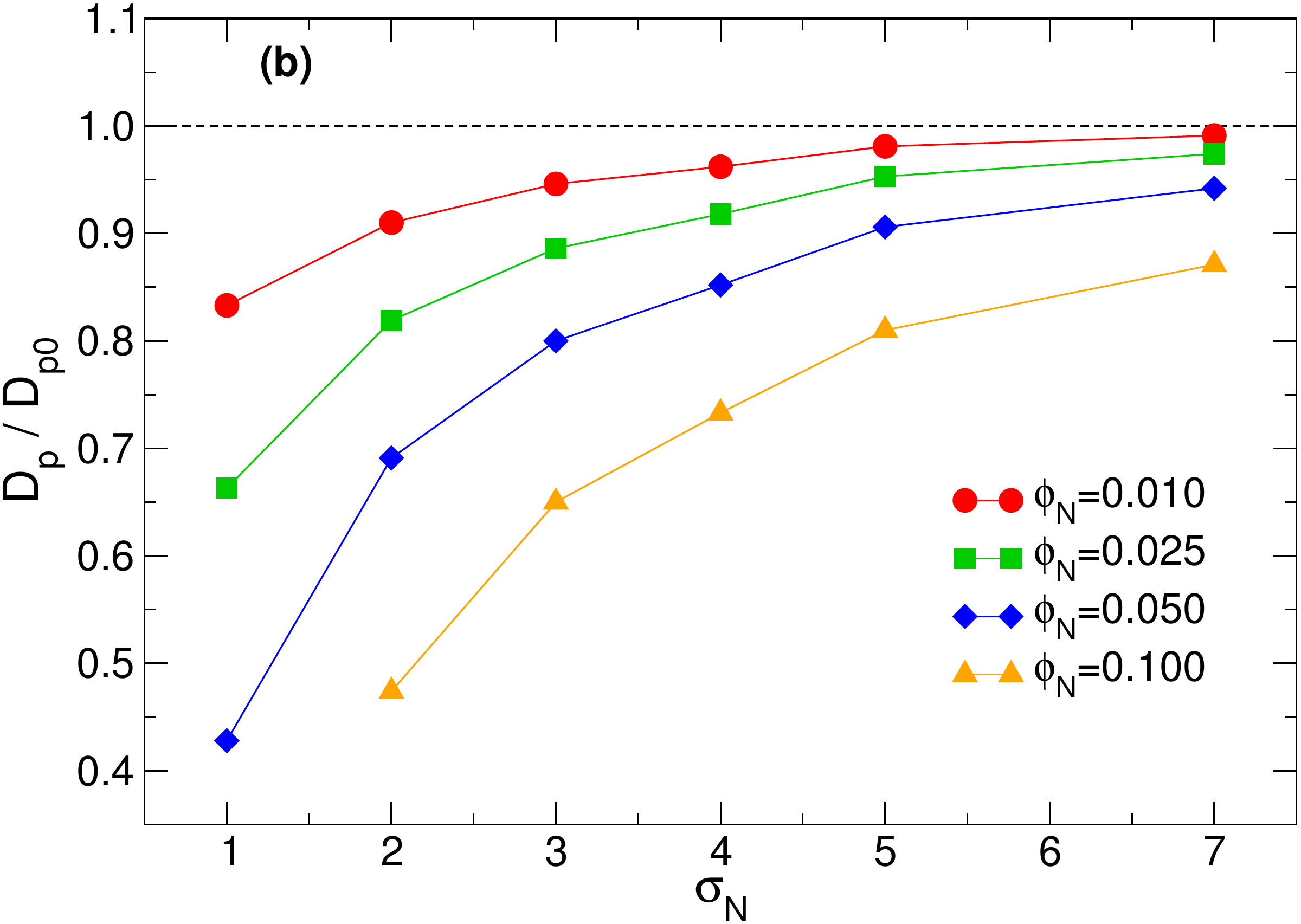}
\caption{(a) Reduced diffusion coefficient of the centers of mass (CM) of the chains $D_p/D_{p0}$ as a function of NP volume fraction, where $D_{p0}=D_p(\phi_N=0)$ . Continuous lines: empirical fits with the functional form  $D_p = D_{p0} [1 - (\phi_N/\phi_{N0})^\alpha]$. (b) $D_p/D_{p0}$ as a function of NP diameter $\sigma_N$.}
\label{dchain}
\end{figure}

In the pure polymer system (${\phi_N=0}$), the diffusion coefficient of the CM of the chains is ${D_{p0}=(1.14 \pm 0.02) \cdot 10^{-2}}$. In Fig.~\ref{dchain}a we plot the reduced diffusion coefficient of the polymer chains $D_p/D_{p0}$ as a function of the NP volume fraction $\phi_N$. We can observe that $D_p/D_{p0}$ decreases with increasing NP volume fraction, with the decrease being stronger, at fixed $\phi_N$, for smaller NPs. The data can be fitted to the empirical functional form ${D_p = D_{p0} [1 - (\phi_N/\phi_{N0})^\alpha]}$, where $\alpha$ increases with NP size (the values of $\phi_{N0}$ and $\alpha$ for the different NP diameters are reported in Tab.~\ref{tab:a_alpha} in the S.I.). We note that this functional form implies that $D_p$ becomes zero at $\phi_N=\phi_{N0}$, i.e., that the dynamics of the polymers is completely arrested. However, for larger values of $\phi_N$ the dependence of $D_p$ on $\phi_N$ changes (see Sec.~\ref{high_nppf}) and thus this dynamic transition is avoided; hence, the above functional form is valid only if $\phi_N/\phi_{N0}$ is small. By using this relation, we can interpolate between the data points and plot $D_p/D_{p0}$ as a function of the NP diameter $\sigma_N$ for different volume fractions (Fig.~\ref{dchain}b) and we observe that $D_p/D_{p0}$ increases monotonically with $\sigma_N$ at fixed $\phi_N$. 

There are two possible causes (or a combination of the two) that can lead to the slowing down of the chains with increasing NP volume fraction: the increase of the number of obstacles to polymer motion and the increase of polymer-NP interfacial area, which, since the interaction between polymers and NPs is attractive, can result in a reduced chain mobility. A predominance of the first effect would imply that the slowing down of the polymers is a mostly entropic effect, while if the second effect is the most important the dynamics of the polymers is mainly controlled by enthalpy.

Composto and coworkers \cite{gam2011macromolecular,gam2012polymer,lin2013attractive,choi2013universal} observed a similar slowing down of chain motion in a series of experimental studies on polymer nanocomposites containing large NPs ($\sigma_N \gtrsim 2R_g$). They found that the reduced diffusion coefficient of the polymers falls on a master curve when plotted versus a \leftquote confinement parameter", defined as $h/2 R_g$, where $h$ is the interparticle distance, which the authors computed using Eq.~\eqref{id} with $\phi_N^\text{\tiny M} = 2 / \pi$. Since the collapse of the data was independent of the strength of the polymer-NP interaction \cite{lin2013attractive}, the authors concluded that the slowing down of the polymers is entropic in origin, caused by the reduction of chain entropy as the chain passes through bottlenecks formed by neighboring NPs (entropic barrier model) \cite{gam2011macromolecular}. An analogous reduction in polymer mobility due to the presence of NPs was observed by Li \emph{et al.} \cite{li2014dynamic} in molecular dynamics simulations of unentangled melt of short chains ($N=35,\rho_m=0.85$) containing repulsive NPs. The slowing down was weaker than that observed by Composto and coworkers in Refs.~\citenum{gam2011macromolecular,gam2012polymer}, an effect which the authors attributed to the absence of chain entanglements. Karatrantos \emph{et al.} \cite{karatrantos2017polymer} also observed a monotonic decrease in the polymer diffusion coefficient with increasing NP volume fraction in molecular dynamics simulations of NPs in unentangled and weakly entangled melts, and attributed this phenomenon to the increase in the polymer-NP interfacial area. Desai \emph{et al.} \cite{desai2005molecular}, on the other hand, have reported that the polymer diffusion coefficient in a simulated lightly entangled melt ($N=80, \rho_m=0.85$) containing repulsive/weakly attractive NPs initially increases with $\phi_N$, reaches a maximum around $\phi_N=5\%$ and decreases for higher values. An enhancement of chain diffusivity at low $\phi_N$ has also been observed in simulations by Kalathi \emph{et al.} \cite{kalathi2014nanoparticle}, possibly because attractive monomer-monomer interactions were considered in their work. It is therefore clear that, despite the fact that some general trends can be identified, the dynamics of the polymers can depend strongly on the details of the simulated systems.

\begin{figure}
\centering
\includegraphics[width=0.45 \textwidth]{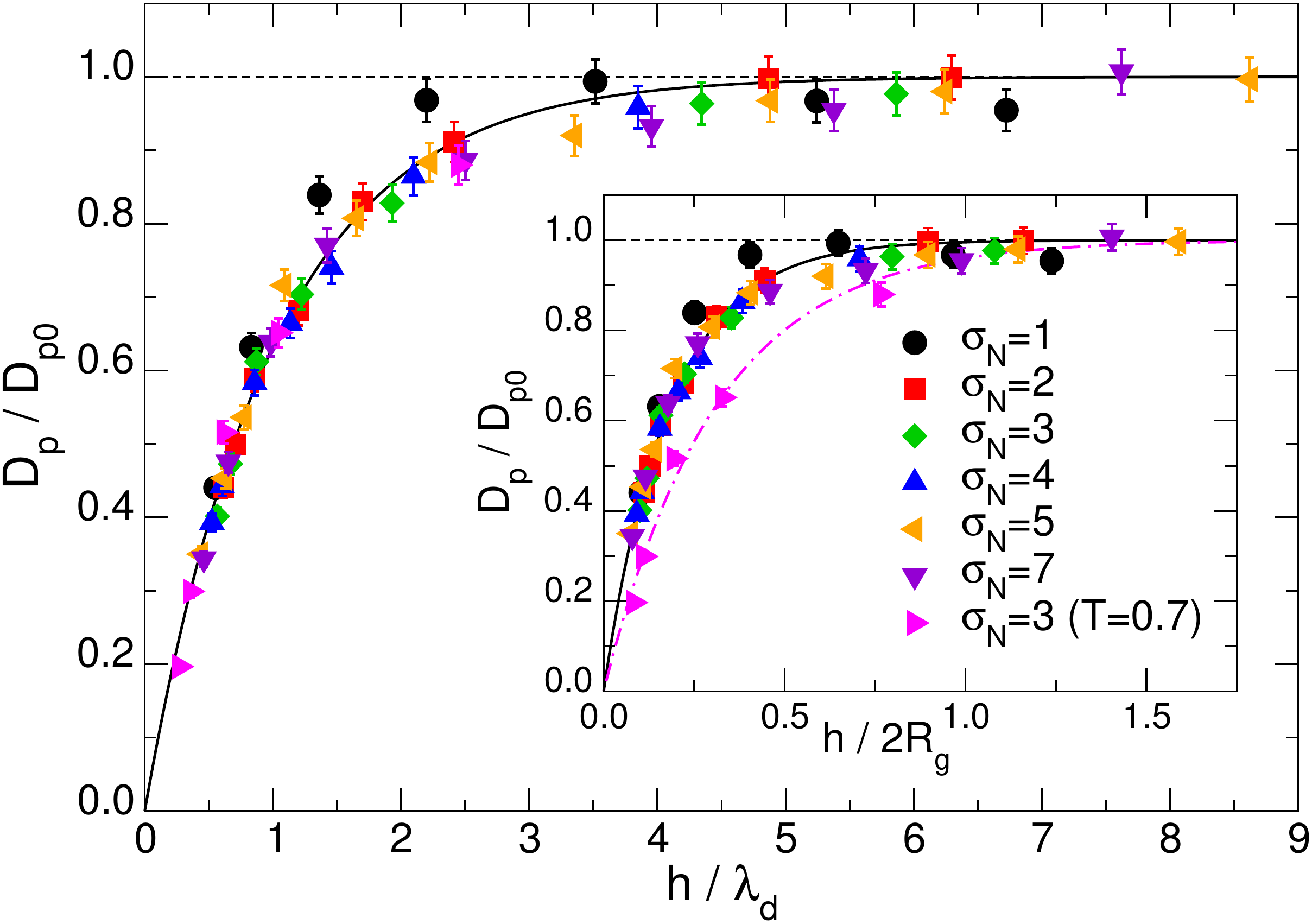}
\caption{Reduced diffusion coefficient of the centers of mass (CM) of the chains  as a function of the dynamic confinement parameter $h/\lambda_d$ for different values of the NP diameter $\sigma_N$. Continuous line: ${D_p=D_{p0}[1-\exp(-h/\lambda_d)]}$. \emph{Inset}: $D_p/D_{p0}$ as a function of $h/2R_g$. Continuous and dash-dotted lines: $D_p=D_{p0}[1-\exp(-a \cdot h/2 R_g)]$, with $a$ respectively equal to $5.44$ and to $3.22$.}
\label{dchain_confinement}
\end{figure}

Following Composto and coworkers \cite{gam2011macromolecular,gam2012polymer,lin2013attractive,choi2013universal}, we plot in the inset of Fig.~\ref{dchain_confinement} the reduced diffusion coefficient of the chains as a function of the confinement parameter $h/2R_g$. We recall that in our case $h$ is not defined by Eq.~\eqref{id}, but rather computed from the pore-size distribution (see Sec.~\ref{sec:pore_size} in the Appendix). In addition to the $T=1.0$ data, we also show the results from simulations at $T=0.7$ with $\sigma_N=3$.

The first observation is that, with the exception of $\sigma_N=1$, all the $T=1.0$ data fall on the same master curve, which is well approximated by the empirical expression ${D_p=D_{p0}[1-\exp(-a h/2 R_g)]}$, with ${a=5.44}$  (continuous line in the inset of Fig.~\ref{dchain_confinement}). The fact that also in our case $D_p/D_{p0}$ is only a function of the confinement parameter is rather surprising, since Composto and coworkers mainly considered NPs of size comparable to that of the polymers or larger, which could be considered as basically immobile \cite{gam2011macromolecular,gam2012polymer,lin2013attractive,choi2013universal}, whereas in our case $\sigma_N<2R_g$ and the NPs diffuse faster than the chains in almost all the systems considered (see Fig.~\ref{d_comparison} in the S.I.). 

We notice, however, two important differences: The first one is that while in our case the diffusion coefficient of the pure polymer solution ($D_p/D_{p0}=1$) is recovered at $h/2R_g \approx 1$, in Refs. \citenum{gam2011macromolecular,gam2012polymer,lin2013attractive,choi2013universal} it is recovered only at much higher values of the confinement parameter, $h/2R_g\approx 20$. Our finding is similar to what observed by Li \emph{et al.} \cite{li2014dynamic}, who attributed the discrepancy between their data and those of Composto and coworkers to the absence of entanglement in their simulated system. The second difference is that the $T=0.7$ data clearly do not fall on the same master curve. Since a decrease in temperature is approximately equivalent to an increase in the strength of the polymer-NP interaction, this result suggests that in our system the polymer-NP interaction plays a relevant role, in contrast with Ref.~\citenum{lin2013attractive}, where the authors concluded that the confinement parameter captures the polymer slowing down independently of the polymer-NP interactions. We propose in the following a possible solution to these discrepancies.

The confinement parameter $h/2R_g$ is a purely static quantity, which only depends on the spatial configuration of the polymers and the NPs in the system. However, there are several cases in condensed matter physics in which two systems with identical structure show a completely different dynamics: A well-known example is that of the glass transition, where a supercooled liquid shows structural properties identical to those of a liquid at higher temperature, but completely different dynamical properties \cite{cavagna2009supercooled,binder2011glassy,berthier2011theoretical}. 

It seems therefore more appropriate to introduce a \emph{dynamic confinement parameter} $h/\lambda_d$, where $\lambda_d$ is a dynamic length scale which will in general depend on temperature, density and on the details of the simulated system. We have already seen that the $T=1.0$ data are well approximated by the function  ${D_p=D_{p0}[1-\exp(-a h/2 R_g)]}$, with $a=5.44$  (continuous line in the inset of Fig.~\ref{dchain_confinement}); the $T=0.7$ data are well approximated by the same functional form, but with a different coefficient, $a=3.22$. In light of what we discussed above, we make the hypothesis that the reduced diffusion coefficient of the polymers can be expressed as

\begin{equation}
D_p = D_{p0} \left[1-\exp \left(-\frac h {\lambda_d (T)} \right)\right],
\label{dchain_dynamic}
\end{equation}

\noindent where we have explicitly reported the dependence of $\lambda_d$ on temperature. Since $R_g$ does not change more than $8\%$ with respect to the pure polymer solution value $R_{g0}$ (see Fig.~\ref{gyr_norm}), we can estimate $\lambda_d$ as $\lambda_d = 2R_{g0}/a$: This gives $\lambda_d(1.0) = 2.31$ and $\lambda_d(0.7)=3.90$. We show $D_p/D_{p0}$ as a function of $h/\lambda_d$ in Fig.~\ref{dchain_confinement}: In this plot, the data for different temperatures fall on the same master curve, showing that the dynamic confinement parameter $h/\lambda_d$ is more successful than the \leftquote static" confinement parameter $h/2R_g$ in capturing the slowing down of the polymers (inset of Fig.~\ref{dchain_confinement}). However, one question remains: What does the dynamic length scale $\lambda_d$ represent exactly, and why does it increase when temperature is decreased? Our answer is that $\lambda_d$ is a \emph{cooperativity length scale}, i.e., it represents the typical length scale of the spatial rearrangement needed for a polymer segment to escape its local cage. Similarly to what happens in a supercooled liquid  \cite{adam1965temperature,cavagna2009supercooled}, this cooperative length scale is expected to increase when $T$ is decreased. In our system, an important role could also be played by the attractive polymer-NP interactions, which become more relevant when $T$ is decreased and could reduce the mobility of polymer segments close to the polymer-NP interface. We also expect $\lambda_d$ to increase with monomer density, since a higher density naturally leads to a locally more constrained dynamics: This could explain why the data of Li \emph{et al.} \cite{li2014dynamic}, who simulate NPs in a dense melt, are compatible with a larger cooperativity length scale. Another factor that is expected to play a major role is the stiffness of the chain, with stiffer chains expected to lead to a larger $\lambda_d$.

To sum up, we propose a modification of the confinement parameter theory of Composto and coworkers \cite{gam2011macromolecular,gam2012polymer,lin2013attractive,choi2013universal}: our hypothesis is that the dynamics of the polymers is controlled by a dynamic confinement parameter $h/\lambda_d$, where $\lambda_d$ is a cooperativity length scale which will depend in general on the thermodynamic parameters and on the details of the model. Further study is required to test the validity of this hypothesis, and to understand how $\lambda_d$ depends on the properties of the physical system.

\subsection{Single nanoparticle diffusion} 

\begin{figure}
\centering
\includegraphics[width=0.465 \textwidth]{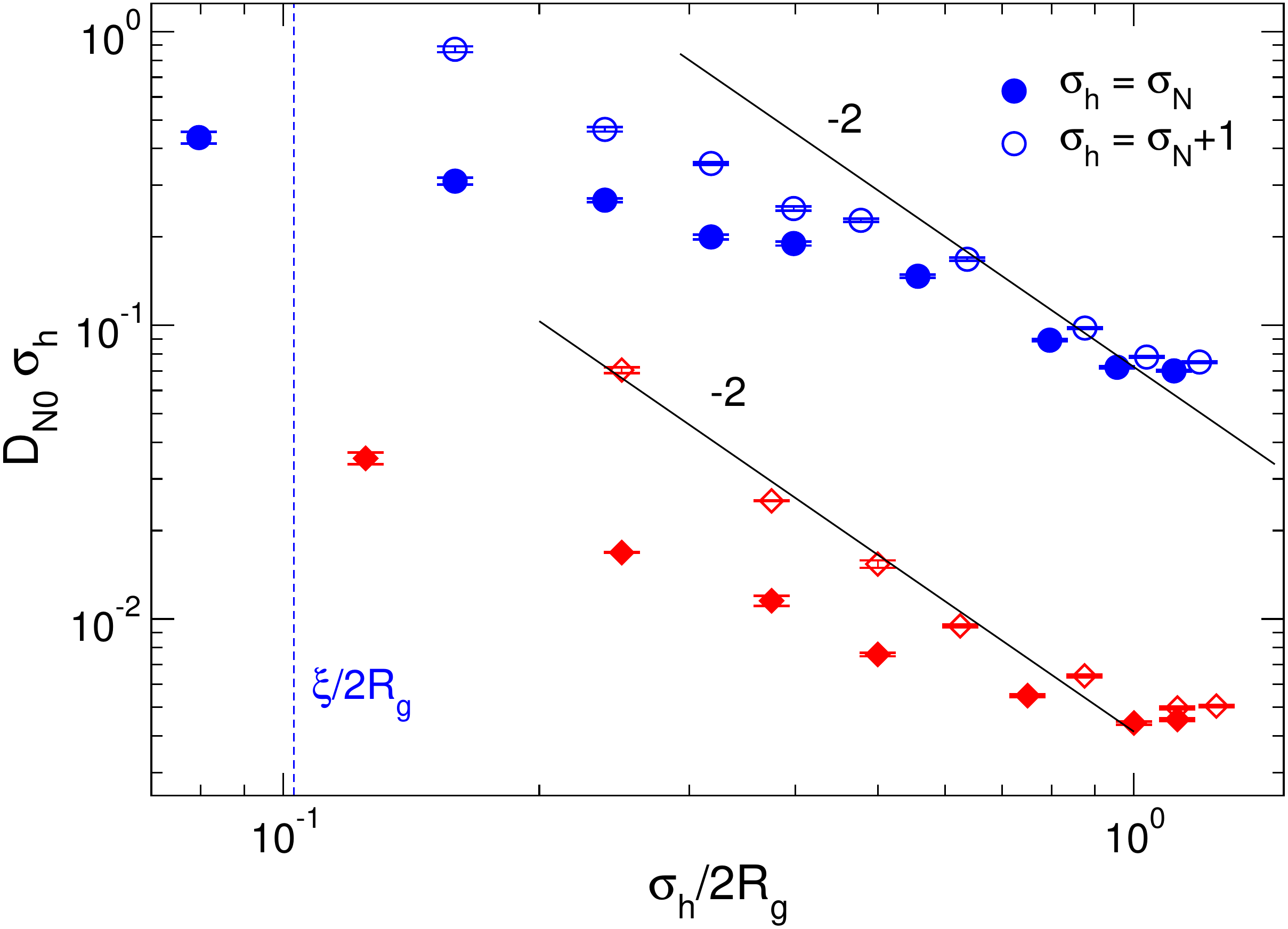}
\caption{Single NP diffusion coefficient multiplied by the effective hydrodynamic diameter of the NP $\sigma_h$ as a function of $\sigma_h/2R_g$ (blue circles), compared with the results from Ref.~\citenum{liu2008molecular} (red diamonds). Filled symbols: $\sigma_h=\sigma_N$. Open symbols: $\sigma_h=\sigma_N+1$. Continuous lines: slope of $-2$. The dashed vertical line marks the value $\xi/2R_g$ for our system.}
\label{dnp0}
\end{figure}

The diffusion coefficient of a hard-sphere probe particle of diameter $\sigma_N$ in a continuum solvent with shear viscosity $\eta$ is given by the Stokes-Einstein equation \cite{hansen1990theory}:

\begin{equation}
D_{N0}=\frac{k_B T}{f \pi \eta \sigma_N},
\label{stokes_einstein}
\end{equation}

\noindent where $f$ is a number between $2$ and $3$ which depends on the choice of the hydrodynamic boundary conditions: $f=2$ for pure slip and $f=3$ for pure stick boundary conditions \cite{felderhof1978diffusion}. If the particle is not a perfect hard sphere, for example because its shape is not perfectly spherical or because there is adsorption of solvent molecules on its surface, $\sigma_N$ must be replaced with an effective hydrodynamic diameter $\sigma_h$ \cite{bocquet1994brownian,schmidt2003hydrodynamic}. It is well-known that Eq.~\eqref{stokes_einstein} is inadequate to describe the motion of particles smaller than the polymer size in a polymer solution/melt \cite{wyart2000viscosity, ould2000molecular, tuteja2007breakdown, liu2008molecular, grabowski2009dynamics, cai2011mobility, kohli2012diffusion, kalathi2014nanoparticle,babaye2014mobility,poling2015size}, because the continuum assumption breaks down when the size of the probe particle becomes comparable to the characteristic length scale of the solvent.

Cai \emph{et al.} \cite{cai2011mobility} have predicted three regimes for the diffusion of a NP of diameter $\sigma_N$ in an unentangled polymer mixture: For $\sigma_N<\xi$, where  $\xi$ is the mesh size, the NP diffusion coefficient should follow the Stokes-Einstein law: $D_{N0}\propto k_B T  /\eta_s \sigma_N$, with $\eta_s$ the viscosity of the pure solvent (\emph{small size regime}). If $\sigma_N>\xi$, the motion of the NPs becomes coupled to the segmental relaxation of the polymer mesh, so that $D_{N0} \propto k_B T / \eta_s \sigma_N^3$ (\emph{intermediate size regime}). The relation $D_{N0} \propto \sigma_N^{-3}$ for intermediate size NPs was originally proposed by Wyart and de Gennes using scaling arguments \cite{wyart2000viscosity}, and was also predicted by Yamamoto and Schweizer using mode-coupling theory \cite{yamamoto2011theory} and subsequently a self-consistent generalized Langevin equation approach \cite{yamamoto2014microscopic}. This prediction has also been confirmed by simulations \cite{liu2008molecular,chen2017coupling}.

For even larger diameters, the Stokes-Einstein law \eqref{stokes_einstein} is eventually recovered, but with the viscosity of the pure solvent replaced by the bulk viscosity of the solution $\eta$: $D_{N0} \propto k_B T / \eta \sigma_N$ (\emph{large size regime}). For this second crossover size, different versions of mode-coupling theory predict different values depending on the level of approximation and on polymer density, all of them of the order of the polymer diameter: $\sigma_N = 2R_g$ (unentangled solutions and melts) \cite{egorov2011anomalous}, $\sigma_N = 3 R_g$ (unentangled melts and unentangled concentrated solutions) \cite{yamamoto2011theory} and $\sigma_N=2R_g^*$ (semidilute solutions) \cite{dong2015diffusion}, where $R_g^*$ is the radius of gyration of an isolated chain. The prediction that the crossover to Stokes-Einstein behavior should occur when the NP diameter is of the order of $2R_g$ has also been confirmed by experiments \cite{grabowski2009dynamics,kohli2012diffusion,poling2015size}. 

%

To test the validity of the Stokes-Einstein formula, Liu \emph{et al.} \cite{liu2008molecular} have used simulations to measure the single particle diffusion coefficient of NPs in simulations of a dense, unentangled melt ($N=60, \rho_m=0.84$). The results of their simulations are shown in Fig.~\ref{dnp0} (red diamonds). The authors argued that the effective hydrodynamic radius of the particle, ${R_h=\sigma_h/2}$, should have the value ${R_h=(\sigma_N+1)/2}$, which corresponds to the contact distance between a NP and a monomer (the same argument can be found in Ref.~\citenum{hynes1979molecular}). By fitting their data in the size range $\sigma_N<2 R_g$ with a power law $D_{N0}\propto R_h^{-\gamma}$, they found ${\gamma \approx 3}$ (open red diamonds in Fig.~\ref{dnp0}), and for diameters ${\sigma_N > 2 R_g}$ they recovered the Stokes-Einstein relation. The results of Ref.~\citenum{liu2008molecular} are therefore in agreement with the prediction that $D_{N0} \propto \sigma_N^{-3}$ \cite{wyart2000viscosity,yamamoto2011theory,cai2011mobility} if one replaces the NP diameter $\sigma_N$ with the effective hydrodynamic diameter $\sigma_h=\sigma_N+1$. However, when plotting $D_{N0}$ as a function of $\sigma_N$ instead of $R_h$ and fitting with a power law $D_{N0}\propto \sigma_N^{-\gamma}$, one obtains instead $\gamma\approx 2$ (filled red diamonds in Fig.~\ref{dnp0}). Hence one must conclude that the value of the exponent $\gamma$ depends on the exact definition of the effective NP diameter, which makes the comparison of simulation data with theoretical predictions a delicate matter,  especially when the size of the NP is of the same order of magnitude as the monomer size (for large NPs, $\sigma_N+1 \approx \sigma_N$).

In order to test these predictions, we have performed additional simulations at low NP volume fraction ($\phi_N<0.015$) for $\sigma_N=10,12$, and $14$. In Fig.~\ref{dnp0} we show $D_{N0}\sigma_h$ as a function of $\sigma_h/2R_g$, with $\sigma_h$ alternatively defined as $\sigma_N$ and $\sigma_N+1$ (blue circles). Also included are the data from Ref.~\citenum{liu2008molecular} (red diamonds). We can see that $D_{N0}$ decreases continuously for $\sigma_h<2R_g$, whereas at $\sigma_h \approx 2R_g$ Stokes-Einstein behavior, $D_{N0} \sigma_h = \text{const.}$, is recovered. Taking $\sigma_h=\sigma_N+1$ we find in the range $0.6 \lesssim \sigma_h/2R_g \lesssim 1$ a slope of approximately $-2$, which agrees with the theoretical predictions. However, the range in which we observe this slope is rather small and hence we cannot claim that our data confirm the theory. In particular, if we use the scaling estimate of Eq.~\eqref{xi} for the mesh size, $\xi \approx 1.27$, we can see that there is no sign of the transition from $D_{N0}\sigma_h = \text{const.}$ to $D_{N0} \sigma_h \propto \sigma_h^{-2}$ at $\sigma_h \approx \xi$ (which corresponds to $\sigma_h/2R_g \approx 0.10$) predicted in Ref.~\citenum{cai2011mobility}. However, a \emph{caveat} is in order: We have verified that, as also reported in previous studies \cite{desai2005molecular,ould2000molecular,liu2008molecular}, the diffusion coefficient of small NPs decreases when the NP mass increases at fixed NP volume, i.e., when the mass density is increased. The effect becomes progressively weaker as $\sigma_N$ is increased, and at $\sigma_N=7$ no mass density dependence is observed. Nevertheless, this effect should be taken into account when comparing the results of simulations to those of experiments or to theoretical predictions. For a detailed discussion, see Sec.~\ref{sec:mass_dependence} in the Appendix. 

\subsection{Nanoparticle diffusion}

\begin{figure}
\centering
\includegraphics[width=0.45 \textwidth]{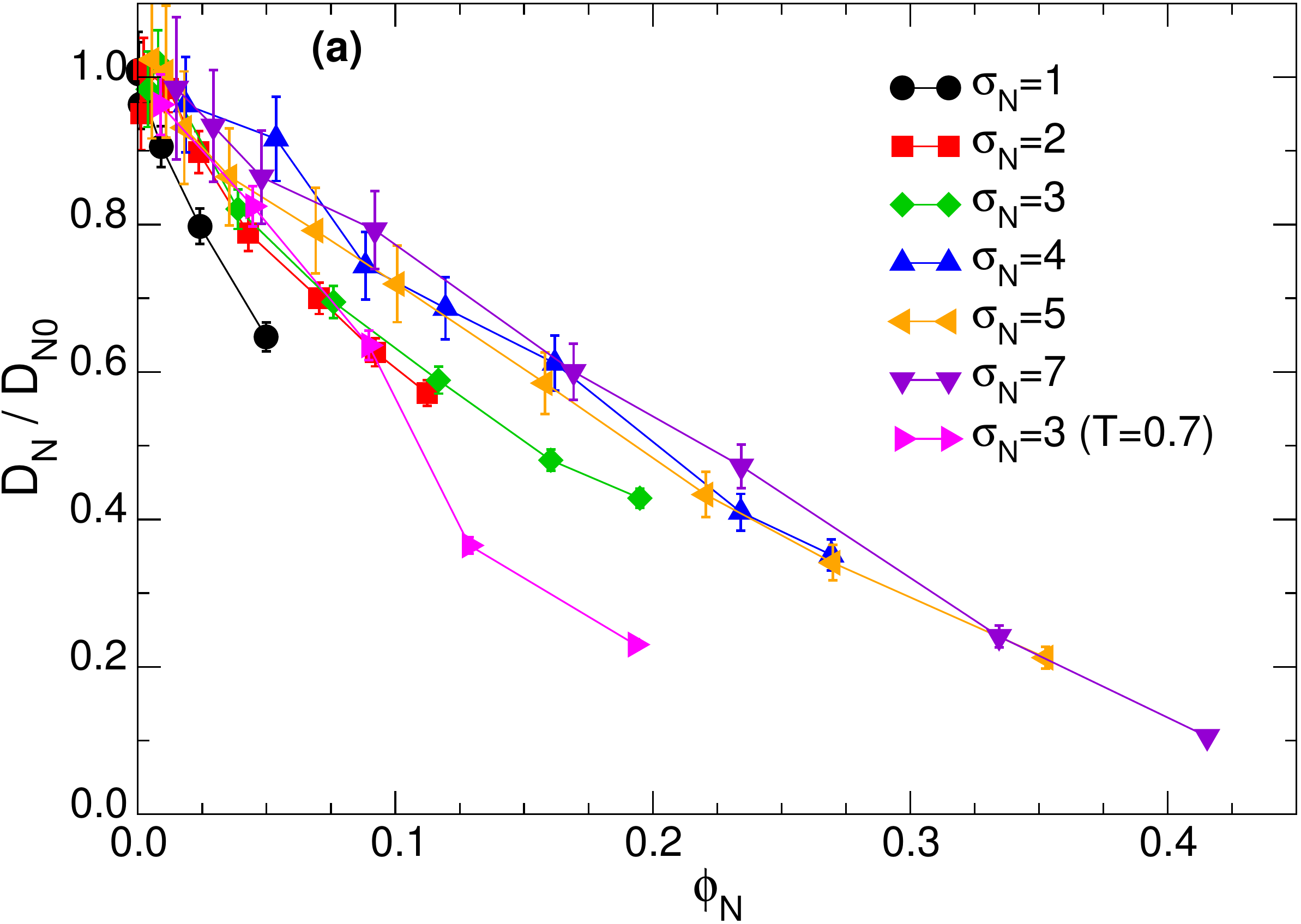}
\includegraphics[width=0.45 \textwidth]{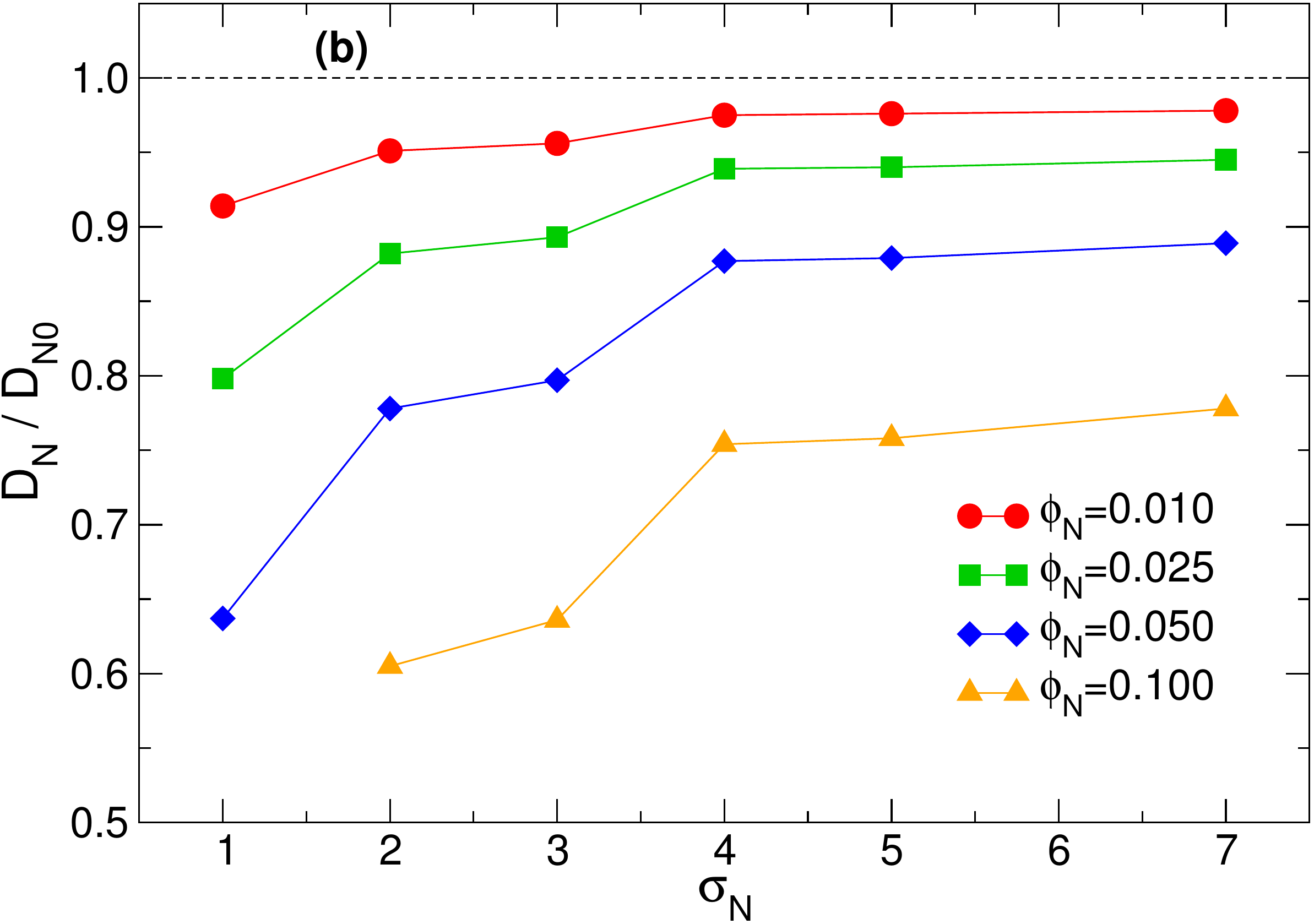}
\caption{Reduced NP diffusion coefficient as a function of the NP volume fraction $\phi_N$ (a) and of NP diameter $\sigma_N$ (b).}
\label{dnp}
\end{figure}

In the previous section, we have dealt with the motion of a single NP in the polymer solution, i.e., we have considered the dilute NP limit: We will now discuss the dynamics of NPs at higher NP volume fraction $\phi_N$. 

Only few simulation studies have considered high NP volume fractions. Liu \emph{et al.} \cite{liu2008molecular} have observed a reduction of the NP diffusion coefficient with increasing $\phi_N$, and attributed the phenomenon to polymer-mediated interactions, i.e., to the formation of chain bridges between neighboring NPs that would hinder NP motion; it is not clear, however, what the typical lifetime of such bridges should be, and thus whether this explanation is correct. Karatrantos \emph{et al.} \cite{karatrantos2017polymer} have observed a similar reduction in NP mobility and argued that it is due \leftquote\emph{to both nanoparticle-polymer surface area and nanoparticle volume fraction}" \cite{karatrantos2017polymer},  implying that pure geometry and polymer-NP attraction both play a role. The importance of polymer-NP interaction in NP dynamics is beyond dispute: Patti \cite{patti2014molecular} showed that the diffusion coefficient of NPs in an unentangled melt decreases monotonically when the strength of the polymer-NP interaction is increased, with the decrease being stronger for smaller NPs. A monotonic decrease of NP diffusivity with the strength of the polymer-NP interaction was also observed by Liu \emph{et al.} \cite{liu2008molecular}. We mention, however, that this trend can be reversed (NP diffusivity \emph{increasing} with increasing interaction strength) in strongly entangled systems, where the dynamics of the NPs is dominated by density fluctuations on length scales of the order of the tube diameter \cite{yamamoto2011theory}.

In Fig.~\ref{dnp}a, we show the reduced diffusion coefficient of the NPs, $D_N/D_{N0}$, where $D_{N0}$ is the diffusion coefficient of a single NP in the polymer solution, as a function of the NP volume fraction $\phi_N$; also shown are data for $\sigma_N=3$ and $T=0.7$. Similarly to the diffusion coefficient of the chains, $D_N$ decreases with increasing NP volume fraction.
The first thing that one can notice is that the decrease of $D_N/D_{N0}$ with the NP volume fraction is rather quick: Already at the modest volume fraction of $\phi_N=0.1$, the diffusion coefficient is reduced by $\approx 40 \%$ for NPs of diameter $\sigma_N=2$ and $3$, and by $\approx 30 \%$ for NPs of diameter $\sigma_N \geq 3$ (Fig.~\ref{dnp}a). The most interesting characteristic of $D_N/D_{N0}$ is however the dependence on $\sigma_N$ at fixed $\phi_N$. To better visualize this, we have interpolated between the points in Fig.~\ref{dnp}a in order to obtain approximately the reduced NP diffusion coefficient as a function of the NP diameter $\sigma_N$ at constant $\phi_N$ (Fig.~\ref{dnp}b). The ratio $D_N(\sigma_N)/D_{N0}$ shows an initial increase with increasing $\sigma_N$, then an inflection point at $\sigma_N\approx 2.5$, and finally it reaches a plateau for $\sigma_N \gtrsim 4$. Such a peculiar behavior can be interpreted in the following manner: At $\sigma_N\approx1$, increasing the NP diameter at fixed volume fraction has the effect of reducing the polymer-NP interface, and therefore decreasing the total interaction energy between polymers and NPs, resulting in an enhanced NP diffusion. When the NP size becomes larger than the mesh size $\xi \approx 1.3$ (the exact value depends on $\phi_m$), the motion of NPs starts to be geometrically hindered by the polymer segments \cite{wyart2000viscosity,cai2011mobility}, and as a result the dependence of $D_N/D_{N0}$ on the NP size weakens. Then, when the NPs become large enough, since the surface-to-volume ratio becomes smaller, the importance of the energetic contribution to the diffusion coefficient starts to decrease, resulting in another increase in the diffusion coefficient. Finally, for large NPs energy becomes irrelevant and $D_N/D_{N0}$ is completely controlled by geometry, and therefore is constant at constant NP volume fraction.

\section{Higher nanoparticle volume fractions} \label{high_nppf}

\begin{figure}
\centering
\includegraphics[width=0.45 \textwidth]{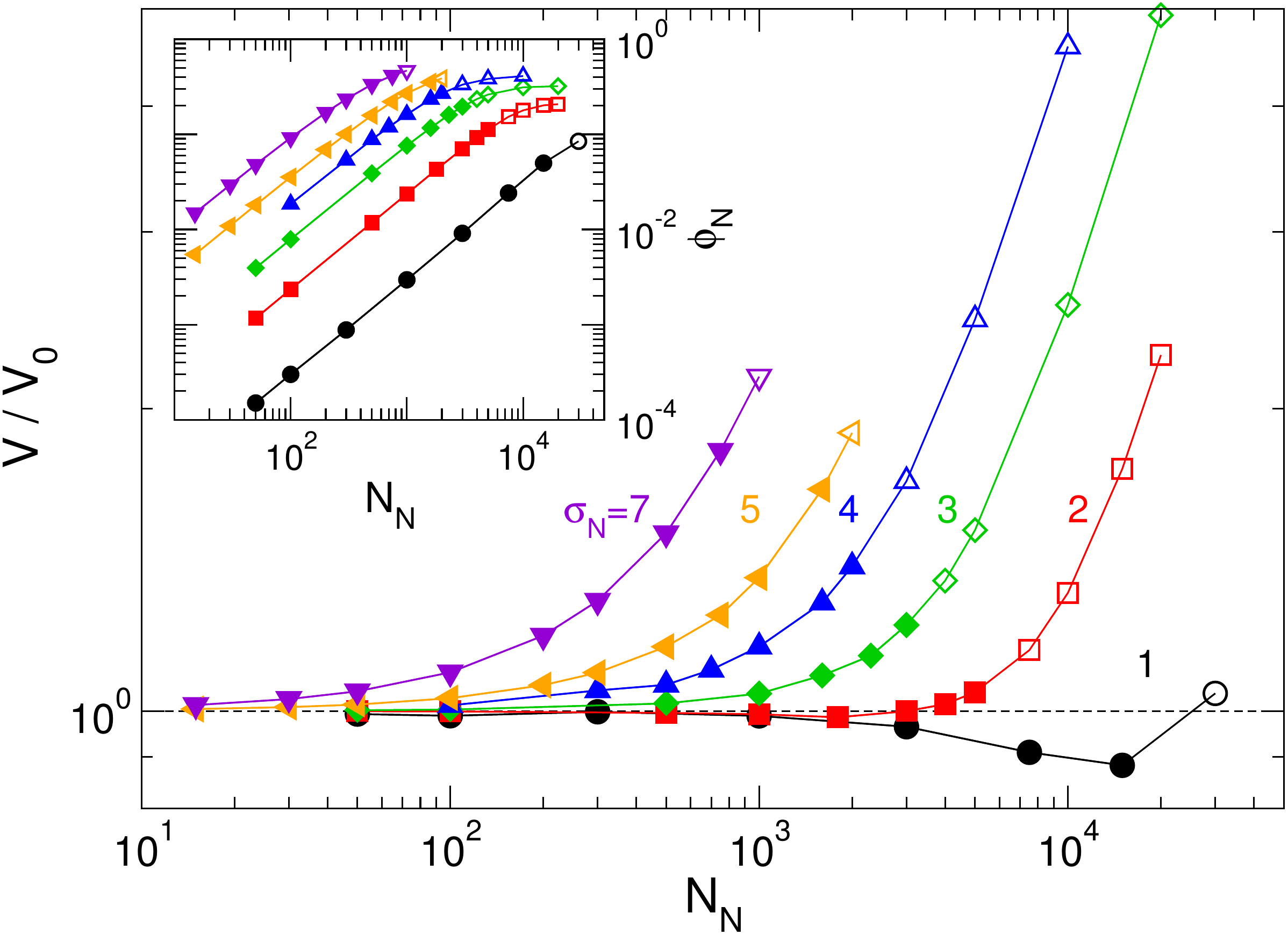}
\caption{Reduced volume of the simulation box versus NP number $N_N$. \emph{Inset}: NP volume fraction $\phi_N$ versus $N_N$. Filled symbols: good NP dispersion. Open symbols: poor NP dispersion.}
\label{nppf_vs_npnumber}
\end{figure}

If we keep increasing the number of NPs $N_N$ while keeping pressure and number of polymers constant, the volume of the simulation box will eventually start to increase proportionally to $N_N$. As a consequence, the NP volume fraction $\phi_N$ will reach a plateau, $\phi_N=\phi_N^\text{max} (\sigma_N)$, which corresponds to the value of $\phi_N$ for a pure NP system at temperature $T=1.0$ and pressure $P=0.1$. This situation corresponds approximately to the one depicted in Fig.~\ref{snapshots}d. If a standard LJ potential was used for the NP-NP interaction, $\phi_N^\text{max}$ would not depend on $\sigma_N$, since the interaction potential would only depend on the ratio $\sigma_N/r$ and all systems would be equivalent apart from a trivial distance rescaling. However, the expanded LJ potential, Eq.~\eqref{uij}, does not simply depend on $\sigma_N/r$; therefore, pure NPs systems with the same $T$ and $P$ are \emph{not} equivalent.

In Fig.~\ref{nppf_vs_npnumber}, we show the reduced volume of the simulation box $V/V_0$, where $V_0=V(\phi_N=0)$, as a function of $N_N$, for different values of $\sigma_N$. For $\sigma_N \geq 3$, the volume increases monotonically with the NP number. For $\sigma_N=1$ and $ 2$, on the other hand, there is a range of $N_N$ values in which we observe a decrease in volume (see Sec.~\ref{polymer_structure}). In the inset of Fig.~\ref{nppf_vs_npnumber} we show the NP volume fraction $\phi_N$ as a function of $N_N$. Initially, the volume is almost constant and therefore as a good approximation $\phi_N=\pi \sigma_N^3 N_N / 6 V_0$. Then, for larger values of $N_N$, the volume starts to increase proportionally to $N_N$ and $\phi_N$ reaches the plateau $\phi_N=\phi_N^\text{max}$.

\begin{figure}
\centering
\includegraphics[width=0.45 \textwidth]{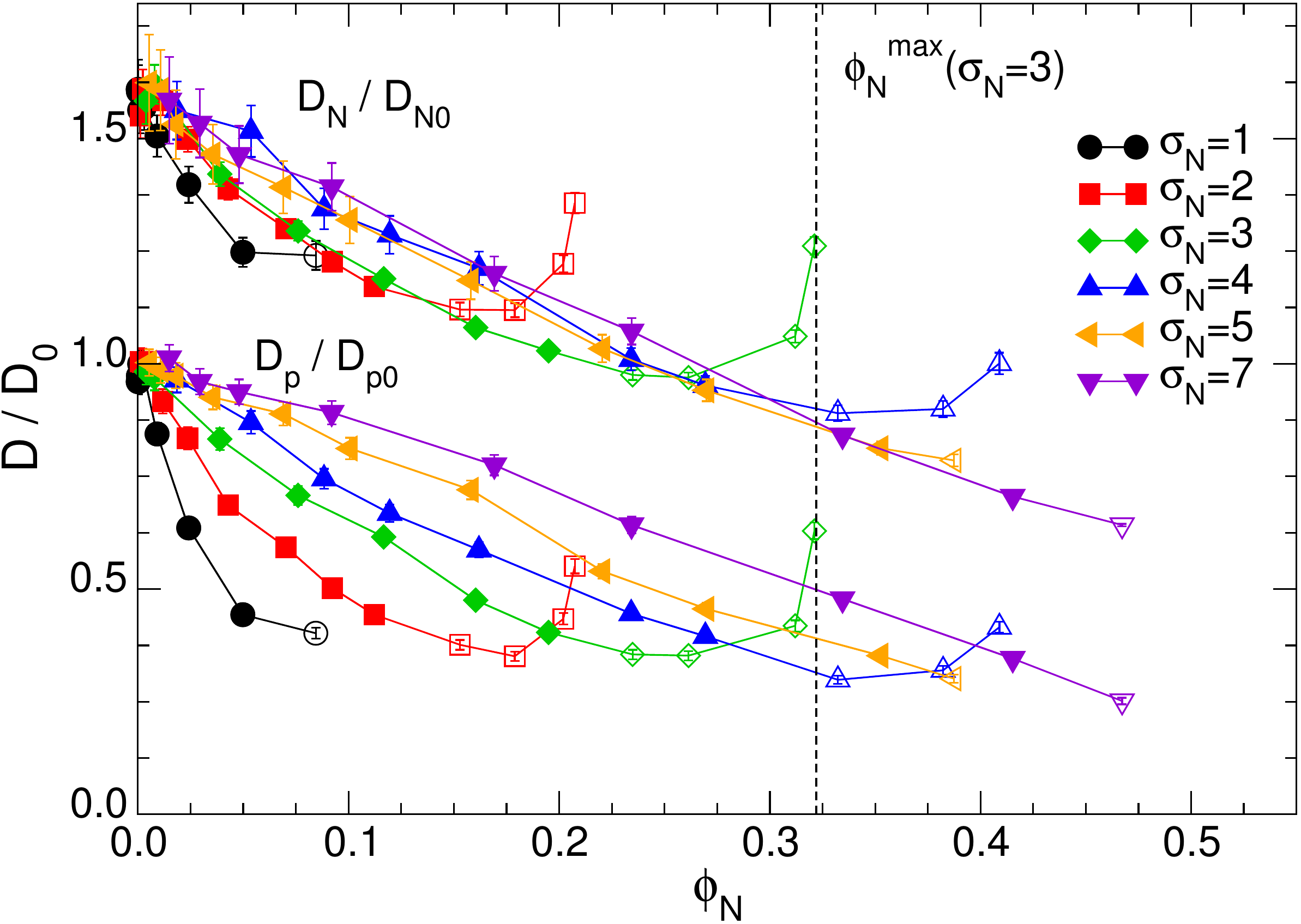}
\caption{Normalized diffusion coefficient of NPs and polymers as a function of NP volume fraction for different values of the NP diameter $\sigma_N$.  Filled symbols: good NP dispersion. Open symbols: poor NP dispersion. Dashed line: approximate value of $\phi_N^\text{max}$ for $\sigma_N=3$. The data for $D_N/D_{N0}$ have been shifted up by $0.6$ for clarity.}
\label{dall}
\end{figure}

Since the number of polymer chains is constant, the increase of the volume at large $N_N$ results in a decrease of polymer volume fraction $\phi_m$, and therefore in a decrease of the polymer-NP interface per unit volume. This in turn causes a weakening of the polymer-mediated attractive interaction between NPs and consequently an increase of the free volume $V_\text{free}=V[1-(\phi_N+\phi_m)]$. Both of these mechanisms result in an increase in the polymer and NP diffusivities. We can observe this effect in Fig.~\ref{dall}, where we show the normalized diffusion coefficient of both polymers and NPs for all the simulated systems, including those where good NP dispersion is not realized (open symbols): $D/D_0$ reaches a minimum corresponding to the value of $\phi_N$ at which the volume starts to increase, and it continues to grow as $\phi_N^\text{max}$ is approached. We note however that  for $\sigma_N=7$, $D/D_0$ shows a monotonic decrease. While in all the other cases we found that the pure NP system at $T=1.0$ and $P=0.1$ is a liquid, for $\sigma_N=7$ it is a crystal, which means that as we approach $\phi_N^\text{max}$ the ratio $D/D_0$ will decrease and eventually settle to a very small value.

\section{Summary and Conclusions} \label{summary}

We have carried out molecular dynamics simulations of a mixture of polymers and spherical nanoparticles (NPs) in a wide range of NP volume fractions ($\phi_N$) and for different NP diameters ($\sigma_N=1,2,3,4,5$, and $7\sigma$, where $\sigma$ is the monomer size). 

We have studied the structural properties of polymers and NPs, identifying the range of values of $\phi_N$ in which the NPs are well dispersed in the solution. In agreement with previous studies \cite{frischknecht2010expanded,karatrantos2015polymer,nakatani2001chain,mackay2006general,tuteja2008polymer}, we have found that the NPs of diameter $\sigma_N<2R_g$ act like a good solvent, swelling the polymers. Surprisingly, however, we have also observed that NPs of the same size as that of the monomers, $\sigma_N=1$, have the opposite effect, acting like a poor solvent. This is due to the high surface-to-volume ratio of small NPs, which causes the energetic contribution (which promotes chain contraction) to become stronger than the excluded volume contribution (which promotes chain expansion). Therefore, this effect is expected to depend on the strength of the monomer-NP interaction.

We have then analyzed the dynamical properties of the system, and in particular the diffusion coefficient of the NPs, $D_N$, and of the centers of mass of the polymer chains, $D_p$. We found that in the presence of good NP dispersion both $D_N$ and $D_p$ decrease monotonically with increasing NP volume fraction. 

The reduction of $D_p$ in the presence of NPs can be well described, at fixed temperature, with a modification of the confinement parameter approach proposed by Composto and coworkers\cite{gam2011macromolecular,gam2012polymer,lin2013attractive,choi2013universal}. Having observed that the conventional \leftquote static" confinement parameter, defined as the ratio between the interparticle distance $h$ and $2R_g$, fails to capture the dynamics of the chains at different temperatures, we propose a \emph{dynamic} confinement parameter  $h/\lambda_d(T)$, where $\lambda_d(T)$ represents a cooperativity length scale which increases with decreasing temperature. When plotted against  $h/\lambda_d(T)$, all the data for $\sigma_N \geq 2$ fall on the same master curve, $D_p=D_{p0}[1-\exp(-h/\lambda_d)]$. Deviations from the master curve only appear at $\sigma_N=1$.

The behavior of the reduced NP diffusion coefficient shows a complex dependence on $\sigma_N$, with an initial increase followed by an inflection point around $\sigma_N=2.5$ and finally a plateau for $\sigma_N \gtrsim 4$.  We speculate that this behavior results from an interplay between energetic and entropic contributions, with the latter depending on the mesh size of the solution, $\xi$. 

We have also studied the single NP diffusion coefficient, $D_{N0} = D_N (\phi_N \to 0)$, performing additional simulations with $\sigma_N=10,12,14$ at low NP volume fraction. For $\sigma_N<2 R_g$ the diffusion is faster than what is predicted by the Stokes-Einstein formula, according to which $D_{N0}\propto \sigma_N^{-1}$. Stokes-Einstein behavior is recovered when $\sigma_N\approx 2R_g$, in agreement with previous studies \cite{egorov2011anomalous,grabowski2009dynamics,kohli2012diffusion,poling2015size,liu2008molecular}. Theoretical studies predicted that for $\sigma_N$ sufficiently smaller than $2R_g$ the NP diffusion coefficient should decrease as  $D_{N0}\propto \sigma_N^{-3}$ \cite{wyart2000viscosity,yamamoto2011theory,cai2011mobility}; Cai \emph{et al.} \cite{cai2011mobility}, in particular, predicted that this behavior should be observed in the range $\xi < \sigma_N < 2 R_g$. However, it is not clear whether our data confirm these predictions or not: The main obstacle to find a conclusive answer is that $D_N$ shows a dependence on the mass density of the NPs for $\sigma_N \lesssim 7$, a dependence which has been reported in previous studies \cite{desai2005molecular,ould2000molecular,liu2008molecular}. This mass dependence becomes stronger with decreasing $\sigma_N$, making the interpretation of the data difficult. In light of this observation, we believe that care has to be taken when comparing the results of simulations to those of experiment and to theory, especially at low monomer volume fractions.

In conclusion, the behavior of a polymer-NP solution is very rich, and it is dictated by the value of many different parameters, most importantly the polymer-NP interaction and the NP size. In particular, since changing the polymer-NP interaction can lead to a qualitatively very different dynamical behavior, we believe that systematic studies should be conducted in order to clarify the role of interactions in the dynamics of nanocomposites. Comparison of simulation data with the available theories is complicated by many factors, such as subtleties in the definition of important quantities, e.g. the effective hydrodynamic radius and the mesh size, and the presence of broad cross-overs and inertial effects (mass dependence of the NP diffusivity). Other fundamental issues that remain to be fully clarified are the relevance of hydrodynamic interactions (recently investigated in Ref.~\citenum{chen2017effect}) and finite-size effects. We thus believe that the investigation of polymer-NP composites will remain an important topic of research also in the future.

\section*{Appendix} 

\setcounter{section}{0}
\renewcommand{\thesection}{A\Roman{section}}

\section{Interparticle distance and pore-size distribution}
\label{sec:pore_size}

In a polymer nanocomposite, the interparticle distance $h$ is the average distance between the surfaces of neighboring NPs in the system. In the literature \cite{gam2011macromolecular,gam2012polymer,lin2013attractive,choi2013universal}, $h$ has often been defined using Eq.~\eqref{id}, which we reproduce here:

\[
h^\text{th.}=\sigma_N\left[\left(\frac{\phi_{N}^\text{\tiny M}}{\phi_N}\right)^{1/3}-1\right],
\]

\noindent where $\phi_{N}^\text{\tiny M}$  is the maximum achievable NP volume fraction \cite{gam2012polymer}. There is however an evident problem with the above expression:  It is \emph{a priori} not clear at all what value should be used for $\phi_{N}^\text{\tiny M}$. Taking the NPs to behave approximately as hard spheres (an approximation that in our case is justified by the very steep NP-NP potential), there are several possibilities, like the close-packing value \cite{hales2005proof} $\phi_\text{cp}=\pi/\sqrt{18}\approx 0.740$, corresponding to an fcc or hcp lattice, or the volume fraction of some other crystal lattice, like the bcc ($\phi_\text{bcc} = \pi \sqrt{3}/8 \approx 0.680$) or the simple cubic ($\phi_\text{sc} = \pi /6 \approx 0.524$). The value $\phi_{N}^\text{\tiny M}= \pi /6 $ was used in Ref.~\citenum{wu1985phase}, one of the first to apply Eq.~\eqref{id} to polymer nanocomposites. More recently, $\phi_\text{rcp} \approx 0.637 \approx 2/\pi$, that should correspond to a random close packing (RCP) of hard spheres, has often been invoked in the definition of $h$ \cite{gam2011macromolecular,gam2012polymer,lin2013attractive,choi2013universal}. However, it has been shown by Torquato \emph{et al.} that the concept of RCP is ill-defined, and that different procedures can result in different values for $\phi_\text{rcp}$, ranging from $0.6$ to $0.68$  \cite{torquato2000random}. This issue could be solved, as suggested by Torquato \emph{et al.}, by replacing the ill-defined concept of RCP with that of \emph{maximally random jammed} (MRJ) structure \cite{torquato2000random,jiao2011maximally}. For monodispere hard spheres, this redefinition should lead to a unique value, $\phi_\text{\tiny{MRJ}} = 0.642$ \cite{jiao2011maximally}. However, the problem of \emph{a priori}  assigning a certain value to $\phi_{N}^\text{\tiny M}$ remains.

We propose therefore a different way to define $h$, which relies on the concept of \emph{pore-size distribution} \cite{torquato2013random}. The pore-size probability density function (PDF) $P(\delta)$ of a system consisting of two phases is defined such that $P(\delta) d\delta$ represents the probability that a randomly chosen point in the phase of interest lies at a distance between $\delta$ and $\delta+d\delta$ of the nearest point on the interface between the two phases \cite{torquato2013random}. It is clear from this definition that the typical interparticle distance $h$ for the NPs in a polymer nanocomposite should correspond approximately to the typical pore size.

There is another clue that suggests the identification of $h$  with some quantity derived from $P(\delta)$. For a system of randomly distributed overlapping spheres of radius $R$ with number density $\rho$, the pore-size PDF can be computed explicitly \cite{rintoul1996structure}:

\begin{equation}
P^\text{os}(\delta) = \frac{3 \phi}{e^{-\phi} R} \left( \frac  \delta R + 1 \right)^2 \exp \left[ - \phi \left( \frac  \delta R + 1 \right)^3 \right].
\label{pore_pdf}
\end{equation}

\noindent In this expression, $\phi=\pi (2R)^3 \rho /6$ is the \leftquote volume fraction" of the spheres (although since the spheres can overlap, this does not correspond to their real volume fraction).

\begin{figure}[t]
\centering
\includegraphics[width=0.45 \textwidth]{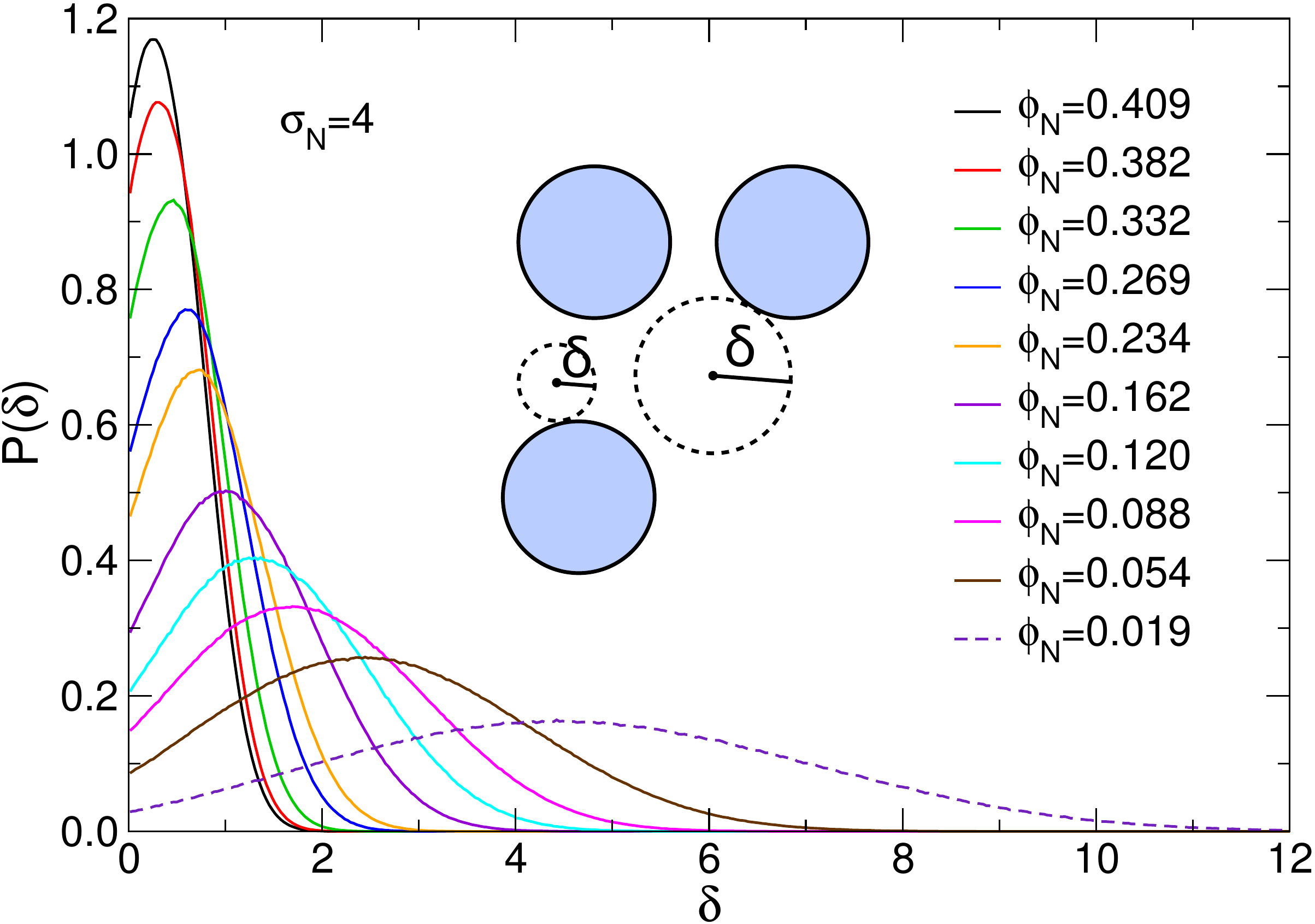}
\caption{(color online) Pore-size probability density function of the NPs for $\sigma_N=4$. Cartoon: schematic representation of how the pore-size distribution is calculated.}
\label{pore_pdf_d4}
\end{figure}

Let us now define $\delta_\text{max}$ as the value of $\delta$ for which $P(\delta)$ is maximized, i.e., $P(\delta_\text{max}) = \max_\delta \{P(\delta)\}$. For this model, we have $\delta^\text{os}_\text{max}=R [(2/3\phi)^{1/3}-1]$, and therefore we can define a \emph{typical pore diameter} as

\begin{equation}
2 \delta^\text{os}_\text{max} = 2R \left[\left( \frac 2 {3 \phi} \right)^{1/3}-1 \right].
\label{pore_diameter}
\end{equation}

\noindent By comparing Eqs.~\eqref{id} and \eqref{pore_diameter} and making the identifications $2R=\sigma_N$ and $\phi=\phi_N$, we note that $2 \delta^\text{os}_\text{max}= h^\text{th.}$, provided that we choose $\phi_{N}^\text{\tiny M}=2/3 \approx 0.667$. Given this quite remarkable connection between $2 \delta^\text{os}_\text{max}$ and $h^\text{th.}$, and given that the definition of $\delta_\text{max}$ does not present the same problems that affect Eq.~\eqref{id}, we are naturally lead to define the interparticle distance as $h = 2 \delta_\text{max}$, where $2 \delta_\text{max}$ is not computed using Eq~\eqref{pore_diameter}, which is strictly valid only for the overlapping spheres system, but rather evaluated directly from the pore-size PDF $P(\delta)$ obtained from the data.

The algorithm employed to obtain $P(\delta)$ from the data is described in Ref.~\citenum{torquato2013random}: (1) A random point $(x,y,z)$ in the NP-free phase is chosen (the point must lay at distance $r>\sigma_N/2$ from every NP). (2) The radius $\delta$ of the largest sphere centered in $(x,y,z)$ that does not intersect any NP is recorded. (3) Step 1 and 2 are repeated many times, and a histogram of the radii is created. (4) The pore-size PDF $P(\delta)$ is obtained by normalizing the histogram. We note that in order for step 2 to be well-defined, we need to approximate the NPs as hard spheres of diameter $\sigma_N$; this approximation is justified by the fact that the NP-NP potential is very steep. 

In Fig.~\ref{pore_pdf_d4} we show the pore-size PDF for $\sigma_N=4$. As already shown in Fig.~\ref{id_vs_nppf}, the interparticle distance obtained from the pore-size PDF and the one calculated using Eq.~\ref{id} are very similar. This means that Eq.~\ref{id} can be used to obtain an estimate of the \leftquote true" interparticle distance, despite the problems affecting its definition. We also mention that the pore-size PDF can also be extracted from experimental data: For example, Rintoul \emph{et al.} used X-ray microtomography to obtain a three dimensional digitalized image of a porous magnetic gel and computed from it the pore-size PDF and other statistical correlation functions \cite{rintoul1996structure}.

\section{Mass dependence of NP and polymer diffusivities}
\label{sec:mass_dependence}

\begin{figure}
\centering
\includegraphics[width=0.45 \textwidth]{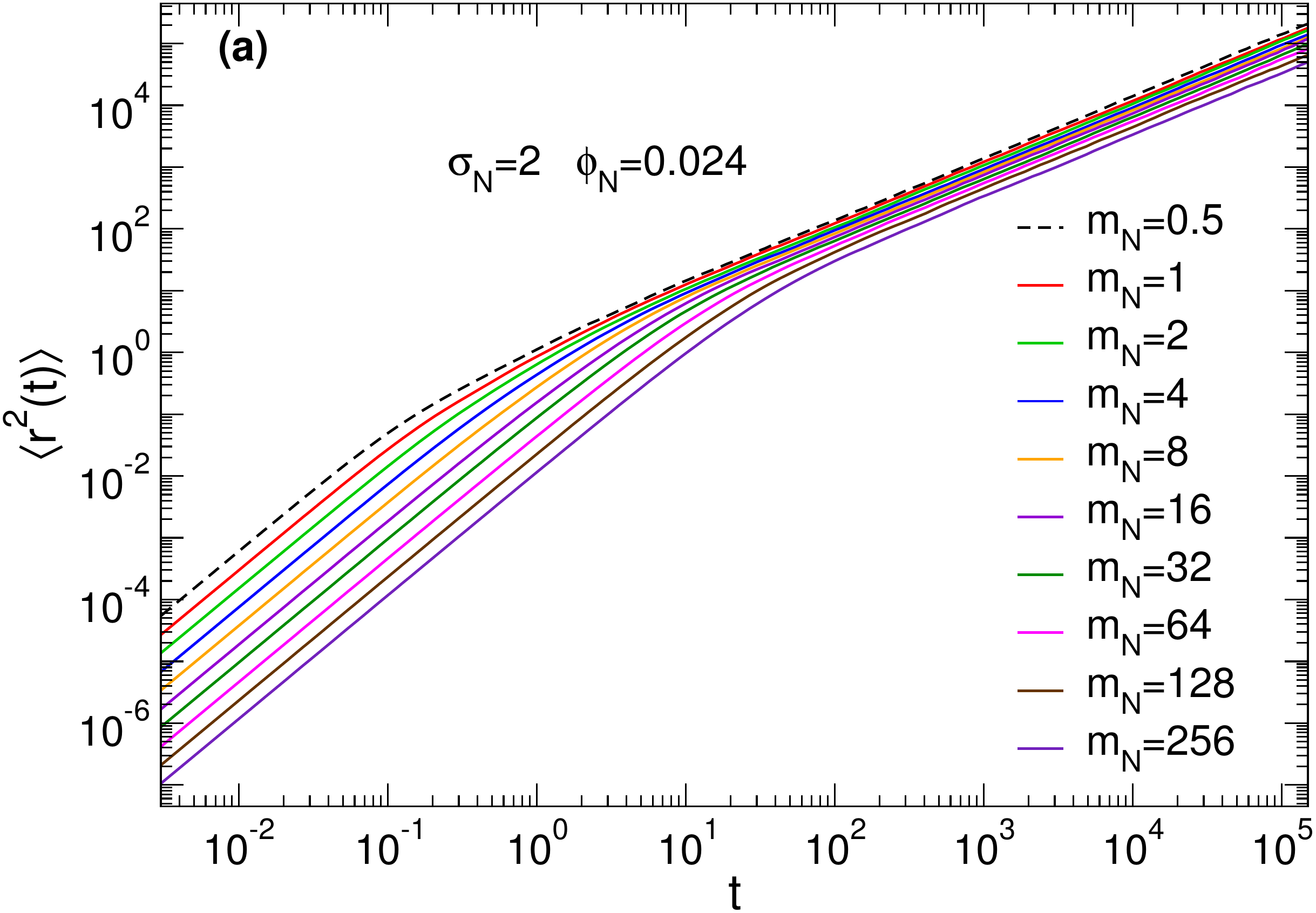}
\includegraphics[width=0.45 \textwidth]{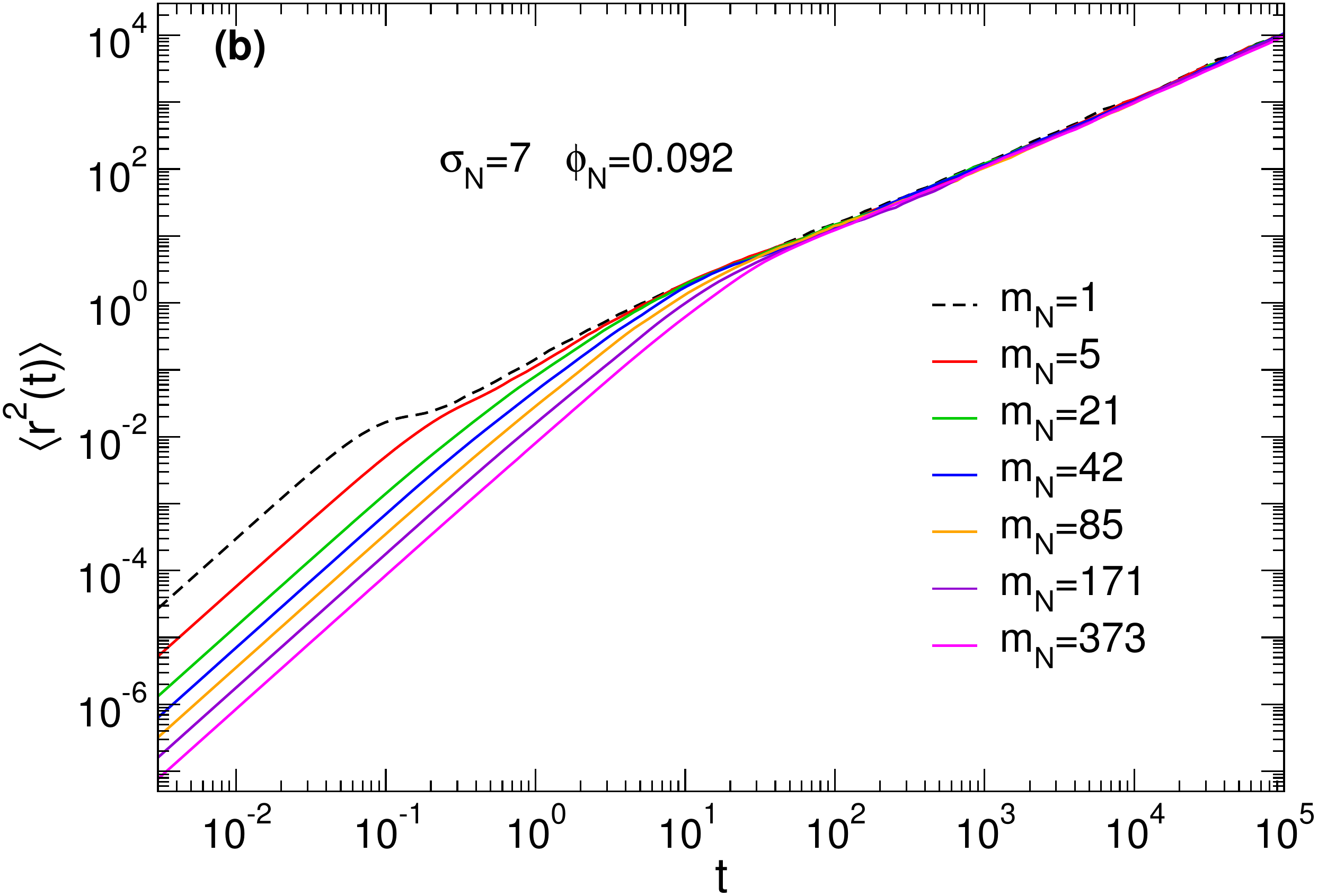}
\caption{(color online) Mean squared displacement of the NPs when varying the NP mass $m_N$ at fixed diameter $\sigma_N$, in the case $\sigma_N=2, \phi_N=0.024$ (a) and  $\sigma_N=7, \phi_N=0.092$ (b).  }
\label{msd_mass}
\end{figure}

In previous studies of polymer nanocomposites \cite{desai2005molecular,liu2008molecular} and binary soft-sphere liquids \cite{ould2000molecular} it has been shown that outside of the Stokes-Einstein regime not only the diameter, but also the mass density of a particle can affect its dynamics. In order to study this effect, we performed some simulations at low NP volume fraction and changed the NP mass while leaving the diameter fixed.

The results for the mean squared displacement (MSD) $\langle r^2(t)\rangle$ of the particles are reported in Fig.~\ref{msd_mass} for the cases $\sigma_N=2, \phi_N=0.024$ and $\sigma_N=7, \phi_N=0.092$. At short times, the motion of the particles is ballistic and the mass dependence of the MSD is trivial:

\begin{equation}
\langle r_N^2(t) \rangle = \langle v_N^2 \rangle t^2 =  \frac {3 k_B T} {m_N} t^2,
\end{equation}

\noindent where $\langle v_N^2 \rangle$ is the mean squared speed of the NPs. At longer times, when the motion becomes diffusive, i.e. $\langle r^2(t)\rangle \propto t$, we can observe a much more interesting effect: While the motion of the larger NPs is unaffected by a change in the mass (Fig.~\ref{msd_mass}b), the motion of smaller NPs presents a clear mass dependence (Fig.~\ref{msd_mass}a).

\begin{figure}
\centering
\includegraphics[width=0.45 \textwidth]{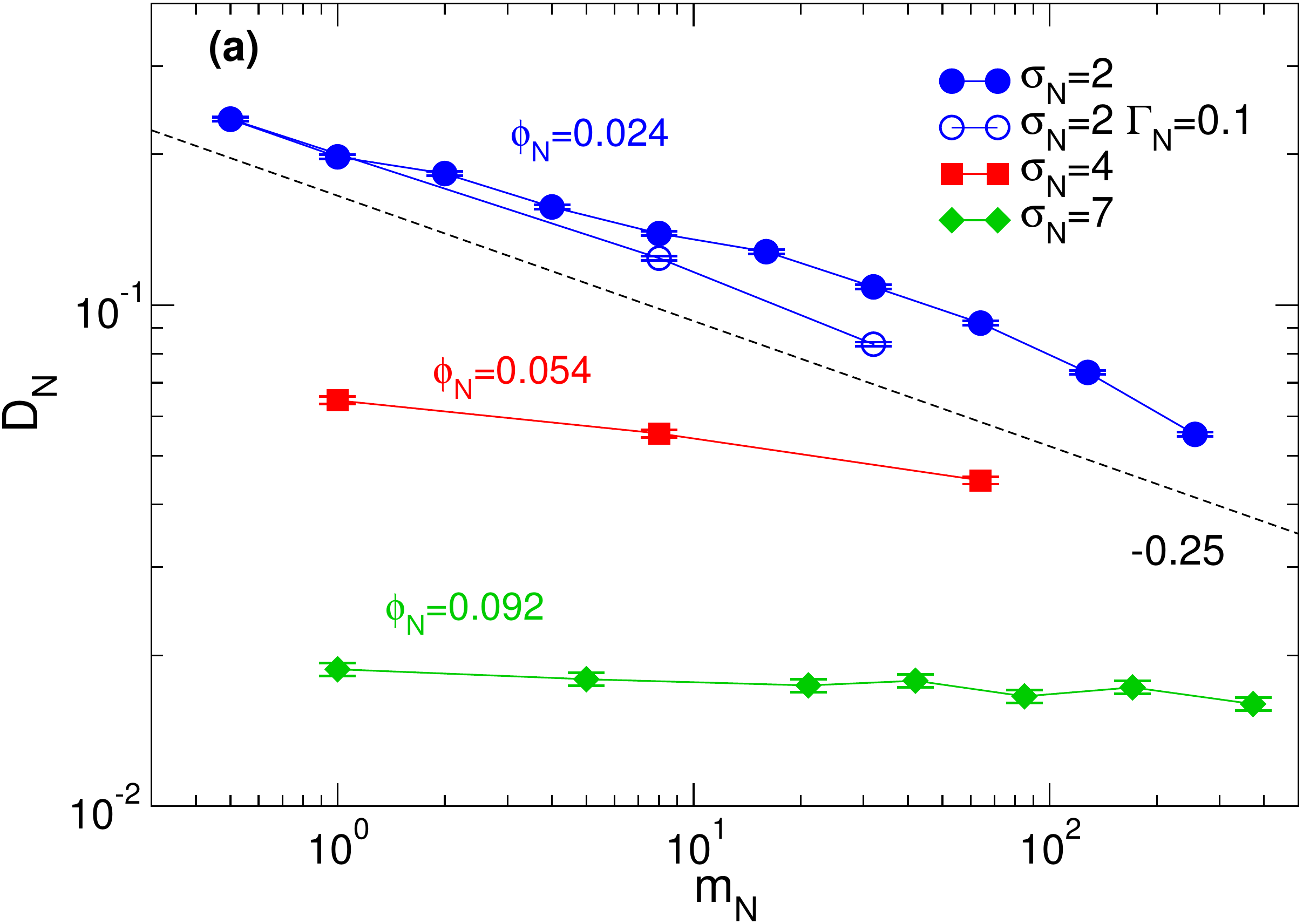}
\includegraphics[width=0.45 \textwidth]{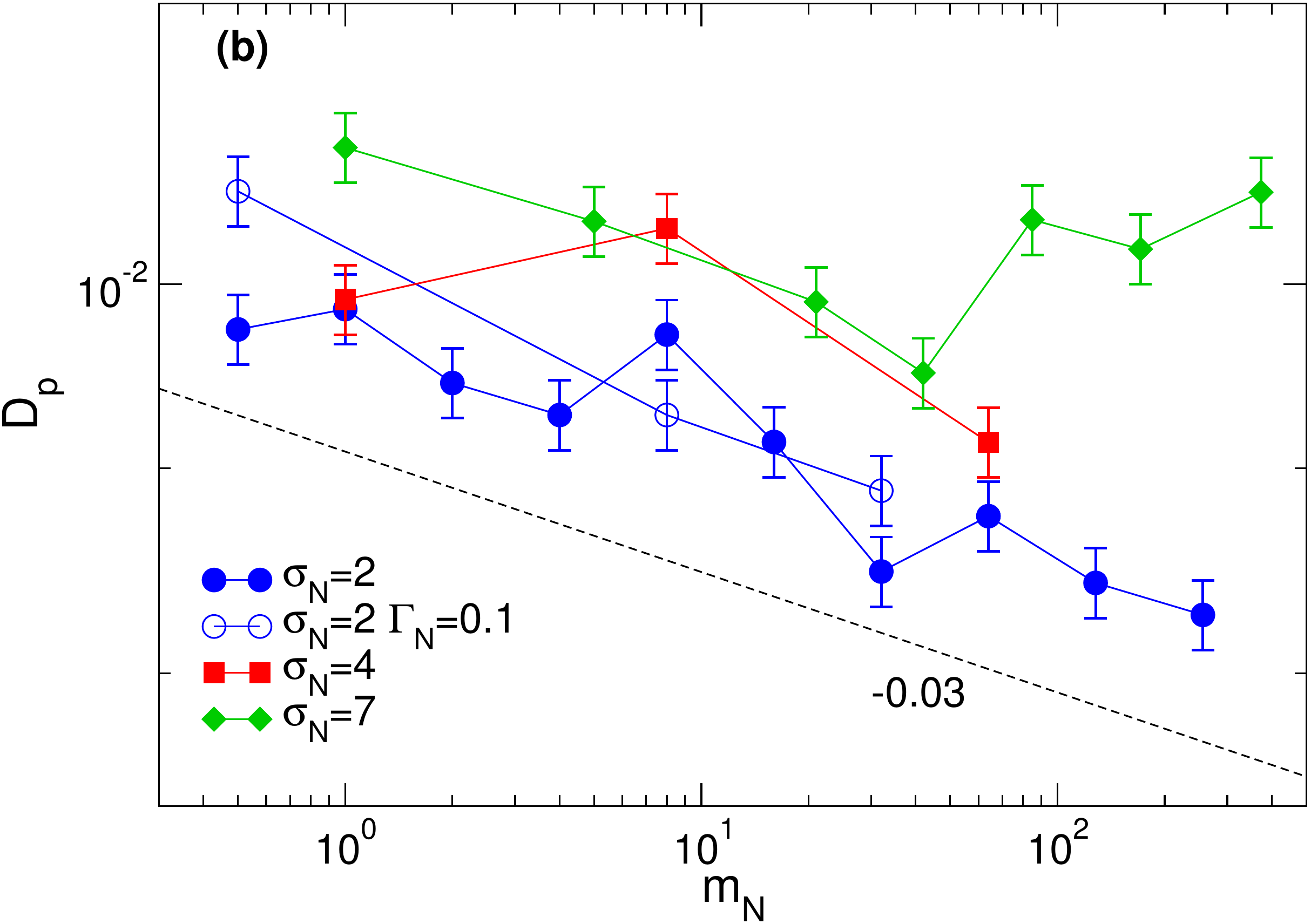}
\caption{Diffusion coefficient of the NPs (a) and of the polymers (b) when varying the NP mass $m_N$ at fixed diameter $\sigma_N$. The values of the NP volume fraction $\phi_N$ are $0.024,0.054$ and $0.092$ respectively for $\sigma_N=2,4$ and $7$. Open symbols: simulation without rescaling of the viscous friction coefficient $\Gamma_N$. Dashed lines: slopes $-0.25$ (a) and $-0.03$ (b) for reference.}
\label{dnp_mass}
\end{figure}

This result can be better appreciated in Fig.~\ref{dnp_mass}a, where we show the NP diffusion coefficient $D_N$ as a function of the NP mass for different values of $\sigma_N$ and $\phi_N$ (here also the case $\sigma_N=4, \phi_N=0.054$ is shown). One recognizes that the mass dependence of the diffusion coefficient becomes weaker when the NP diameter is increased, and for $\sigma_N=7$ it has disappeared almost completely.  The result that the mass dependence of the long time diffusion coefficient is stronger for smaller particles is in agreement with previous studies \cite{ould2000molecular,liu2008molecular}. 

Our interpretation of this result is that large particles are forced to wait for the polymers to relax in order to diffuse through the solution: The constraints on their motion are purely geometric, and mass plays no role. Smaller particles, on the other hand, can \leftquote slip" between the polymers; the probability that they find a passage to slip through in a certain time interval increases with their average velocity, which at equilibrium decreases as $m_N^{-1/2}$, as it follows from the equipartition of energy: $\langle v_N^2 \rangle = 3 k_B T/m_N$. We expect this effect to be strongly suppressed when the polymer density is increased, and to be negligible for the motion of NPs of diameter $\sigma_N \gtrsim \sigma$ in a melt. 

Although in Refs.~\citenum{ould2000molecular,liu2008molecular} it was claimed that the mass dependence disappears for high values of $m_N$, in our case for $\sigma_N=2$ there is no hint that the mass dependence vanishes for larger mass values. In the case of Ref.~\citenum{ould2000molecular}, however, the disappearance of the mass dependence for high values of the mass is only apparent, as it becomes clear once the data are plotted in double logarithmic instead of linear scale (not shown). We believe that this could also be the case for Ref.  \citenum{liu2008molecular}, since also there the data are only reported in linear scale. From the analysis of our data and of those of Ref.~\citenum{ould2000molecular}, it seems that not only the mass dependence does not disappear when $m_N$ is increased, but on the contrary it becomes stronger.

The diffusion coefficient of the polymer chains $D_p$ is almost unaffected by changes in the NP mass density, as we can observe from Fig.~\ref{dnp_mass}b, although it is possible to see a very weak decrease of $D_p$ for $\sigma_N=2$.  It is possible that the slowing down of the NPs has an effect on the dynamics of the polymers, but this effect is very weak at low NP concentrations. Further study should be dedicated to clarifying this point.

Finally, we have made some tests to determine whether the observed mass dependence is an artifact resulting from the scaling of the friction coefficient of the Langevin thermostat (see Sec.~\ref{sec:thermostat} in the S.I.). To this aim, we ran some simulations where the friction coefficient of the NPs, $\Gamma_N$, was kept constant and equal to that of the monomers: $\Gamma_N=\Gamma_m=0.1$. The result is included in Fig.~\ref{dnp_mass} (open blue circles). One sees that the effect of mass density is still present, but using $\Gamma_N=0.1$ (fixed) has the effect of reducing the diffusion coefficient of the NPs, as one expects since this means that more massive NPs experience a higher solvent viscosity (we recall that $\eta_s \propto \Gamma_N m_N / \sigma_N$). 

In conclusion, we have shown that the effect of mass density on the dynamical properties should be taken into careful consideration when performing molecular dynamics simulations of multi-component systems, such as polymer mixtures, binary fluids and solutions with explicit solvents. We think that the dependence of this effect on polymer and NP density and on polymer-NP interaction should be more thoroughly investigated in order to gain a better understanding of polymer-NP mixtures from the point of view of molecular dynamics simulations. 

\begin{acknowledgements}

We thank K. Schweizer, J. Oberdisse, L. Cipelletti, and M. Ozawa for fruitful discussions. This work has been supported by LabEx NUMEV (ANR-10-LABX-20) funded by the ‘‘Investissements d'Avenir’’ French Government program, managed by the French National Research Agency (ANR). Part of the simulations were performed at the Center of High Performance Computing MESO@LR in Montpellier. The snapshots in Fig.~\ref{snapshots} have been realized with VMD \cite{vmd}. 

\end{acknowledgements}

\section*{SUPPLEMENTARY INFORMATION}

\setcounter{equation}{0}
\setcounter{figure}{0}
\setcounter{table}{0}
\setcounter{section}{0}

\renewcommand{\theequation}{S\arabic{equation}}
\renewcommand{\thefigure}{S\arabic{figure}}
\renewcommand{\thetable}{S\arabic{table}}
\renewcommand{\thesection}{S\Roman{section}}

\section{Scaling of the friction coefficient of the thermostat} \label{sec:thermostat}

\begin{figure}[h]
\centering
\includegraphics[width=0.45 \textwidth]{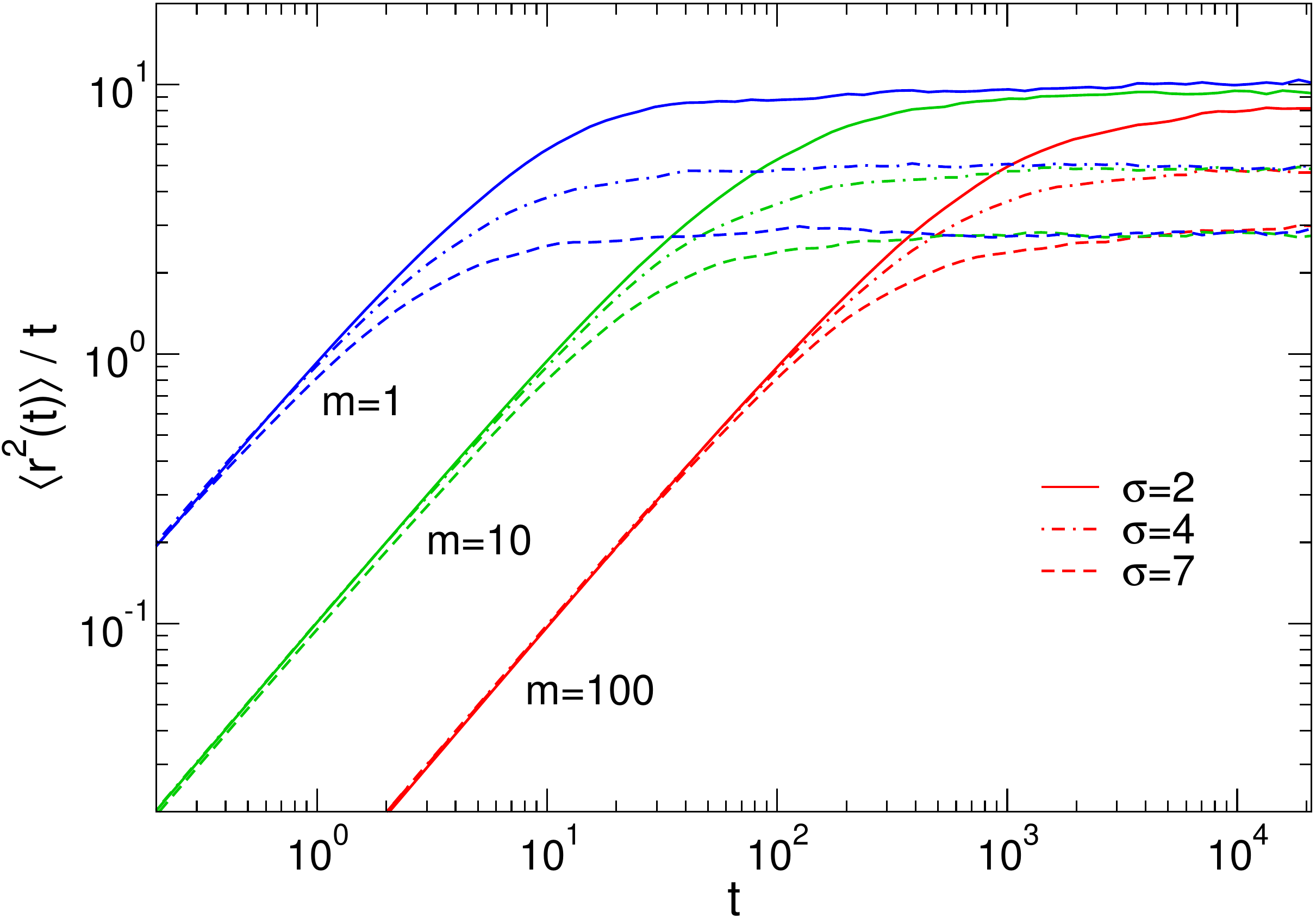}
\caption{(color online) Mean-squared displacement (divided by time) of a particle of diameter $\sigma$ and mass $m$ diffusing in pure (implicit) solvent. The long-time diffusion coefficient depends only on $\sigma$.}
\label{viscosity}
\end{figure}

As discussed in the main text, we have used for our simulations a Langevin thermostat, so that the force experienced by particle $i$ is

\begin{equation}
m_\alpha \ddot {\mathbf r}_{\alpha,i} = - \mathbf \nabla_i U (\{ \mathbf r_{\alpha,k} \}) - m_\alpha \Gamma_\alpha \dot {\mathbf r}_{\alpha,i} + \sqrt{2m_\alpha \Gamma_\alpha k_B T} \ \boldsymbol{\zeta}(t),
\label{langevin}
\end{equation}

\noindent where $\alpha$ denotes the particle type (${m=\text{monomer}}$, ${N=\text{NP}}$). The mass and diameter of the monomers are respectively $m_m=m$ and $\sigma_m=\sigma$. $ U (\{ \mathbf r_{\alpha,k} \}) $ is the total interaction potential acting on the particle, with $\{ \mathbf r_{\alpha,k} \}$ representing the set of coordinates of all the particles in the sysem. The term $\sqrt{2m_\alpha \Gamma_\alpha k_B T} \boldsymbol \zeta(t)$ is a random force which represents collisions with solvent molecules, while $\Gamma_\alpha$ is a viscous friction coefficient, which is related to the viscosity of the implicit solvent $\eta_s$ by

\begin{equation}
\Gamma_\alpha = \frac{C \eta_s \sigma_\alpha}{m_\alpha} ,
\label{stokes}
\end{equation}

\noindent where $\sigma_\alpha$ is the diameter of the particle and $C$ a coefficient which depends on the hydrodynamic boundary conditions.

In our simulations we made the assumption that every particle (monomer or NP) interacts with the implicit solvent as if it was a continuum with fixed $\eta_s$. Therefore, from Eq.~\eqref{stokes} we get

\begin{equation}
\eta_s =\frac{ \Gamma_m m}{C \sigma} = \frac{ \Gamma_N m_N}{C \sigma_N} = \text{const.} \Rightarrow \Gamma_N = \Gamma_m \cdot \left(\frac{\sigma_N m}{m_N\sigma}\right).
\end{equation}

\noindent With this choice of the friction coefficient, the long-time diffusion coefficient of a particle in the pure (implicit) solvent $D_{s,\alpha}$ follows the Stokes-Einstein law with viscosity $\eta_s$:

\begin{equation}
D_{s,\alpha} = \frac{k_B T}{\Gamma_\alpha m_\alpha} = \frac{k_B T}{C \eta_s \sigma_\alpha}.
\end{equation}

\noindent This way, $D_{s,\alpha}$ depends only on the diameter of the particle, and not on its mass, as we can see from Fig.~\ref{viscosity}, where the mean-squared displacement (divided by time) of particles of various masses and diameters in the pure solvent is shown. We note that since $\Gamma_m=0.1$ and $m,\sigma=1$, the numerical value of the solvent viscosity is $\eta_s=0.1/3 \pi = 1.06 \cdot 10^{-2}$.

\section{Monomer-monomer radial distribution function}

\begin{figure*}
\centering
\includegraphics[width=0.45 \textwidth]{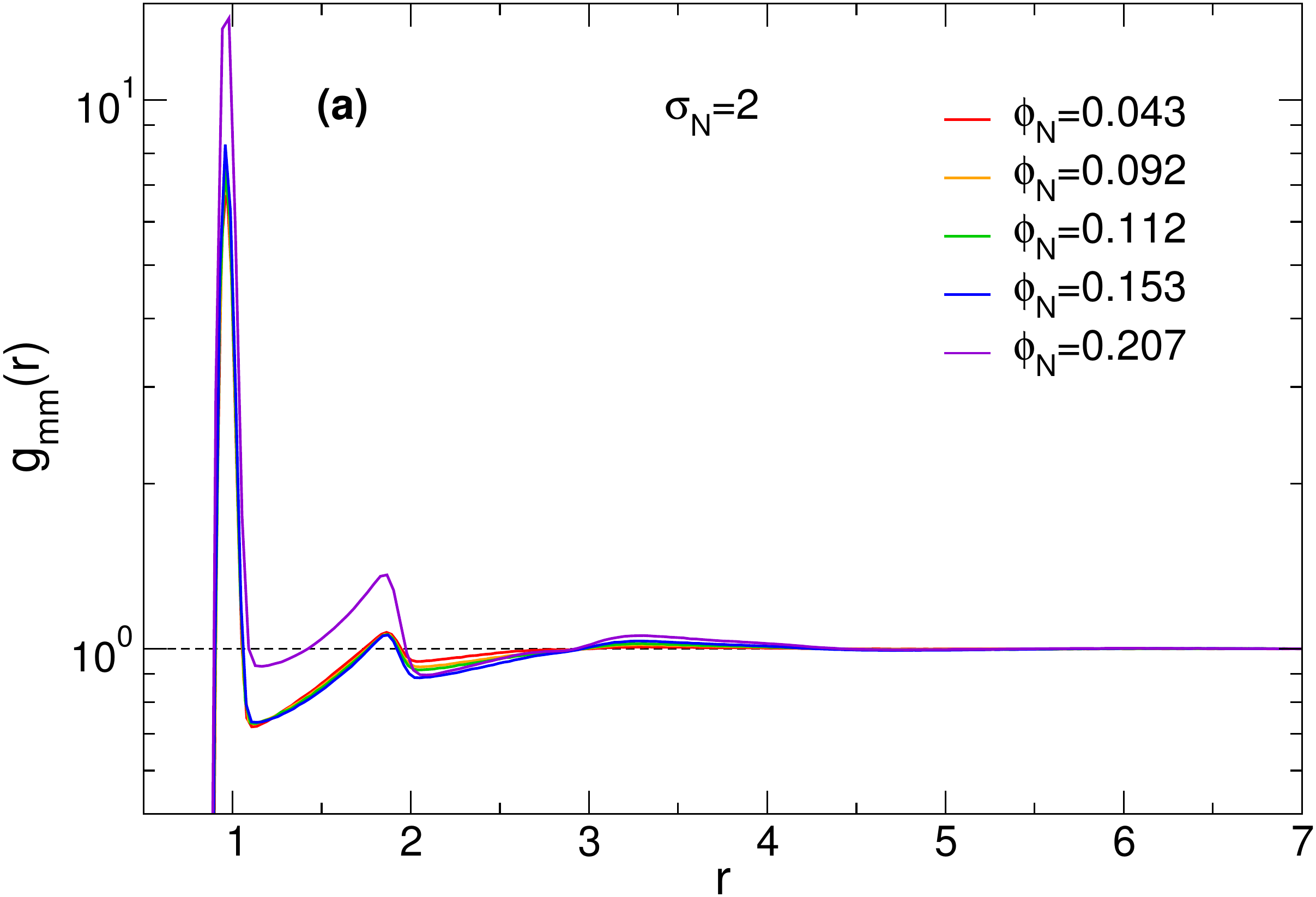}
\includegraphics[width=0.45 \textwidth]{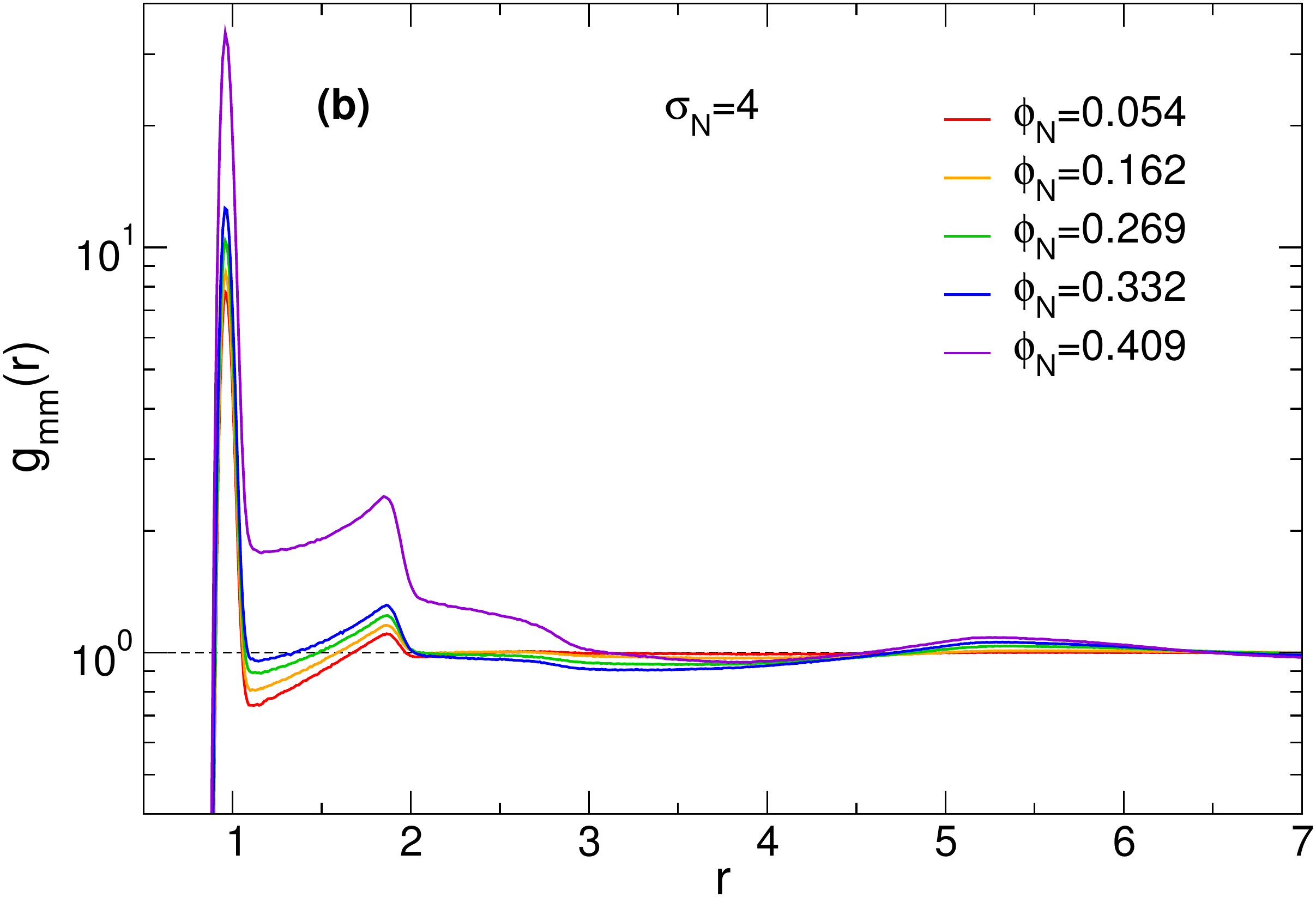}
\includegraphics[width=0.45 \textwidth]{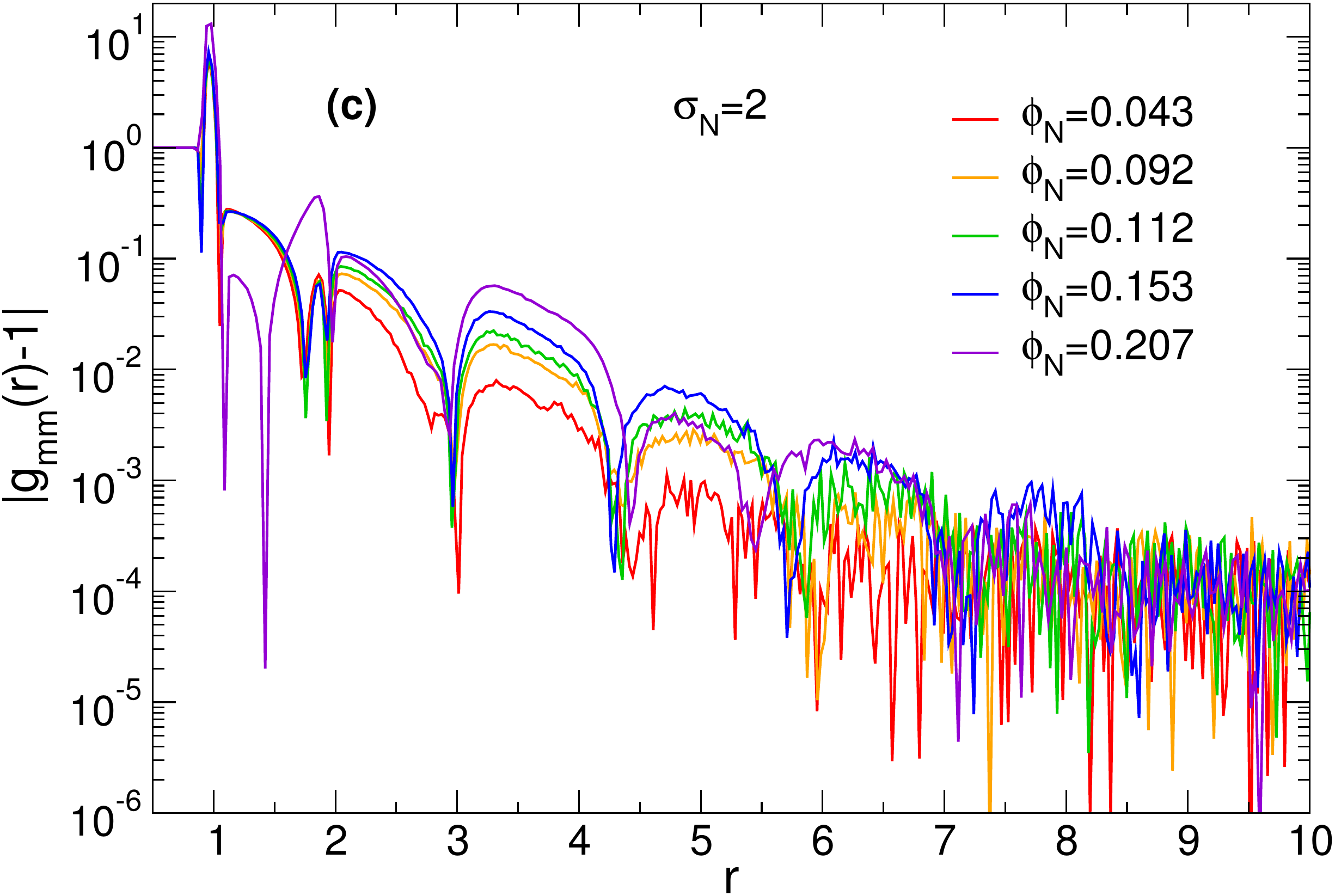}
\includegraphics[width=0.45 \textwidth]{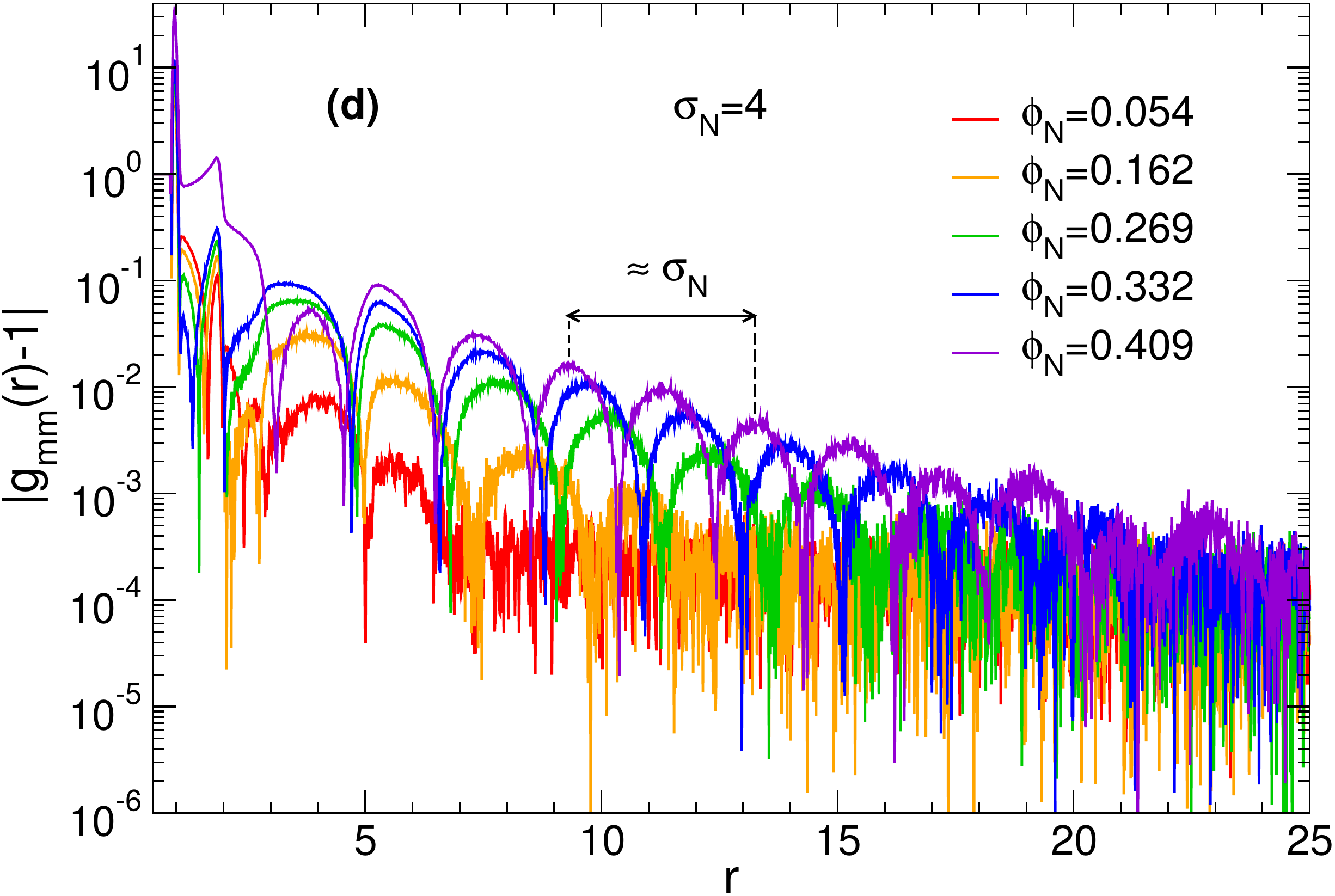}
\caption{(color online) Monomer-monomer radial distribution function for $\sigma_N=2$ (a) and $\sigma_N=4$ (b) and different values of the NP volume fraction $\phi_N$.}
\label{rdf_mm}
\end{figure*}

In Figs.~\ref{rdf_mm}a-b, we report the monomer-monomer radial distribution function $g_{mm}(r)$ for $\sigma_N=2$ and $4$. One sees that at low NP volume fraction the most prominent features of $g_{mm}(r)$ are a sharp peak at $r \approx r_b =0.96$ (first nearest neighbor distance in a chain) and a smaller one at $r\approx 2 r_b = 1.92$ (second nearest neighbor). When the NP volume fraction is increased, the height of these two peaks increases. The reason is that, while the structure of the chain at the length scale $r\lesssim 2 r_b$ remains almost unchanged when $\phi_N$ increases, the monomer density $\rho_m$ decreases, because the volume increases and the number of monomers is fixed. Since $g_{mm}(r)$ contains a factor $\rho_m^{-1}$ (see Eq.~\eqref{rdf}  in the main text), this results in an increase of this function for $r\lesssim 2 r_b$. Incidentally, this is why the radial distribution function of a single polymer, used to derive the function $p(r)$ (Sec.~\ref{polymer_structure} in the main text) is a more sensible quantity when studying chain conformation. 

Since $g_{mm}(r)$ is dominated by the two peaks at $r_b$ and $2r_b$ and shows only very small fluctuations around $g_{mm}(r) = 1$ at larger $r$, we plot in Figs.~\ref{rdf_mm}c-d the function $|g_{mm}(r)-1|$, hence allowing to detect more easily the structure at large $r$. For $\sigma_N=4$, Fig.~\ref{rdf_mm}d, $|g_{mm}(r)-1|$ shows very clearly at intermediate and high $\phi_N$ a long-range modulation with typical wavelength $\sigma_N$, due to the presence of the NPs. Notice that, since we have taken the absolute value, the wavelength must be calculated as the distance between the $n$th peak and the $(n+2)$th. For  $\sigma_N=2$, the presence of this modulation is less clear, because the size of the NPs is close to the monomer size and as a consequence the signal coming from the NPs cannot be distinguished well from the one coming from the monomers themselves.

\section{Systematic study of the different dynamical regimes} \label{sec:beta}

\begin{figure}[h]
\centering
\includegraphics[width=0.45\textwidth]{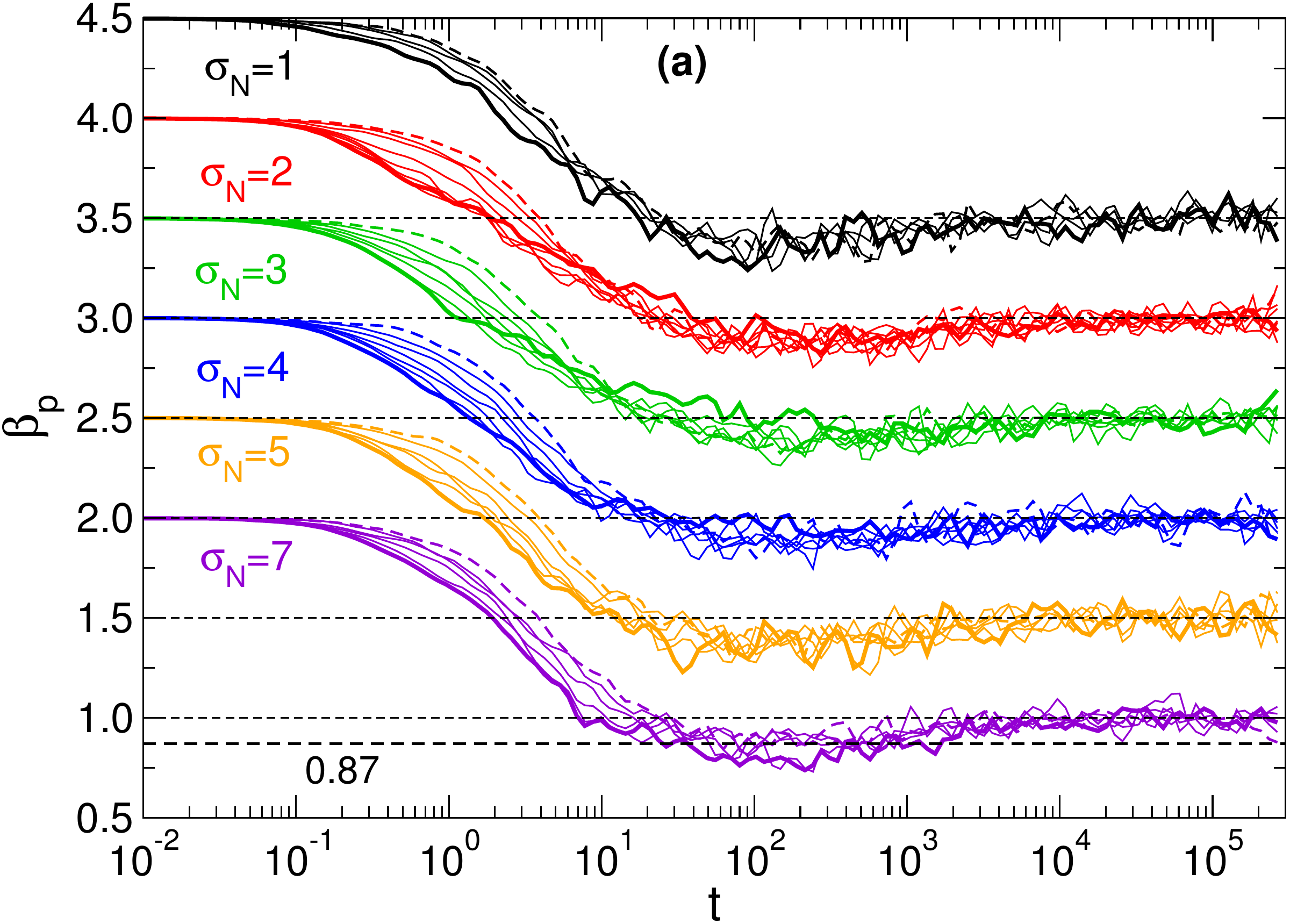}
\includegraphics[width=0.45\textwidth]{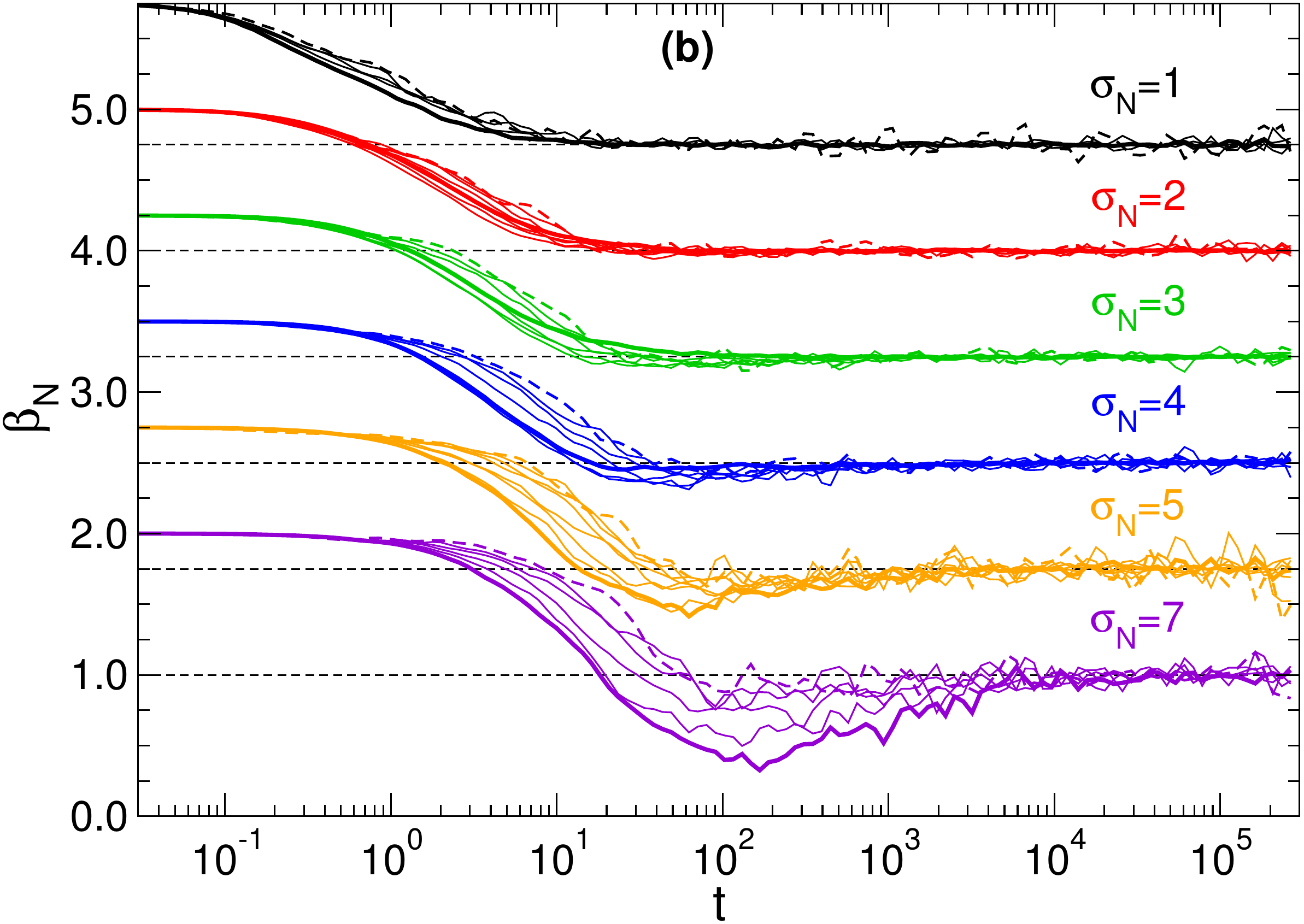}
\caption{Slope of the MSD in log-log scale, $\beta_\alpha$ (Eq.~\eqref{beta}), for the centers of mass of the polymers (a) and for the NPs (b). For the sake of clarity, for $\sigma_N<7$ every set of curves has been shifted with respect to the previous one, by $0.5$ in (a) and by $0.75$ in (b). Different curves in the same set correspond to different NP volume fractions $\phi_N$. The dashed curves correspond to the lowest $\phi_N$, while the thick continuous curves correspond to the highest $\phi_N$.}
\label{msd_slope}
\end{figure}

In Sec.~\ref{sec:poly_diff} of the main text we have studied the mean-squared displacement (MSD) ${\langle r^2(t) \rangle}$ of the NPs and of the centers of mass of the polymers, identifying three different regimes: a short-time ballistic regime, ${\langle r^2(t) \rangle \propto t^2}$, an intermediate-time subdiffusive regime, ${\langle r^2(t) \rangle \propto t^\beta}$, with ${\beta<1}$, and a long-time diffusive regime, ${\langle r^2(t) \rangle \propto t}$. In order to study in a more systematic way these regimes and the transition between them, we introduce the function

\begin{equation}
\beta_\alpha (t) = \frac{d \log \langle r^2_\alpha(t)\rangle }{d \log t}
\label{beta}
\end{equation}

\noindent where $\alpha=N,p$ respectively for the NPs and for the polymers. This quantity represents the slope of the MSD in a log-log scale, and it is a generalization of the above mentioned subdiffusive exponent $\beta$.

In Fig.~\ref{msd_slope}a we show $\beta_p$ for the simulated systems. The presence of the transient subdiffusive regime with $\beta_p\approx 0.87$, already mentioned in the main text, is clearly visible. The dynamics of the polymers becomes again diffusive ($\beta_p = 1$) after a time $t \approx 10^4$, which we identify with the relaxation time of the chain \cite{rubinstein2003polymer}. We can observe that $\beta_p$ is not much affected by the NP volume fraction $\phi_N$ and size $\sigma_N$.

In Fig.~\ref{msd_slope}b we report $\beta_N$. As already discussed in the main text, the dynamics of the NPs is more strongly influenced by the values of $\phi_N$ and $\sigma_N$ than that of the polymers. For $\sigma_N=5,7$, we can observe a subdiffusive transient ($\beta_N<1$) appearing very clearly at intermediate times. For $\sigma_N<5$, the effect is much smaller. 

\section{van Hove function} \label{sec:vanhove}

The van Hove function $G(\mathbf r, t) $ \cite{hansen1990theory} can be written as the sum of a self and a distinct part: $G(\mathbf r, t) = G_s(\mathbf r,t)+ G_d (\mathbf r, t)$, where

\begin{equation}
G_s (\mathbf r, t) = \frac 1 M \sum_{i=1}^M \langle \delta [\mathbf r - \mathbf r_i (t) + \mathbf r_i (0) ] \rangle 
\label{eq:g_self}
\end{equation}

\noindent and

\begin{equation}
G_d (\mathbf r, t)  = \frac 1 M \sum_{\substack{i=1\\ j\neq i}}^M \langle \delta [\mathbf r - \mathbf r_j (t) + \mathbf r_i (0) ] \rangle.
\label{eq:g_distinct}
\end{equation}

\noindent  The self part of the van Hove function represents the time-dependent spatial autocorrelation of a particle, while the distinct part represents the time-dependent spatial pair correlation. As usual, we will consider the spherical average of these two quantities: $G_s(r,t)$ and $G_d(r,t)$. The function $4 \pi r^2 G_s(r,t)$ represents the probability to find a particle at time $t$ a distance $r$ from its original position. We note that $G_s(\mathbf r, 0) = \delta (\mathbf r)$ and $G_d (\mathbf r, 0) = \rho g(\mathbf r)$, where $ g(\mathbf r)$ is the pair correlation function \cite{hansen1990theory}.

\noindent For both small and large values of $t$, the self part of the van Hove function is a Gaussian \cite{hansen1990theory}:

\begin{equation}
G_s(r,t) = \Gamma_s(r,t) = \left( \frac 3 {2 \pi \langle r^2(t) \rangle} \right)^{3/2} \exp \left( - \frac {3 r^2}{2 \langle r^2(t)\rangle}\right).
\label{eq:g_self_Gaussian}
\end{equation}

\noindent We can therefore define a rescaled self van Hove function which also preserves the probability with the following change of variables:

\begin{equation}
\begin{split}
r & \rightarrow r'=r \cdot \left(\frac 3 {2 \langle r^2(t)\rangle}\right)^{1/2} \\
G_s & \rightarrow G_s' =G_s \cdot \left(\frac 3 {2 \langle r^2(t)\rangle}\right)^{-3/2}
\end{split}
\label{eq:gs_change}
\end{equation}
 
\noindent If $G_s$ is Gaussian, the result of the transformation \eqref{eq:gs_change} is

\begin{equation}
\Gamma_s' (r') =  \pi^{-3/2} e^{-r'^2},
\label{eq:g_rescaled}
\end{equation}

\noindent i.e., the distribution is independent of time.

\begin{figure*}
\centering
\includegraphics[width=0.462 \textwidth]{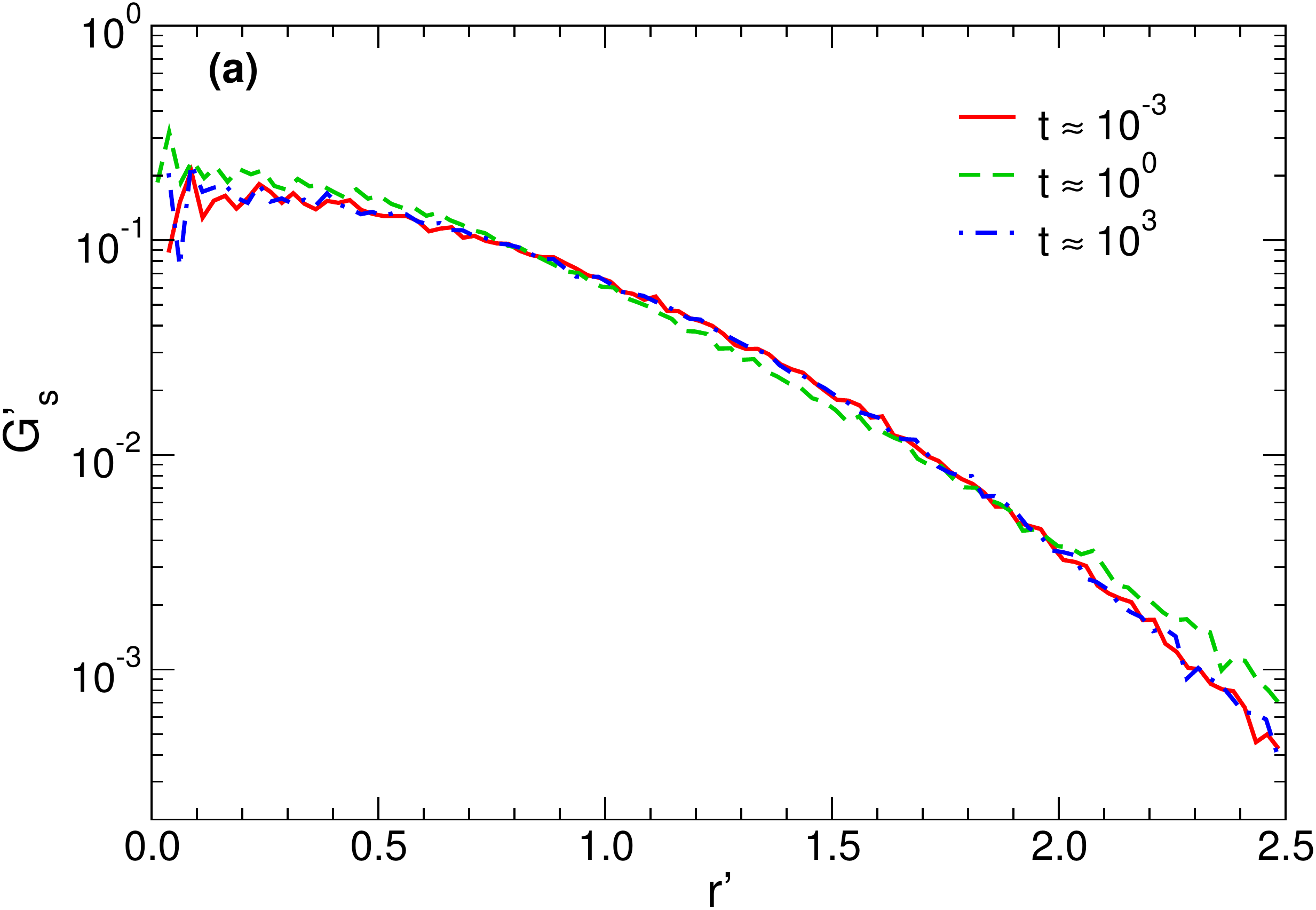}
\includegraphics[width=0.45 \textwidth]{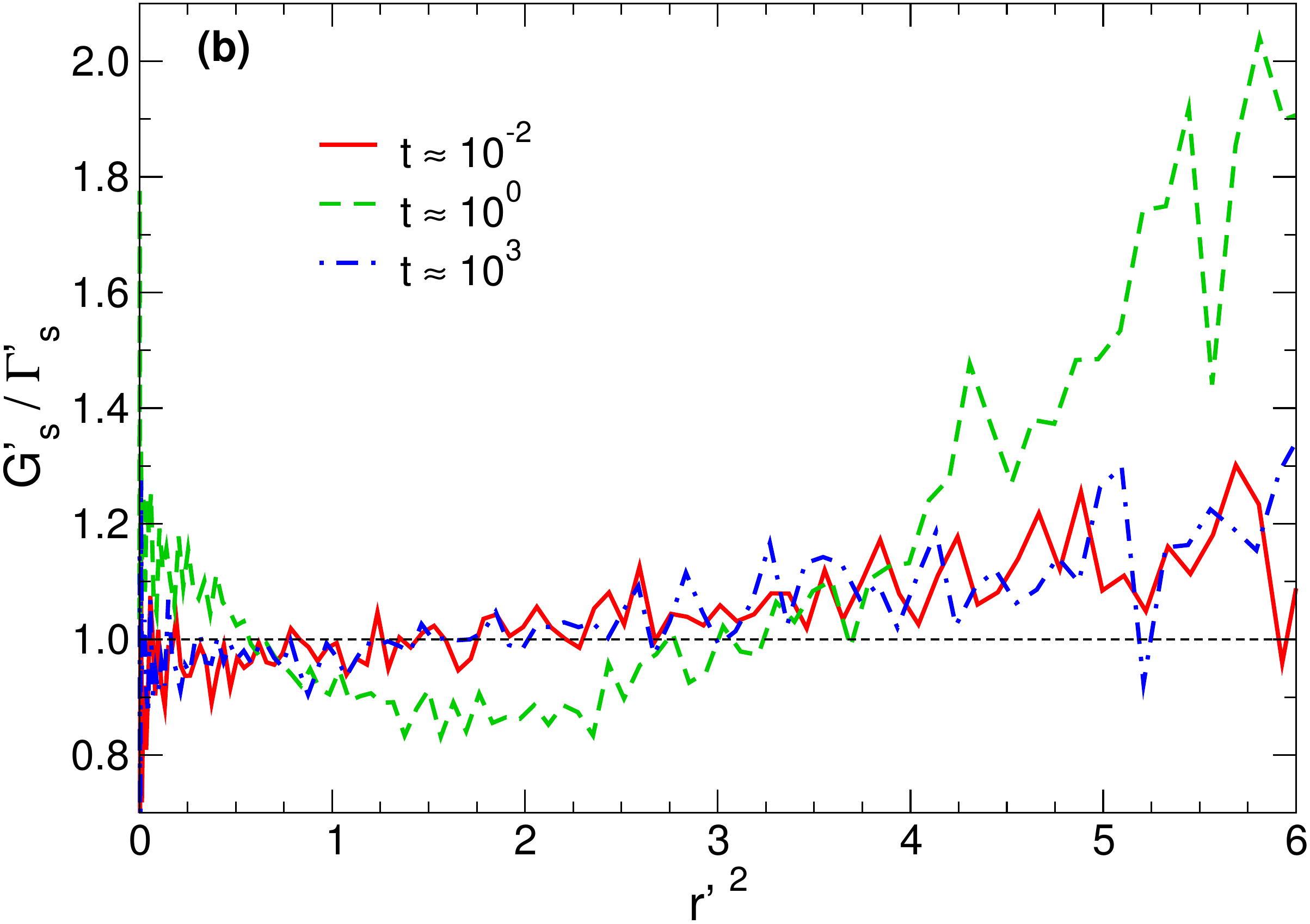}
\caption{(a) Rescaled self part of the van Hove function (Eqs.\eqref{eq:g_self}-\eqref{eq:gs_change}) for $\sigma_N=1, \phi_N=0.05$ (b) Ratio between the rescaled self part of the van Hove function  and the same quantity in the Gaussian approximation (Eq.~\eqref{eq:g_rescaled}), as a function of $r'^2$, for the same system. }
\label{coll_vhs}
\end{figure*}

In Fig.~\ref{coll_vhs}a, we show the rescaled van Hove function for the case $\sigma_N=1, \phi_N=0.05$ (but for other parameters we find the same qualitative behavior). In Fig.~\ref{coll_vhs}b, we report the ratio between the rescaled self van Hove function $G'_s(r,t)$ of the NPs and the same quantity in the Gaussian approximation $\Gamma'_s(r,t)$, for the same system. We observe that the shape of $G'_s(r,t)$ at short and long times is indeed very close to a Gaussian ($G'_s(r,t)=\Gamma'_s(r,t)$), and that the largest deviation from Gaussian behavior occurs when the dynamics of the NPs starts to be diffusive (in this case, at $t \approx 1$). These deviations are found to be most pronounced at large $r$, i.e., the NPs move a bit further than expected from a Gaussian approximation.

\noindent To better quantify how dissimilar $G_s(r,t)$ is from a Gaussian, it is customary to define a non-Gaussian parameter \cite{kob1995testing}, 

\begin{equation}
\alpha_2 (t)= \frac{3 \langle r^4(t) \rangle}{5 \langle r^2(t) \rangle^2} -1,
\label{eq:nongaussian}
\end{equation}

\noindent where 

\begin{equation}
\begin{split}
\langle r^n(t)\rangle & = \frac 1 M  \sum_i^M \langle |\mathbf r_i(t) - \mathbf r_i(0)|^n\rangle  \\
& = 4 \pi \int_0^\infty G_s(r,t) \ r^{2+n} dr.
\end{split}
\end{equation}

\noindent If $G_s(r,t)$ is Gaussian, Eq.~\eqref{eq:g_self_Gaussian}, we have $\alpha_2=0$. Therefore, high values of $|\alpha_2|$ indicate a significant non-Gaussian behavior.

\begin{figure*}[t]
\centering
\includegraphics[width=0.45 \textwidth]{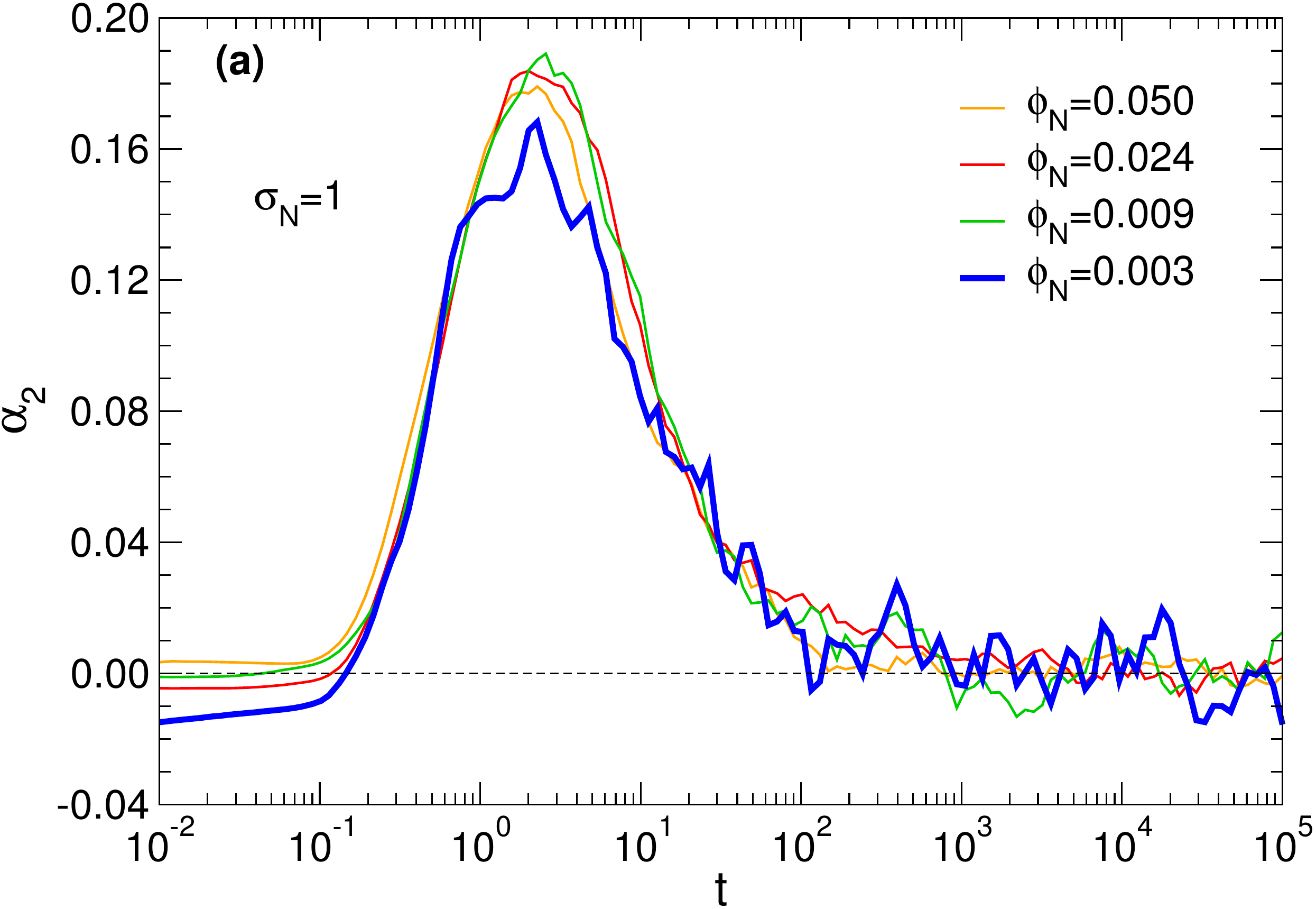}
\includegraphics[width=0.45 \textwidth]{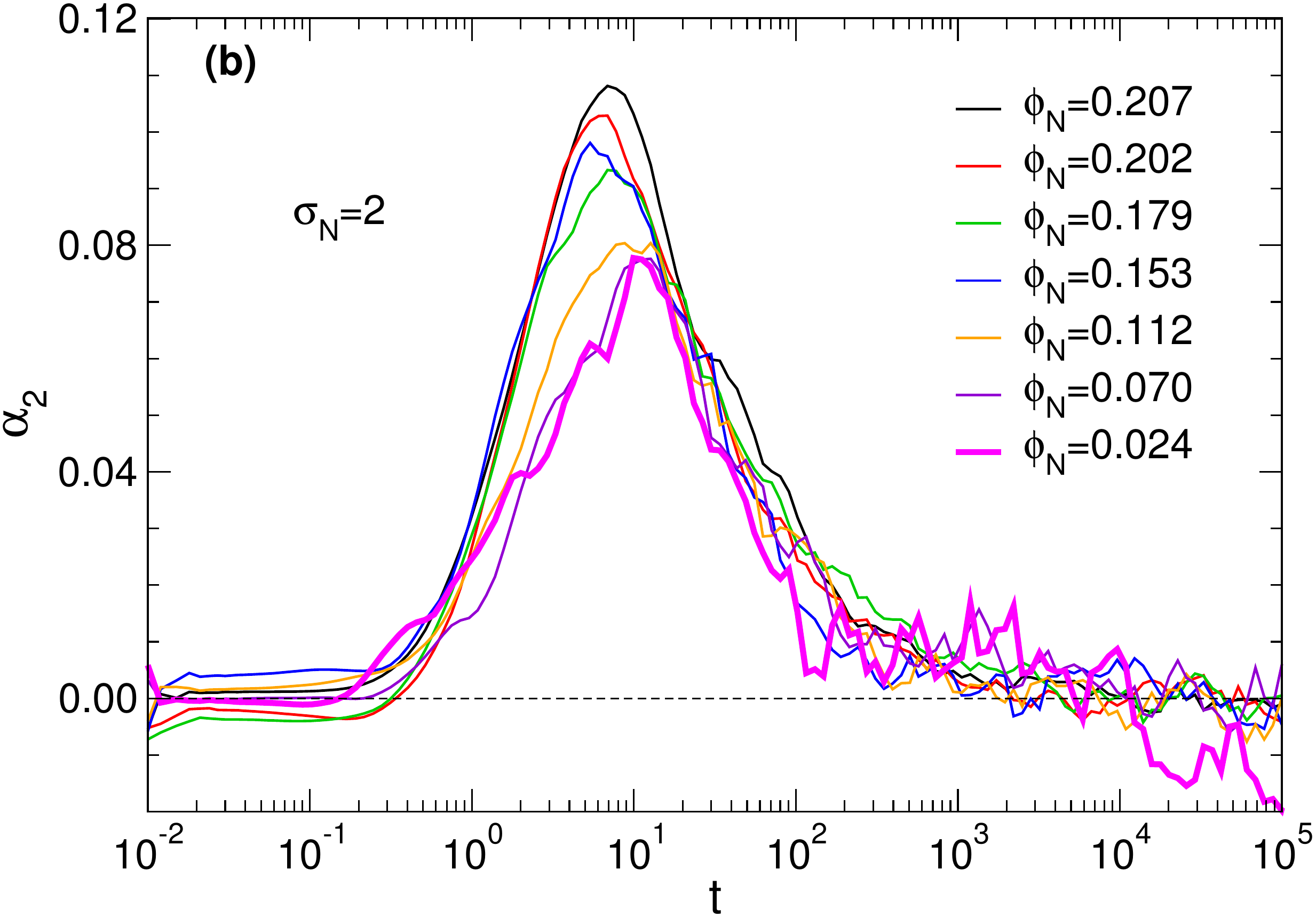}
\includegraphics[width=0.45 \textwidth]{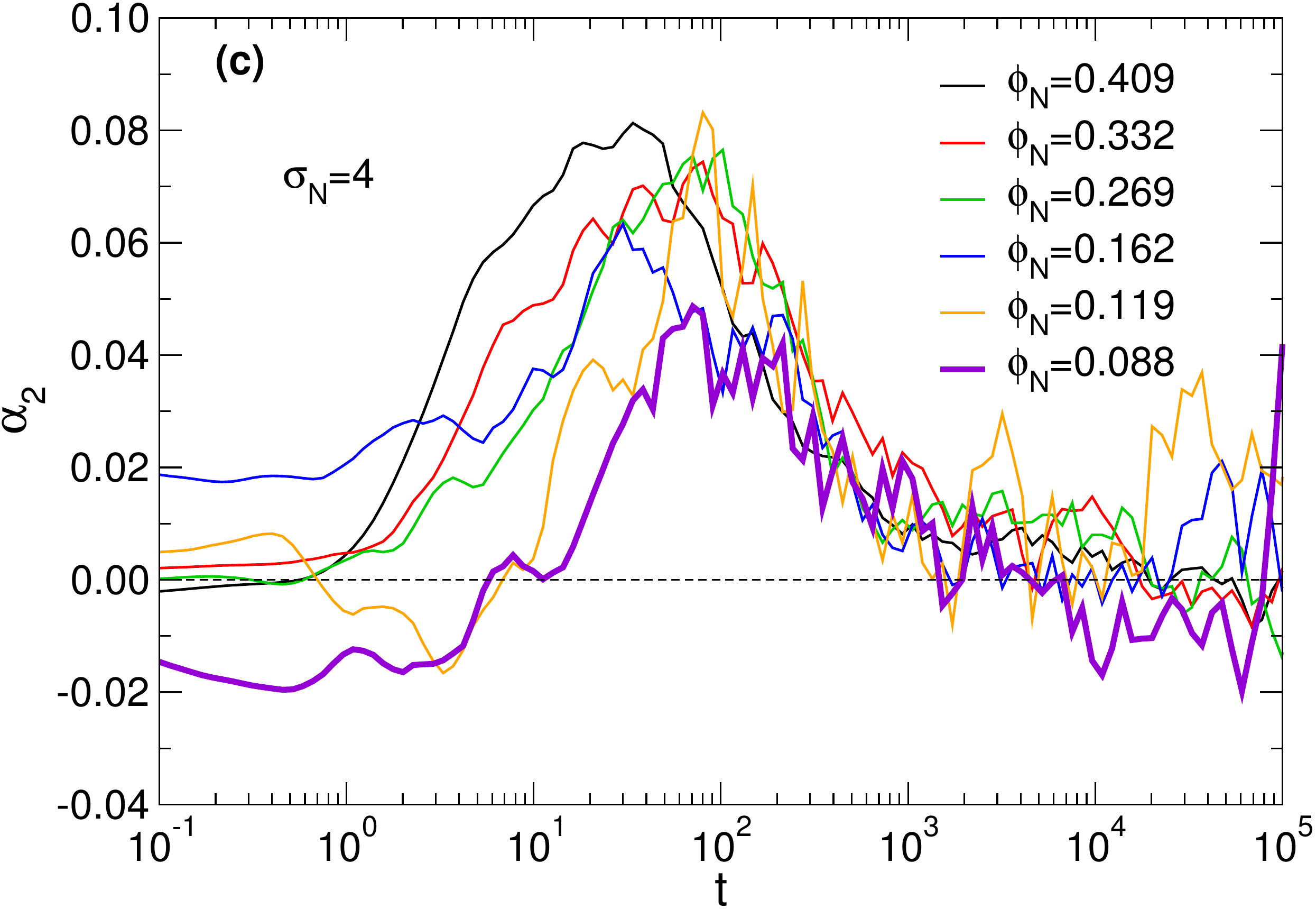}
\includegraphics[width=0.45 \textwidth]{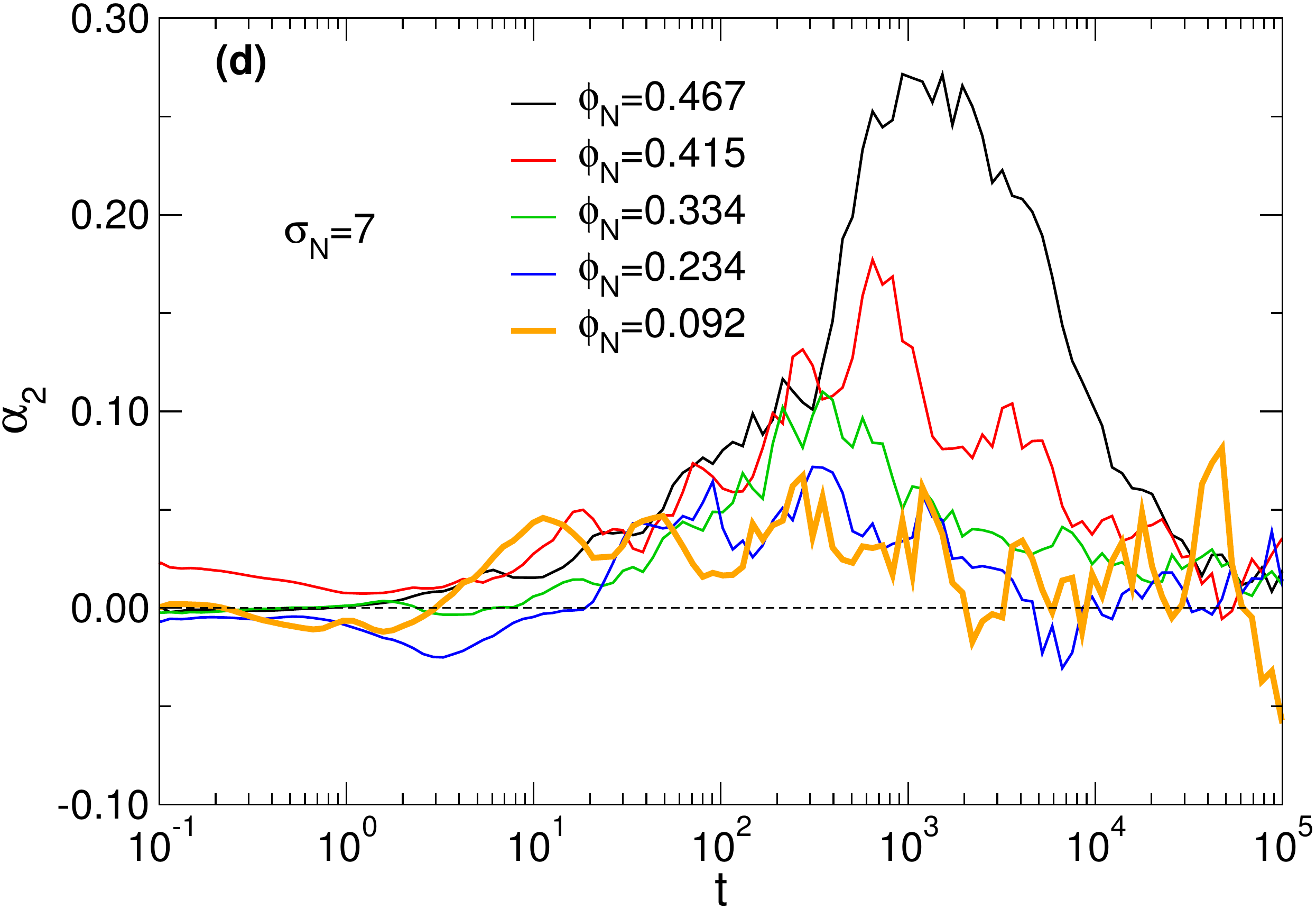}
\caption{(color online) Non-Gaussian parameter $\alpha_2$, Eq.~\eqref{eq:nongaussian}, for various NP diameters $\sigma_N$ and NP volume fractions.}
\label{nongaussian}
\end{figure*}

In Fig.~\ref{nongaussian}, we show $\alpha_2(t)$ of the NPs for several values of the NP diameter $\sigma_N$ and of the NP volume fraction $\phi_N$. The largest departure from Gaussian behavior happens when the dynamics of the NPs starts to be diffusive, in agreement with what is observed from $G'_s(r,t)$. Both at short and long times $\alpha_2 \approx 0$, as expected. We notice that the maximum deviation from Gaussian behavior (the maximum of the curves in Fig.~\ref{nongaussian}) becomes larger when $\phi_N$ is increased. This trend shows that the structure of the surrounding polymer mesh and the presence of nearby NPs both contribute to the non-Gaussian behavior. Moreover, increasing the NP size at fixed $\phi_N$ generally reduces the magnitude of $\alpha_2$. This is reasonable, since a large NP interacts with a large number of monomers and other NPs and thus \leftquote feels" an averaged interaction, which results in a reduction of the dynamical fluctuations and therefore of $\alpha_2$. One exception to this trend is $\sigma_N=7$ at high NP volume fraction (Fig.~\ref{nongaussian}d). The reason for this is likely that the system is approaching crystallization. Apart from the case $\sigma_N=7,\phi_N=0.467$, we always have $|\alpha_2| < 0.2$, and we can therefore state that the dynamics of the NPs is, to a good approximation, Gaussian. The non-Gaussian parameter of the polymer chains (not shown) always satisfies $|\alpha_2|<0.1$, therefore also the dynamics of the polymers is approximately Gaussian. We also mention that in this case no clear dependence of $\alpha_2$ on $\phi_N$ and $\sigma_N$ is observed.

\begin{figure*}[t]
\centering
\includegraphics[width=0.45 \textwidth]{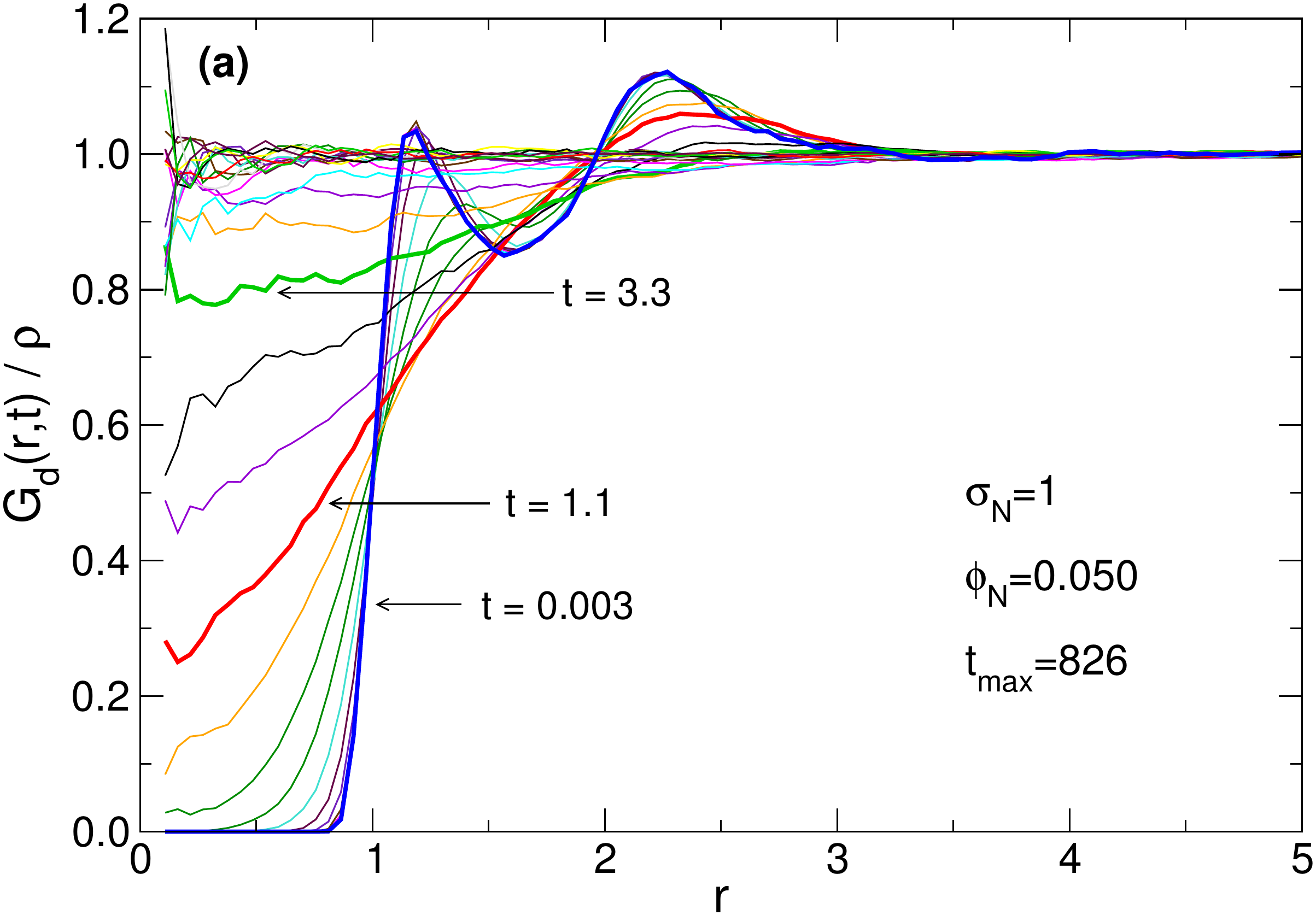}
\includegraphics[width=0.45 \textwidth]{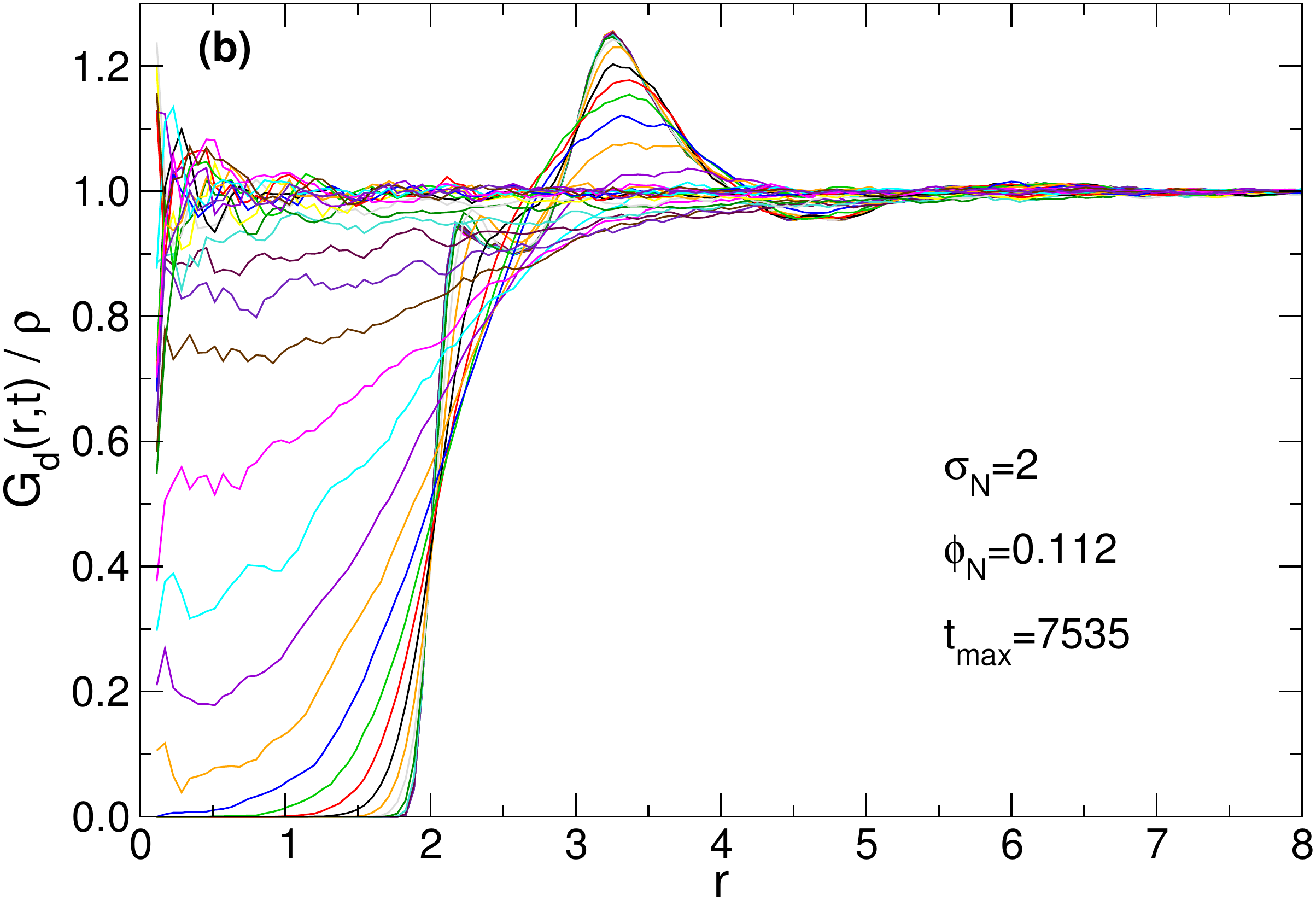}
\includegraphics[width=0.45 \textwidth]{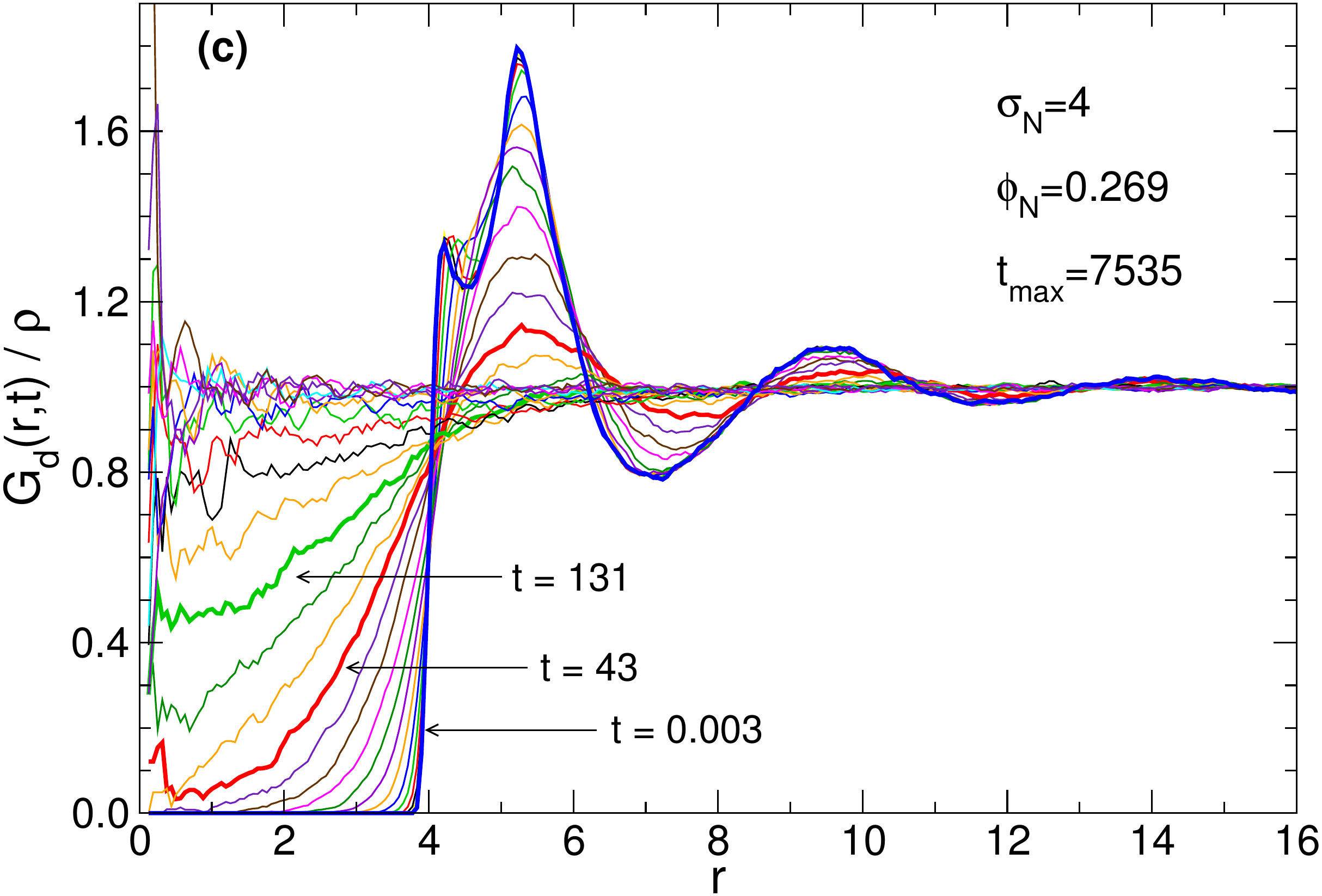}
\includegraphics[width=0.45 \textwidth]{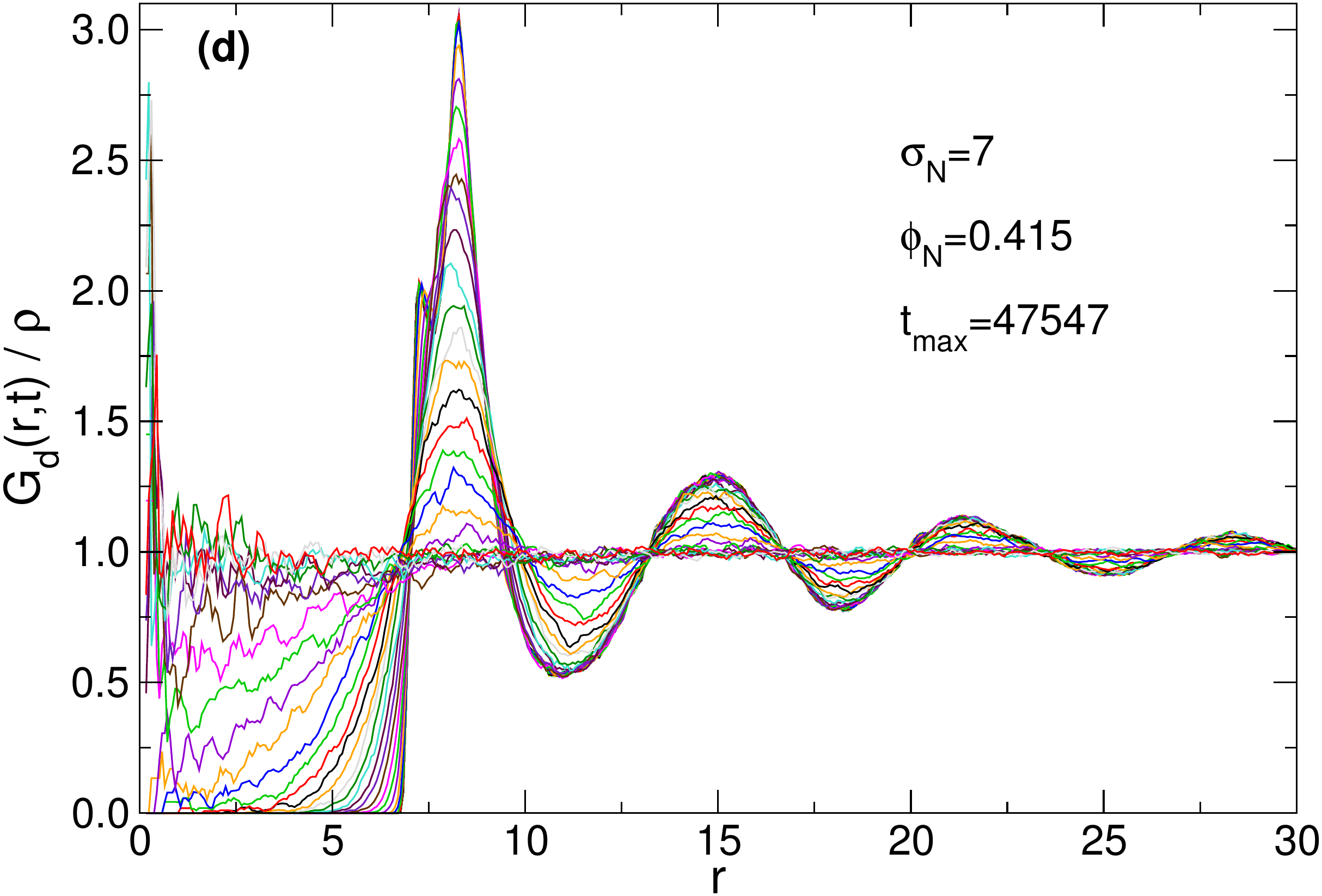}
\caption{(color online) Distinct part of the van Hove function, Eq.~\eqref{eq:g_distinct}, for some selected systems. The curves are equally spaced in logarithmic time ($t$). In (a) and (c), we have indicated some of the $t$ values associated to the curves. The minimum value of $t$ is for all the figures $t_\text{min}=\delta t=0.003$ (integration time step), while the maximum value, $t_\text{max}$, is reported in the legend.}
\label{coll_vhd}
\end{figure*}

Finally, in Fig.~\ref{coll_vhd} we show the distinct part of the van Hove function, $G_d(r,t)$, of the NPs for some selected systems. The relaxation happens in a way which is very similar to that observed in simple, non-supercooled liquids \cite{kob1995testing}, in that we observe that in all cases the correlation hole at $r=0$ is slowly filled as $t$ is increased. Since there is no evidence for the presence of a peak at $r=0$, we conclude that hopping dynamics is absent in the studied systems \cite{horbach1999static}.

\section{Polymer diffusion coefficient as a function of NP volume fraction}

In Sec.~\ref{sec:poly_diff} of the main text we have seen how, in the regime of good NP dispersion, the decrease of the polymer diffusion coefficient $D_p$ relative to the diffusion coefficient in the pure polymer solution $D_{p0}$ can be empirically described by the function $D_p = D_{p0} [1 - (\phi_N/\phi_{p0})^\alpha]$ (see Fig.~\ref{dchain}a in the main text). In Table \ref{tab:a_alpha}, we report the fit parameters for this empirical relation.   

\begin{table}[h]
\caption{Fit parameters for  $D_p = D_{p0} [1 - (\phi_N/\phi_{p0})^\alpha]$.}
\label{tab:a_alpha}
\begin{tabular}{@{\hspace{5em}} c @{\hspace{5em}} c @{\hspace{5em}} c @{\hspace{5em}}}
\toprule
$\sigma_N$ & $\phi_{p0}$ & $\alpha$\\
\colrule
1 & 0.104 & 0.763\\
2 & 0.231 & 0.768\\
3 & 0.364 & 0.812\\
4 & 0.471 & 0.851\\
5 & 0.519 & 1.006\\
7 & 0.594 & 1.150\\
\botrule
\end{tabular}
\end{table}

\section{Comparison of polymer and NP diffusivities}

\begin{figure*}
\centering
\includegraphics[width=0.45 \textwidth]{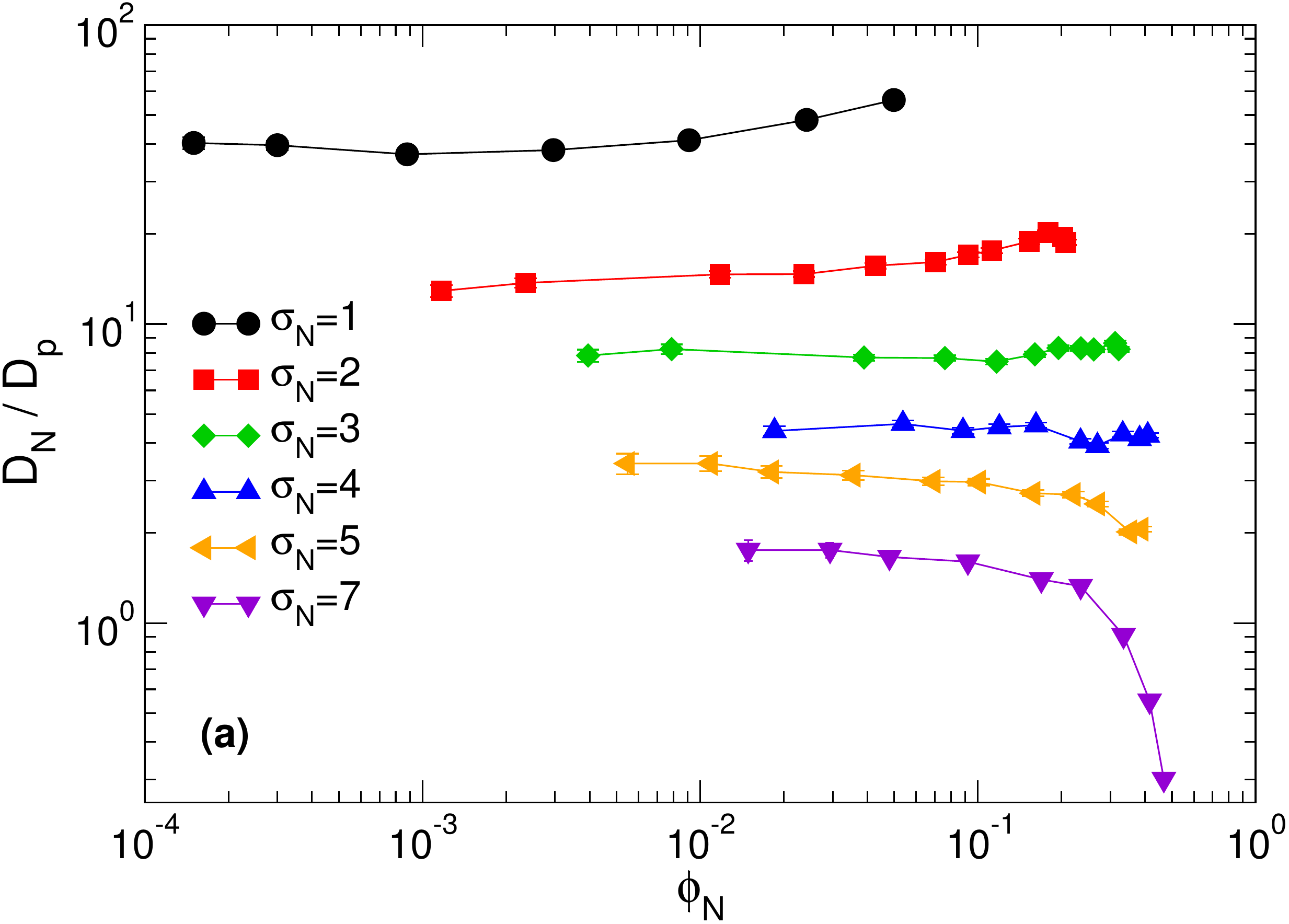}
\includegraphics[width=0.45 \textwidth]{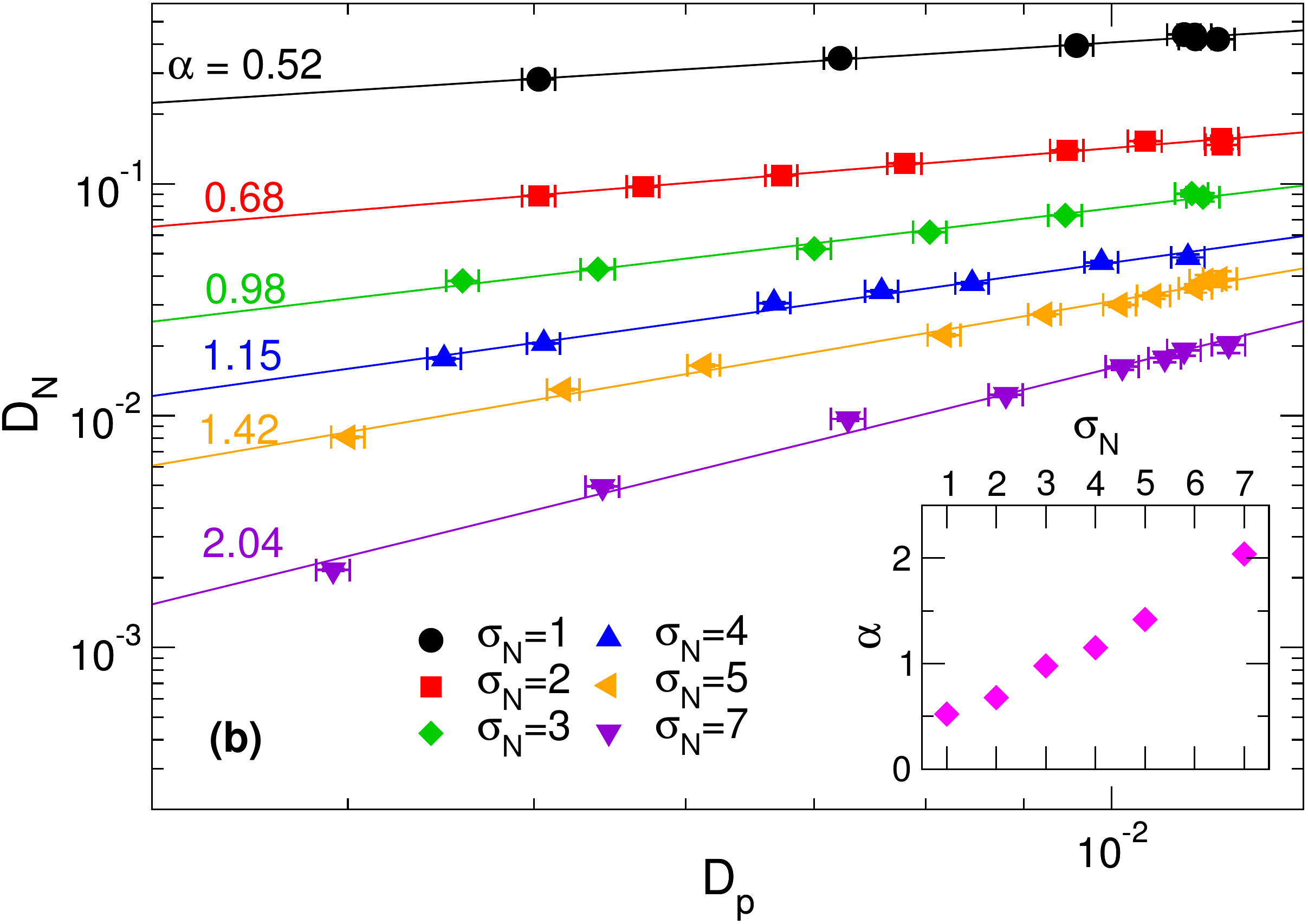}
\caption{(a) Ratio between the NP and polymer diffusion coefficients as a function of the NP volume fraction and for different values of the NP diameter $\sigma_N$. (b) NP diffusion coefficient $D_N$ versus the chain CM diffusion coefficient $D_p$. The correlation between $D_N$ and $D_p$ takes the form of a power law, $D_N\propto D_p^\alpha$,  where $\alpha$ increases with the NP diameter. \emph{Inset}: $\alpha$ as a function of the NP diameter. }
\label{d_comparison}
\end{figure*}

In Fig.~\ref{d_comparison}a, we compare the diffusion coefficient of the NPs, $D_N$, and of the CM of the polymers, $D_p$. One sees that, in almost all the systems we considered, $D_N>D_p$, i.e. the NPs move faster than the polymer chains. An exception is $\sigma_N=7$ at high densities; however, we know that at higher NP volume fraction the NPs form in this case a crystal and therefore $D_N$ becomes very small.

Figure~\ref{d_comparison}b shows the NP diffusion coefficient $D_N$ as a function of the polymer center of mass diffusion coefficient $D_p$. One can observe that there is a strong correlation between  $D_N$ and $D_p$, which can be empirically described via a power law: $D_N\propto D_p^\alpha$, where $\alpha$ increases with increasing NP diameter (inset of Fig.~\ref{d_comparison}b). This correlation suggests that there is a coupling between the long time diffusivities of NPs and the centers of mass of the polymers, as it was recently proposed by Chen \emph{et al.} \cite{chen2017coupling}.

\section{Additional details on the simulated systems}
In Tab. \ref{tab:data1} we report some additional details on the simulated systems. $N_N$: number of NPs; $\sigma_N$: NP diameter; $L$: side of the cubic simulation box; $\phi_m$: monomer volume fraction; $\phi_N$: NP volume fraction; $R_g$: average radius of gyration of the polymers; $h$: interparticle distance. 

\clearpage
\onecolumngrid

\begin{table}[b]
\centering
\caption{Details of the simulated systems. Systems in which there is poor NP dispersion (p.d.) are denoted by an asterisk in the last column.}
\label{tab:data1}
\begin{tabular}{l @{\hspace{2em}}  c @{\hspace{2em}} c @{\hspace{2em}} c @{\hspace{2em}} c @{\hspace{2em}} c @{\hspace{2em}} c @{\hspace{2em}} c @{\hspace{2em}} c @{\hspace{2em}} c}
\toprule
$N_N$ & $\sigma_N$ & $L$ & $\phi_m$ & $\phi_N$ & $D_N$ & $D_p$ & $R_g$ & $h$ & p.d.\\

0 & - & 56.32 & 0.1466 & 0.0000 & - & 0.0114 & 6.282 & - & -\\ 
\colrule
30000 & 1 & 57.08 & 0.1408 & 0.0845 & 0.2793 & 0.0046 & 6.4681 & 0.8980 & *\\ 
15000 & 1 & 54.03 & 0.1660 & 0.0498 & 0.2824 & 0.0050 & 6.2133 & 1.2760 & \\ 
7500 & 1 & 54.57 & 0.1611 & 0.0242 & 0.3477 & 0.0072 & 6.2358 & 1.9194 & \\ 
3000 & 1 & 55.64 & 0.1520 & 0.0091 & 0.3949 & 0.0096 & 6.2452 & 3.1421 & \\ 
1000 & 1 & 56.11 & 0.1482 & 0.0030 & 0.4221 & 0.0111 & 6.2688 & 5.0730 & \\ 
300 & 1 & 56.28 & 0.1468 & 0.0009 & 0.4199 & 0.0114 & 6.2535 & 8.1108 & \\ 
100 & 1 & 56.11 & 0.1482 & 0.0003 & 0.4380 & 0.0111 & 6.2621 & 12.0997 & \\ 
50 & 1 & 56.18 & 0.1476 & 0.0001 & 0.4400 & 0.0109 & 6.2639 & 15.5146 & \\ 
\colrule
20000 & 2 & 73.92 & 0.0648 & 0.2074 & 0.1174 & 0.0063 & 7.1314 & 0.7926 & *\\ 
15000 & 2 & 67.76 & 0.0842 & 0.2020 & 0.0964 & 0.0049 & 6.9420 & 0.8338 & *\\ 
10000 & 2 & 61.64 & 0.1118 & 0.1789 & 0.0805 & 0.0040 & 6.6559 & 0.9663 & *\\ 
7500 & 2 & 59.00 & 0.1275 & 0.1530 & 0.0807 & 0.0043 & 6.4846 & 1.1229 & *\\ 
5000 & 2 & 57.13 & 0.1404 & 0.1123 & 0.0886 & 0.0050 & 6.3783 & 1.4177 &\\ 
4000 & 2 & 56.61 & 0.1443 & 0.0923 & 0.0971 & 0.0057 & 6.3390 & 1.6361 &\\ 
3000 & 2 & 56.31 & 0.1466 & 0.0704 & 0.1085 & 0.0067 & 6.3218 & 1.9785 &\\ 
1800 & 2 & 56.05 & 0.1487 & 0.0428 & 0.1222 & 0.0078 & 6.2599 & 2.7735 &\\ 
1000 & 2 & 56.18 & 0.1477 & 0.0236 & 0.1392 & 0.0095 & 6.2879 & 3.9275 &\\ 
500 & 2 & 56.23 & 0.1472 & 0.0118 & 0.1526 & 0.0104 & 6.2609 & 5.5744 &\\ 
100 & 2 & 56.30 & 0.1467 & 0.0023 & 0.1566 & 0.0114 & 6.2632 & 11.2290 &\\ 
50 & 2 & 56.31 & 0.1466 & 0.0012 & 0.1473 & 0.0114 & 6.2605 & 14.5222 &\\ 
\colrule
20000 & 3 & 95.84 & 0.0297 & 0.3212 & 0.0589 & 0.0071 & 7.6393 & 0.6161 & *\\ 
10000 & 3 & 76.81 & 0.0578 & 0.3120 & 0.0410 & 0.0048 & 7.2650 & 0.6705 & *\\ 
5000 & 3 & 64.66 & 0.0968 & 0.2614 & 0.0329 & 0.0040 & 6.7890 & 0.9414 & *\\ 
4000 & 3 & 62.21 & 0.1087 & 0.2348 & 0.0334 & 0.0040 & 6.5920 & 1.0718 & *\\ 
3000 & 3 & 60.14 & 0.1204 & 0.1950 & 0.0382 & 0.0046 & 6.5102 & 1.3057 & \\ 
2300 & 3 & 58.75 & 0.1291 & 0.1604 & 0.0428 & 0.0054 & 6.4311 & 1.5429 & \\ 
1600 & 3 & 57.87 & 0.1351 & 0.1167 & 0.0524 & 0.0070 & 6.3742 & 2.0149 & \\ 
1000 & 3 & 57.07 & 0.1408 & 0.0761 & 0.0618 & 0.0080 & 6.3179 & 2.8246 & \\ 
500 & 3 & 56.64 & 0.1441 & 0.0389 & 0.0731 & 0.0095 & 6.2917 & 4.4527 & \\ 
100 & 3 & 56.37 & 0.1461 & 0.0079 & 0.0908 & 0.0110 & 6.2921 & 10.0294 & \\ 
50 & 3 & 56.35 & 0.1463 & 0.0040 & 0.0875 & 0.0112 & 6.2624 & 13.5295 & \\ 
\colrule
10000 & 4 & 93.57 & 0.0320 & 0.4090 & 0.0200 & 0.0047 & 7.7135 & 0.5071 & *\\ 
5000 & 4 & 75.96 & 0.0597 & 0.3823 & 0.0150 & 0.0036 & 7.2264 & 0.6233 & *\\ 
3000 & 4 & 67.13 & 0.0865 & 0.3323 & 0.0145 & 0.0034 & 6.8100 & 0.8860 & *\\ 
2000 & 4 & 62.90 & 0.1052 & 0.2694 & 0.0176 & 0.0045 & 6.6007 & 1.2091 & \\ 
1600 & 4 & 61.18 & 0.1143 & 0.2342 & 0.0205 & 0.0051 & 6.5100 & 1.3929 & \\ 
1000 & 4 & 59.15 & 0.1265 & 0.1619 & 0.0306 & 0.0067 & 6.3726 & 1.9771 & \\ 
700 & 4 & 58.12 & 0.1334 & 0.1195 & 0.0343 & 0.0076 & 6.3358 & 2.6185 & \\ 
500 & 4 & 57.45 & 0.1381 & 0.0884 & 0.0372 & 0.0085 & 6.3197 & 3.3626 & \\ 
300 & 4 & 57.21 & 0.1398 & 0.0537 & 0.0458 & 0.0099 & 6.3056 & 4.8379 & \\ 
100 & 4 & 56.56 & 0.1447 & 0.0185 & 0.0481 & 0.0110 & 6.2800 & 8.8842 & \\ 
\colrule
2000 & 5 & 69.65 & 0.0775 & 0.3874 & 0.0071 & 0.0034 & 6.9309 & 0.8357 & *\\ 
1600 & 5 & 66.72 & 0.0882 & 0.3526 & 0.0081 & 0.0040 & 6.7538 & 1.0106 & \\ 
1000 & 5 & 62.36 & 0.1080 & 0.2699 & 0.0130 & 0.0052 & 6.5122 & 1.4406 & \\ 
750 & 5 & 60.60 & 0.1176 & 0.2206 & 0.0165 & 0.0061 & 6.4122 & 1.7888 & \\ 
500 & 5 & 59.15 & 0.1265 & 0.1581 & 0.0222 & 0.0082 & 6.3587 & 2.5069 & \\ 
300 & 5 & 57.99 & 0.1342 & 0.1007 & 0.0273 & 0.0092 & 6.3333 & 3.8062 & \\ 
200 & 5 & 57.44 & 0.1381 & 0.0691 & 0.0301 & 0.0101 & 6.3126 & 5.1301 & \\ 
100 & 5 & 56.87 & 0.1423 & 0.0356 & 0.0329 & 0.0105 & 6.2973 & 7.7382 & \\ 
50 & 5 & 56.59 & 0.1444 & 0.0181 & 0.0354 & 0.0111 & 6.2820 & 11.2616 & \\ 
30 & 5 & 56.48 & 0.1453 & 0.0109 & 0.0383 & 0.0112 & 6.2879 & 14.4058 & \\ 
15 & 5 & 56.40 & 0.1459 & 0.0055 & 0.0389 & 0.0114 & 6.2622 & 19.9009 & \\
\colrule
1000 & 7 & 72.71 & 0.0681 & 0.4672 & 0.0009 & 0.0029 & 6.9195 & 0.7881 & *\\ 
750 & 7 & 68.71 & 0.0807 & 0.4153 & 0.0022 & 0.0039 & 6.7000 & 1.0633 & \\ 
500 & 7 & 64.51 & 0.0975 & 0.3345 & 0.0049 & 0.0054 & 6.5015 & 1.5040 & \\ 
300 & 7 & 61.26 & 0.1139 & 0.2344 & 0.0097 & 0.0073 & 6.3621 & 2.2690 & \\ 
200 & 7 & 59.65 & 0.1234 & 0.1692 & 0.0123 & 0.0088 & 6.3070 & 3.2795 & \\ 
100 & 7 & 58.00 & 0.1341 & 0.0920 & 0.0162 & 0.0101 & 6.2774 & 5.7758 & \\ 
50 & 7 & 57.17 & 0.1401 & 0.0481 & 0.0177 & 0.0107 & 6.2980 & 9.1238 & \\ 
30 & 7 & 56.83 & 0.1426 & 0.0294 & 0.0191 & 0.0109 & 6.2728 & 12.4074 & \\ 
15 & 7 & 56.57 & 0.1446 & 0.0149 & 0.0202 & 0.0115 & 6.2612 & 17.5854 & \\
\botrule
\end{tabular}
\end{table}
\clearpage

\twocolumngrid
\bibliography{biblio_abbrev}

\end{document}